\documentclass[journal]{IEEEtran}

\ifCLASSINFOpdf

\else
\fi

\usepackage{amsmath,amsfonts}
\usepackage{algorithmic}
\usepackage{algorithm}
\usepackage{array}
\usepackage{textcomp}
\usepackage{stfloats}
\usepackage{url}
\usepackage{verbatim}
\usepackage{graphicx}
\usepackage{subfigure}
\usepackage{subfig}
\usepackage{amsmath}
\usepackage{color}
\usepackage{booktabs}
\usepackage{enumitem}
\usepackage{nicefrac}     
\usepackage{microtype}  
\usepackage{lipsum}	
\usepackage{hhline}
\usepackage{graphicx} 
\usepackage{amsmath,amssymb}
\usepackage{graphicx}
\usepackage{float}
\usepackage{array}
\usepackage[export]{adjustbox}
\usepackage{xcolor,colortbl}
\usepackage{multirow}
\usepackage{mathtools}
\usepackage{longtable}
\usepackage{threeparttable}
\usepackage{wrapfig}
\usepackage{xspace}
\usepackage{url}
\usepackage{array,multirow,graphicx}
\usepackage{float}
\usepackage{amsmath,amssymb,amsfonts}
\usepackage{algorithmic}
\usepackage{graphicx}
\usepackage{textcomp}
\usepackage{xcolor}
\usepackage{algorithm}

\usepackage{amsmath,amssymb}
\usepackage{url}

\usepackage[subtle,tracking=normal]{savetrees}

\newcommand{\sysname}{ERUDITE\xspace}

\newcommand{\eg}{{\it e.g.,}\xspace}

\newcommand{\etal}{{\it et al.}\xspace}

\definecolor{blond}{rgb}{0.98, 0.94, 0.75}
\definecolor{grannysmithapple}{rgb}{0.66, 0.89, 0.63}
\definecolor{lightgray}{rgb}{0.83, 0.83, 0.83}

\newcommand{\cellone}{\cellcolor{blond}$1$}  
\newcommand{\cellzero}{\cellcolor{grannysmithapple}$0$}  
\newcommand{\cellgrey}{\cellcolor{lightgray}}

\begin{document}

\title{\sysname: Human-in-the-Loop IoT for an Adaptive Personalized Learning System }

\author{Mojtaba Taherisadr, Mohammad Abdullah Al Faruque~\IEEEmembership{Senior,~IEEE,} Salma Elmalaki 
}


\author{Mojtaba~Taherisadr,
        Mohammad Abdullah Al Faruque~\IEEEmembership{Senior,~IEEE,}
       Salma Elmalaki 
       
\thanks{M. Taherisadr, MA. Al Faruque, and S. Elmalaki are with the Department
of Electrical Engineering and Computer Science, University of California Irvine, Irvine,
CA, 92697 USA. e-mail: \{taherisa,alfaruqu,selmalak\}@uci.edu} %

\markboth{IEEE Internet of Things Journal, March~2023}%
{Taherisadr \MakeLowercase{\textit{et al.}}: adaPARL: Human-in-the-Loop IoT for an Adaptive Personalized Learning System }}

\maketitle

\begin{abstract}
Thanks to the rapid growth in wearable technologies and advancements in machine learning, monitoring complex human contexts becomes feasible, paving the way to develop human-in-the-loop IoT systems that naturally evolve to adapt to the human and environment state autonomously. Nevertheless, a central challenge in designing many of these IoT systems arises from the requirement to infer the human mental state, such as intention, stress, cognition load, or learning ability.  
While different human contexts can be inferred from the fusion of different sensor modalities that can correlate to a particular mental state, the human brain provides a richer sensor modality that gives us more insights into the required human context. This paper proposes \sysname, a human-in-the-loop IoT system for the learning environment that exploits recent wearable neurotechnology to decode brain signals. Through insights from concept learning theory, \sysname can infer the human state of learning and understand when human learning increases or declines. By quantifying human learning as an input sensory signal, \sysname can provide adequate personalized feedback to humans in a learning environment to enhance their learning experience. \sysname is evaluated across $15$  participants and showed that by using the brain signals as a sensor modality to infer the human learning state and providing personalized adaptation to the learning environment, the participants' learning performance increased on average by $26\%$. Furthermore, to evaluate \sysname practicality and scalability, we showed that \sysname can be deployed on an edge-based prototype consuming $75$~mW power on average with $100$~MB memory footprint. 
\end{abstract}


\IEEEpeerreviewmaketitle

\section{Introduction}\label{sec:intro}

The disruption of face-to-face exchanges and the urgent digital transformation of our everyday activities, abide the COVID-19 hit have spurred us to shift many working paradigms to incorporate the new reality of the online and remote collaborative environment. In particular, many sectors, including education, workforce training, and the healthcare systems, had to integrate new technology to continue their services even with remote interactions~\cite{taiwo2020smart,mantena2021strengthening}. Nevertheless, this forced digital transformation of many sectors is expected to stay even in the post-COVID-19 era. In particular, in the education and workforce training sector, the e-learning market worldwide is forecast to surpass $243$ billion U.S. dollars~\cite{elearningStatista}. This paradigm shift needs a new approach to rethink the future learning environment to fit this new reality for remote and online learning environments without sacrificing the efficiency of the learning outcomes~\cite{awais2020lstm}. Moreover, even with regular in-person interaction, everyday life and complex work and learning environments make it difficult for humans to show extended focus, engagement, or attention on a specific learning task, directly affecting their performance in everyday tasks~\cite{leroy2009so}.
Hence, we need to rethink the future of the learning environment to incorporate the human learning state (focus, engagement, attention) as an integral part of its design. 

Measuring and evaluating the learning outcomes is a cornerstone in the design of any learning environment. For example, traditional measures can take different forms in an educational environment, such as quizzes, exam scores, and teacher evaluation~\cite{hong2017review}. Different approaches have been proposed in the workforce environment, such as the learning-transfer evaluation model (LTEM), to measure if employees learned and whether they can or will perform the desired behavior on the job~\cite{LTEM}.      

With the recent advances in IoT, machine learning models, and context-aware computing~\cite{elmalaki2015caredroid}, we envision that the future of the learning environment has to be redesigned to incorporate a \emph{real-time} measurement of the human learning state with \emph{real-time} assessment and provide adequate feedback to the human and the instructor (whether remote or online or in-person learning) to improve the learning outcomes~\cite{taherisadr2021future,zhang2018internet}. 
Accordingly, in this paper, we ask the following questions: \textbf{Q1. Using IoT technologies (wearable sensors and edge-cloud computing), can we understand the learning process and infer the human learning state at run-time? Q2. Can we integrate this learning state signal as a sensor modality in the IoT-based learning environment to make a real-time adaptation and improve the learning state of a human? Q3. Can we measure the effectiveness of these adaptations and tune them to provide a more personalized learning experience?} 

These questions have multifold challenges. A fundamental challenge is understanding how the human brain learns a new concept. Indeed, recent advances in neuroscience have opened the gate to unveil fundamental processes in the human brain, such as the ability to generate emotions, memories, and actions~\cite{zheng2019multiplexing}.  
In particular, Electroencephalography (EEG) and functional magnetic resonance imaging (fMRI) were the two primary approaches used to record and measure the human brain state under learning processes~\cite{berka2007eeg, yoo2012brain}. While these approaches provide necessary insights into the human brain, the main drawbacks are that they must be conducted in a laboratory setting, are computationally costly, and are inadequate in a natural learning environment, such as education or workforce training where IoT-based systems are more adequate. Hence, other approaches used real-time signals from physiological sensors available in commodity wearables to correlate learning outcomes with different physiological states and hence can use machine learning models to predict the learning experience~\cite{giannakos2020fitbit, gao2020n, wilson2021objective}. However, these approaches measure the learning indirectly as a correlation to other physiological states, such as engagement, attention, stress, focus, or comparatively under different stimuli. Moreover, these approaches rarely close the loop by providing real-time adaptation to improve human learning.

Nevertheless, another challenge to answering the aforementioned questions is that even if we can quantify the human learning state outside the laboratory settings using IoT devices (such as wearables) and independently from other physiological states, learning varies from time to time and from person to person. The state of learning or memory formation varies over time for any individual, partly due to moment-to-moment fluctuation of brain state. Moreover, varying levels of alertness, attention, and motivation likely contribute to fluctuating brain states for learning~\cite{yoo2012brain}. In addition, recent research in the human-in-the-loop IoT systems has shown that humans have intrinsic inter-and intra- variation. Hence, we can not have one model-fits-all approach in designing real-time adaptations to improve the human learning~\cite{elmalaki2021fair, elmalaki2021towards, elmalaki2018sentio, elmalaki2022maconauto, taheri2023}.

In this paper, we propose \textbf{\sysname}, an IoT-based system to quantify the human learning state using a non-invasive wearable technology by decoding the brain signals under different learning approaches. We then use the human learning state signal as another sensor modality in the IoT-based learning environment to design human-in-the-loop adaptations to enhance human learning.

\paragraph*{\textbf{Contribution}}
Our contributions in this paper can be summarized as follows: 
\begin{itemize}[leftmargin=*, topsep=0pt,itemsep=0pt,parsep=0pt,partopsep=0pt]
    \item Exploring two concept-learning-based experiments to determine the human learning state as a sensor modality (\textbf{Q1}). 
    
    \item Exploiting recent IoT technology (wearable EEG devices) to find the correlations between EEG signal dynamics and human learning states  (\textbf{Q2}).    
    
    \item Designing \sysname, an IoT system that integrates the wearable EEG device as a sensor modality with an edge-based prototype to provide real-time adaptation for the learning environment and online feedback to the human to improve the learning  (\textbf{Q3}).  

\end{itemize}

\section{Background and Related Work}\label{sec:related}
In this section, we briefly discuss some background related to the philosophy of learning that we will use as a basis to design \textbf{\sysname}. Afterward, we list recent research approaches in quantifying learning using brain signals. 

\subsection{Concept Learning Theory} 
Concept learning is defined by Bruner~\cite{ozdem2020discovery} as ``the search for and listing of attributes that can be used to distinguish exemplars from non-exemplars of various categories.'' Thus, concept learning is a strategy that requires a learner to compare and contrast groups or categories that contain concept-relevant features with groups or categories that do not contain concept-relevant features. Concept learning must be distinguished from ``learning by reciting'' something from memory (recall) or ``discriminating between two things'' by refining concepts (discrimination).
In this paper, we borrow from philosophy and psychology literature to exploit two types of concept learning, namely ($1$) Rule-based concept learning, and ($2$) Explanation-based concept learning. We briefly explain them below.

\noindent \textbf{(1) Rule-based Concept Learning} has initiated with cognitive psychology and early learning models that could be implemented with a computer language using conditional statements such as if: then, which produces rules. Computer models utilize training data and a rule-based approach as input, which results from a rule-based learner to produce a more accurate data model. A rule-based system can be used in concept learning when the stimuli are confusable instead of simple. When rules are used in learning, decisions are made based on properties alone and rely on simple criteria that do not require a lot of memory~\cite{goodman2008rational}. For instance, a radiologist uses the rule of differences in brightness to categorize the X-ray images. Using the rule, the radiologist decides which X-ray image belongs to what category.  
    
\noindent \textbf{(2) Explanation-based Concept Learning} suggests that a new concept is acquired by experiencing examples of it and forming a basic outline. In particular, by observing or receiving the qualities of a thing, the mind forms a concept that possesses those qualities~\cite{dejong1986explanation}. The revised explanation-based learning model integrates four mental processes – generalization, chunking, operationalization, and analogy. \textit{Generalization} is the process by which the characteristics of a concept that are fundamental to it are recognized and labeled. For example, birds have feathers and wings. Hence, anything with feathers and wings will be identified as a `bird.' When information is grouped mentally, whether, by similarity or relatedness, the group is called a \textit{chunking}. An example of chunking occurs in phone numbers; a phone number sequence of $4-7-1-1-3-2-4$ would be chunked into $471-1324$ to remember them more manageably. 

A concept is \textit{operationalized} when the mind can turn abstract concepts into measurable observations. Operationalization is a process of defining the measurement of a phenomenon that is not directly measurable, though its existence is inferred by other phenomena. For example, in medicine, the phenomenon of health might be operationalized by one or more indicators like body mass index or tobacco smoking. \textit{Analogy} is the recognition of similarities among potential examples. For example, an atom is like a miniature solar system, and a database is like a filing cabinet.  

In this paper, we will borrow from the rule-based concept learning theory due to its deterministic approach in identifying if a rule is learned or not to get insights into how to measure the human learning state. Afterward, we will use these insights to analyze the explanation-based concept learning since its concept is more applied in learning environments.

\subsection{Learning Quantification}
Two bodies of work use brain signals to quantify learning.  
One approach uses functional magnetic resonance imaging (fMRI), and the other is Electroencephalography signals (EEG). We summarize them below.  

\noindent \textbf{(1) fMRI}: Yoo \etal{}~\cite{yoo2012brain} designed two experiments to measure the learning state. One experiment examined whether such brain states that were good or bad for learning concepts could be identified. In another experiment, they examined whether such brain states, detected by real-time fMRI on a moment-moment basis, could be used to trigger concept presentation with the hypothesis that concepts triggered by good brain states would be better remembered than concepts triggered by bad brain states. Such a finding would offer evidence of the ability to monitor online whether a person is optimally prepared to learn and the ability to use fMRI to causally enhance human learning in the sense that the real-time fMRI-measured brain state caused the concept presentation. Toni \etal{}~\cite{toni1998time} exploited fMRI to measure changes in blood oxygen level-dependent (BOLD) signals throughout learning. 
Their analysis found linear and nonlinear changes of BOLD signal over time in prefrontal, premotor, and parietal cortex areas.

\noindent \textbf{(2) EEG}: has been employed in numerous studies to monitor different human states such as brain diagnostics~\cite{idrees2022edge},  sleep~\cite{mandekar2021earable}, emotion~\cite{nie2020spiders},  attention~\cite{kosmyna2018attentivu}, and learning state~\cite{shadiev2017enhancing, babini2020physiological}. In the field of education and learning, Xu \etal{}~\cite{xu2018review} reviewed the usage of portable EEG technologies and categorized them into seven research topics including reading context, presentation patterns of learning materials, interactive behavior, edutainment, e-learning, motor skill acquisition, and promoting learning performance. In the context of learning, Wang \etal{}~\cite{wang2014exploratory} utilized the Neurosky Mindset headset to gauge the concentration of learners while engaged in computer-based instructional learning. In this study, participants underwent three lessons with varying difficulty levels, including easy, medium, and difficult. Chen \etal{}~\cite{chen2017assessing} also assessed the students' focus and concentration using the NeuroSky MindSet headset while engaged in online synchronous instruction and learning tasks. Ghergulescu \etal{}~\cite{ghergulescu2016totcompute} utilized the Emotiv EPOC neuroheadset to gauge students' level of involvement in game-based e-learning. In addition, the NeuroSky MindWave headset was used to detect the attentiveness of students in the context of e-learning~\cite{chen2018effects,lin2016construction}.

In particular, Babini \etal{}~\cite{babini2020physiological}, using an EEG device, designed an experiment to analyze the effect of virtual reality (VR) conditions on students' learning ability and physiological state. Their results showed that students' learning ability was increased in the three-dimensional condition compared to that in the two-dimensional condition. In this study, they compared the engagement of the brain in different presentation modes to correlate it to the learning outcome. Xu \etal{}~\cite{xu2018review} reviewed portable EEG devices in educational research and envisioned that portable EEG would be employed extensively in the education field in the near future. 

In this paper, as we envision a human-in-the-loop IoT learning environment, it is crucial that we consider wearable devices that can be integrated with edge devices. Hence, we build upon recent work in the literature that uses EEG from wearable devices for learning measurement.

\subsection{Human-in-the-Loop Approaches for Learning} 
Changing the learning environment to incorporate the human and improve the human learning experience has been addressed in the area of Smart Classrooms. Chamba-Eras \etal{}~\cite{chamba2017augmented} proposed the utilization of augmented reality (AR) and an agent that provides services of AR to display and design augmented scenarios in a smart classroom. In addition, they designed an experimental environment to evaluate the utilization and impact of AR in it. Their results showed that the students' criteria of motivation, learning curve, and memorization were substantially improved in the experimental group. 
Ahuja \etal{}~\cite{ahuja2019edusense} proposed ``EduSense'', a system that captures both audio and video streams of the instructor and students to analyze effective instruction, including hand raises, body pose, body accelerometry, and speech acts and provides analytical feedback to the instructor. While the education sector has already taken technological leaps over the last decades with the introduction of smart classroom concept~\cite{kwet2020smart, ahuja2019edusense,sutjarittham2019experiences}, efficient real-time measurement of the learning outcomes that incorporate the student's mental state has to be considered~\cite{elmalaki2021towards, taherisadr2021future}. 

~\\

\begin{figure}[!t]
  \centering
  {\includegraphics[scale=0.42]{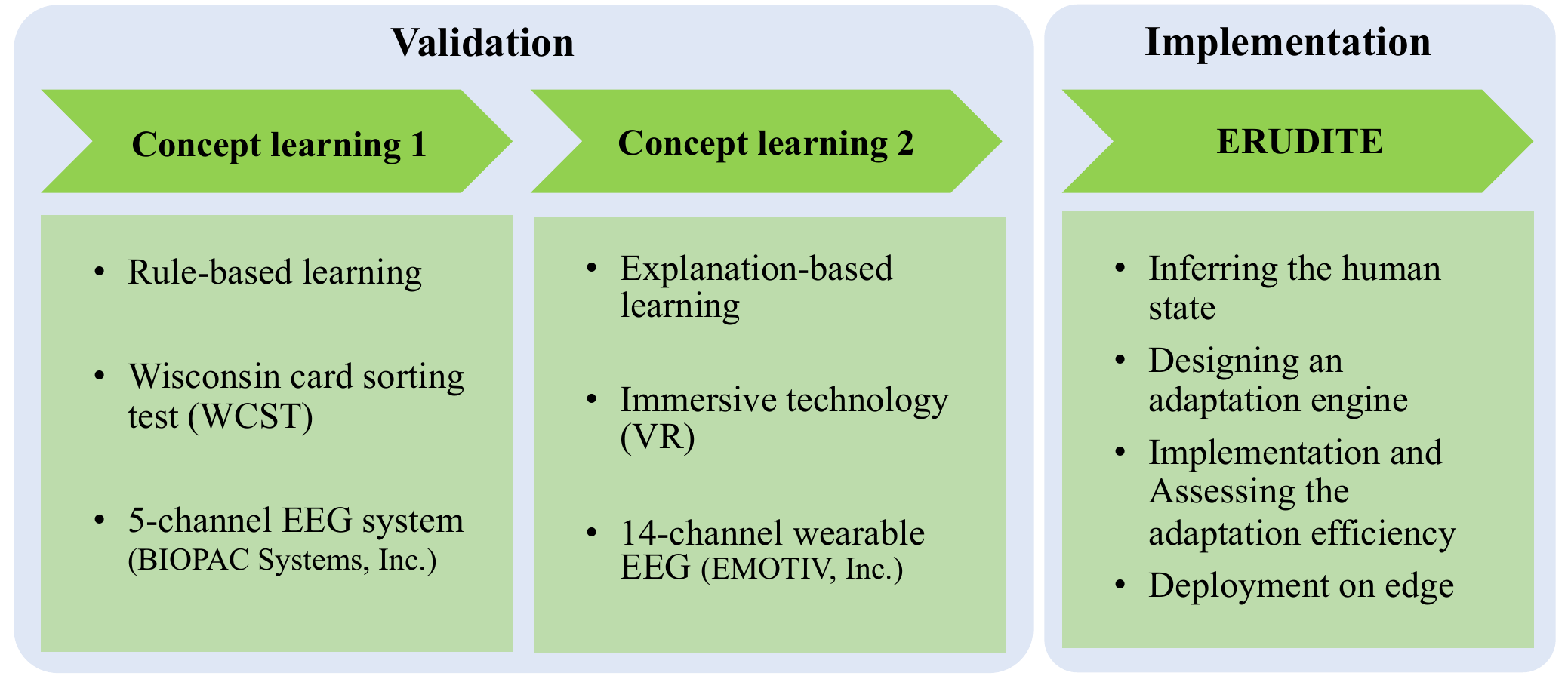}}
  \caption{Roadmap of ERUDITE validation and implementation. Two concept learning-based experiments are conducted for validation and then implementation  
  is done based on the insights gained from those experiments.} 
  \label{fig:roadmap}
\end{figure}

This paper is organized as follows; first, we explore two concept learning approaches, rule-based and explanation-based, in Sections~\ref{sec:rule} and~\ref{sec:exp-based}, respectively, to get insights into how to measure the learning state of the human. In Section~\ref{sec:erudite}, we describe \sysname framework for a human-in-the-loop IoT system for a learning environment. We evaluated \sysname on $15$ human subjects and discussed the deployment of \sysname on an edge-based prototype in Section~\ref{sec:evaluation}. Finally, we conclude the paper with some discussion and future work in Section~\ref{sec:disc}. 
Figure \ref{fig:roadmap} illustrates the roadmap of \sysname, where in the validation phase we exploit two concept-learning experiments as discussed in Sections~\ref{sec:rule} and~\ref{sec:exp-based} to gain insights into the correlations between learning and brain signals. Then we utilize these insights in the implementation phase to design \sysname as discussed in Section~\ref{sec:erudite}.

\begin{figure}
  \centering
  \subfigure[EEG cap worn by a participant while performing the WCST.]{\includegraphics[scale=0.075]{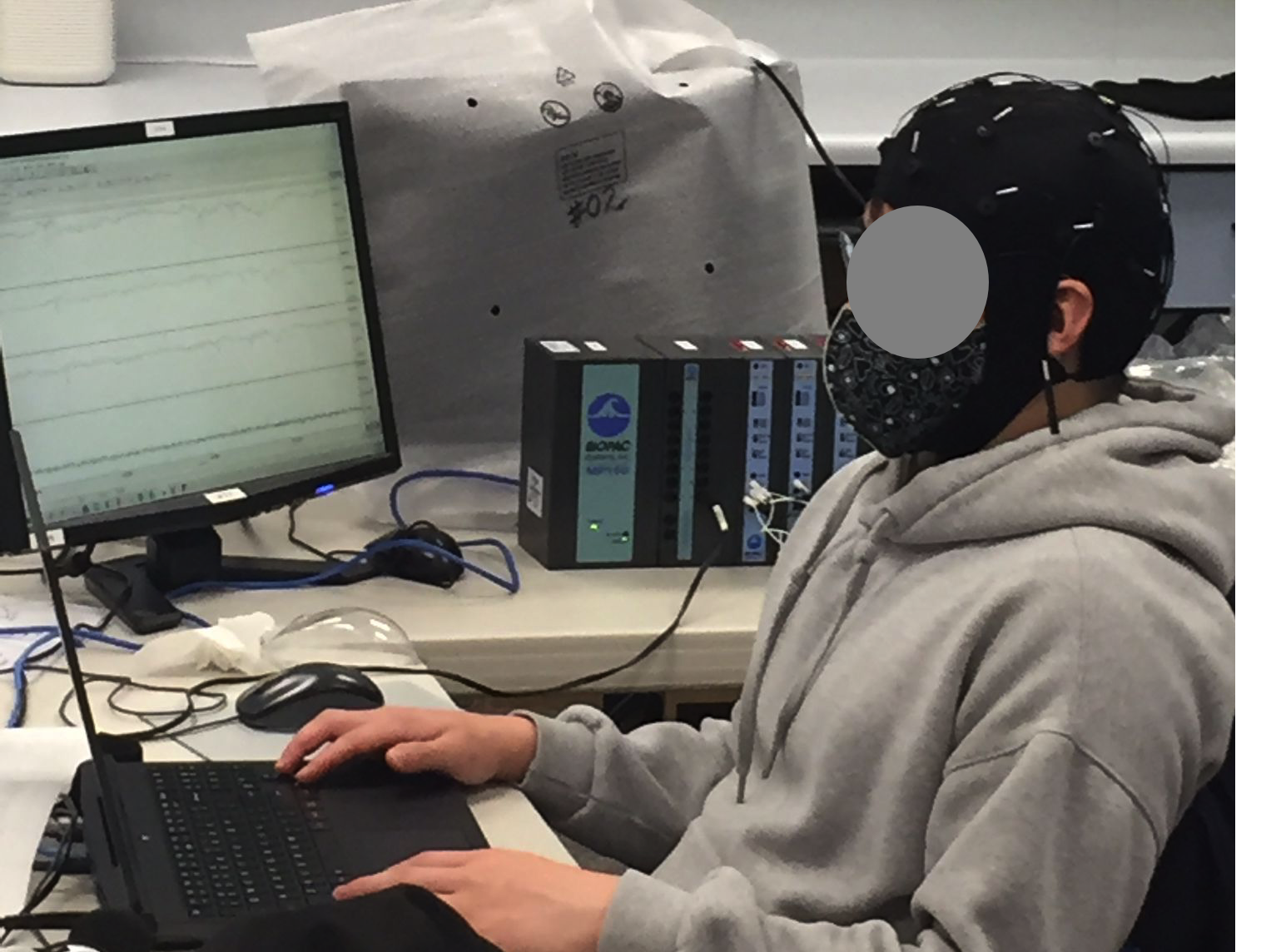}}\hspace{3mm}
  \subfigure[EEG amplifier, data acquisition unit, and  accessories.]{\includegraphics[scale=0.094]{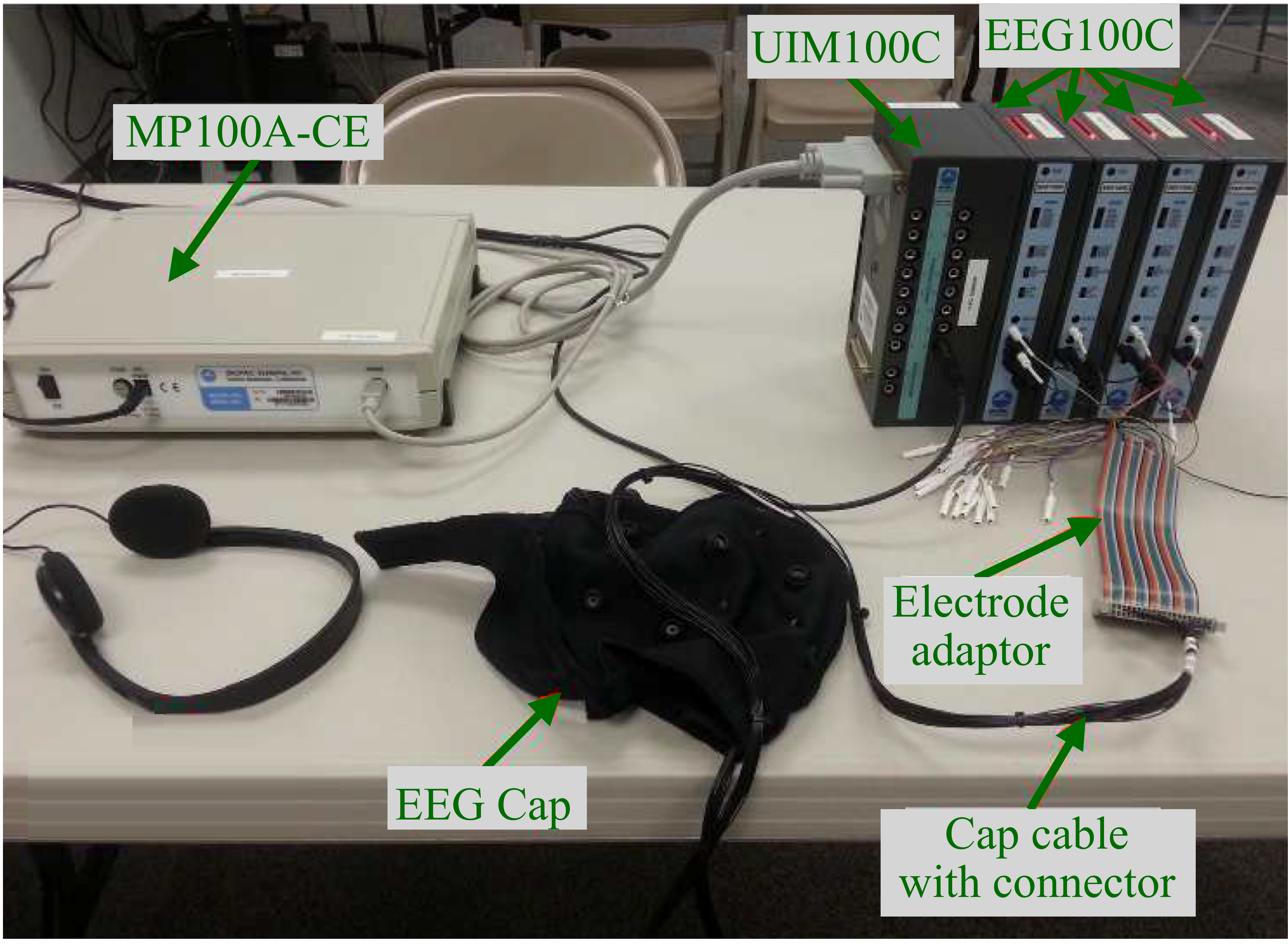}}\quad
    \subfigure[An instance of the WCST environment. In this figure, the rule is color (green).]{\includegraphics[scale=0.2]{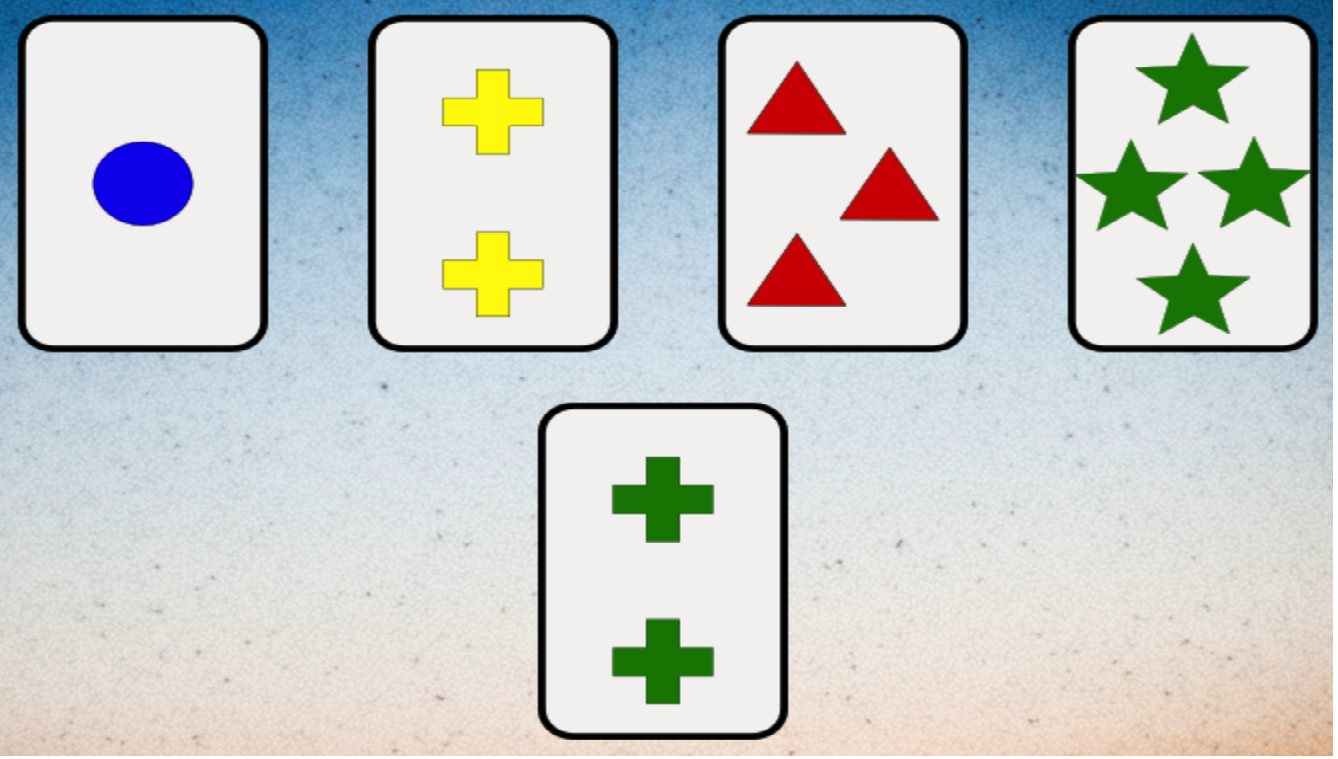}}\hspace{3mm}
  \subfigure[EEG channels' names and locations.]
  {\includegraphics[scale=0.12]{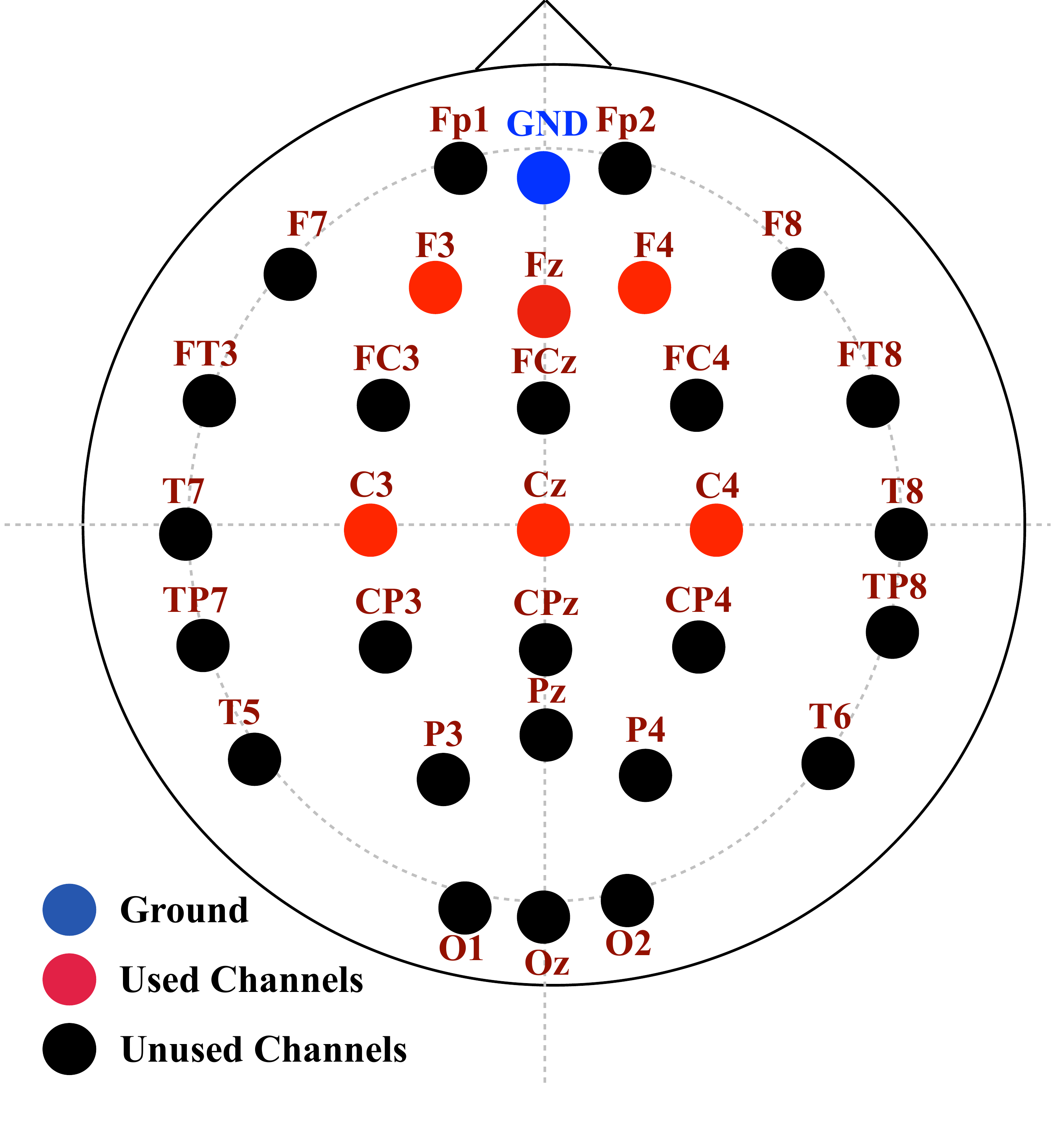}}
  \caption{WCST experiment setup and EEG device.}
     \label{fig:EEGdevice}
\end{figure}

\section{Rule-Based Concept Learning} \label{sec:rule} %
As we highlighted in the background in Section~\ref{sec:related}, cognitive psychologists and philosophers have paid particular attention to concepts that identify kinds of things—those that classify or categorize objects—and such concepts that are named rule-based learning~\cite{goodman2008rational}. In this section, we utilize the rule-based learning concept to explore the ability to understand and quantify learning through EEG signals. In particular, we choose the Wisconsin Card Sorting Test (WCST) to represent the rule-based learning~\cite{grant1993wisconsin}.

\subsection{The Wisconsin Card Sorting Test (WCST)}\label{sec:wcst}
The Wisconsin Card Sorting Test (WCST) is a neuropsychological test frequently used to measure higher-level cognitive processes such as perseverance, attention, abstract thinking, and learning \cite{grant1993wisconsin}. The Wisconsin Card Sorting Test (WCST) is a widely recognized and standardized neuropsychological assessment tool. It has a rich historical background, serving as a benchmark for comparing our research results with existing literature~\cite{grant1993wisconsin,eling2008historical}. The test's reliability and accessibility make it a solid choice for evaluating cognitive and executive functions. This availability of data supports the reproducibility of our findings and facilitates further research~\cite{ozonoff1995reliability,kopp2021reliability}. Cognitive flexibility, the ability to switch between mental tasks, is a crucial aspect of human learning, and the WCST objectively assesses it, minimizing the influence of bias or subjective judgment~\cite{miles2021considerations}.  
The WCST consists of two card packs, each containing four stimulus cards and $128$ response cards. 
The cards are of various geometric shapes in different colors and patterns. The participants are expected to accurately sort every response card with one of four stimulus cards through the feedback (right or wrong) given based on a predefined rule that is oblivious to the participant. Among various versions, the version of WCST with $128$ cards developed by Heaton is what we adopted~\cite{heaton1993wisconsin}. 

Figure~\ref{fig:EEGdevice}c depicts the computerized version of the WCST environment. The first row is the four stimulus cards that the player has to choose among them, and the second row is the test card that has to be matched with one of the stimulus cards based on some rule that the participant needs to learn. For example, if the rule is ``Matching Color'' and the participant chooses the pattern card, the feedback will be ``wrong.''  Participant has five chances to find out (\emph{learn}) the rule and choose the right card. If the participant selects the right card in any of the five moves, the move is called the ``correct move'', and one of the $128$ card rounds ends. If, after five tries, the participant cannot find out the right rule, the current round ends without any right answer. Then the sorting rule changes discreetly from i.e. color to form or pattern of figures. This rule change does not follow any pattern, and it happens randomly without the participants being informed. The number of the correct moves represents the times the participant could learn the correct rule out of $128$ rounds of the game~\cite{grant1993wisconsin}.

\subsection{EEG Data Collection} 
As the purpose of this experiment is to get insights on how to quantify the learning event, we used a $5-$channel research-grade EEG system (BIOPAC Systems, Inc. \cite{Biopack} shown in Figure~\ref{fig:EEGdevice}b with a sampling rate ($Fs$) of $200$ $Hz$ for data collection. Channel $Fpz-1$ was selected as the ground for data acquisition setup, and the electrodes' impedance was clamped at less than $5$ $K\Omega$. Electrode locations were based on the $5-$ channel subset of $24-$channel $10-20$ systems and were chosen to be over the frontal areas of the skull because it is the area in the brain that is responsible for cognitive processes, such as executive function, attention, memory, and language~\cite{chayer2001frontal}. The electrodes are attached to $Fz$, $F3$, $F4$, $C3$, and $C4$ channels. Figure~\ref{fig:EEGdevice}a illustrates the setup attached to one of the participants. Figure~\ref{fig:EEGdevice}b shows the EEG device setup and associated accessories. Figure~\ref{fig:EEGdevice}d presents the EEG channels' locations. We used the $AcqKnowledge$ recording platform to record the EEG signals from these channels~\cite{Biopack}.

We conducted the WCST experiment on ten healthy participants, including six males and four females, all between $20$–$30$ years old. The average and standard deviation of the age of the participants were $27$ and $2.36$, respectively. Each participant was asked to do two separate sessions for a total of $20$ sessions across all participants. Each session was done on different days and at different times of the day. 
Before the experiment, the color blindness test is used (\eg{} by the Ishihara Plate Test~\cite{de1992new}) to ensure it does not influence the test result.

\subsection{Analysis and Correlation of EEG Signal to Learning}\label{sec:analysis_rule}

\begin{figure}[!t]
  \centering
  {\includegraphics[trim={2cm 11.5cm 2cm 2cm},clip, scale=0.35]{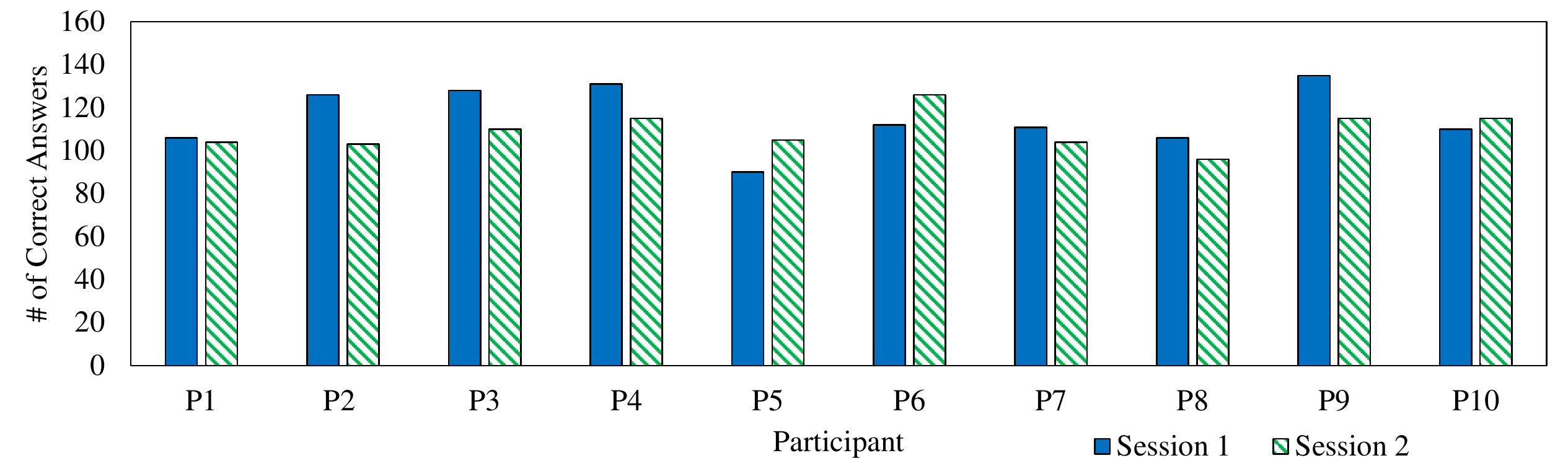}}
  \caption{Total number of correct answers for each participant in WCST across two different sessions. 
  }
  \label{fig:demog}
\end{figure}

\subsubsection{Memorization Effect} A central assumption in rule-based concept learning is that there is no effect of memory or experience in the learning process. Hence, no matter the number of times a participant conducts the WCST, the performance is independent of the previous sessions. Thus, in rule-based concept learning, the participant only follows the current rule of the game. 
Figure~\ref{fig:demog} illustrates the number of correct moves each participant took in two separate sessions. Figure~\ref{fig:demog} represents the number of correct moves for each participant in session $1$ and session $2$ of the experiment. The total correct move for all participants in session $1$ and session $2$ are $1154$ and $1105$, respectively. Overall, the correct moves change by $\pm 4\%$ between sessions. As Figure~\ref{fig:demog} shows, from session $1$ to session $2$, the number of correct moves does not follow a particular pattern for participants --- decrease for some participants and increase for others. This indicates that repeating the WCST has no effect on the result (as experience or memorization).

\subsubsection{Effect of Correct Moves on EEG Dynamics} We analyzed the collected EEG data to evaluate the correlation between the EEG dynamics and the rule-based concept learning experiment. In particular, how to use the EEG signal to infer whether the participant learned the rule or not. 
An overall pictorial figure of our analysis approaches is shown in Figure~\ref{fig:flowwcsg}. Below we describe the details of this analysis to detect this correlation:

\begin{figure*}[!t]
\centering
{\includegraphics[scale=0.21]{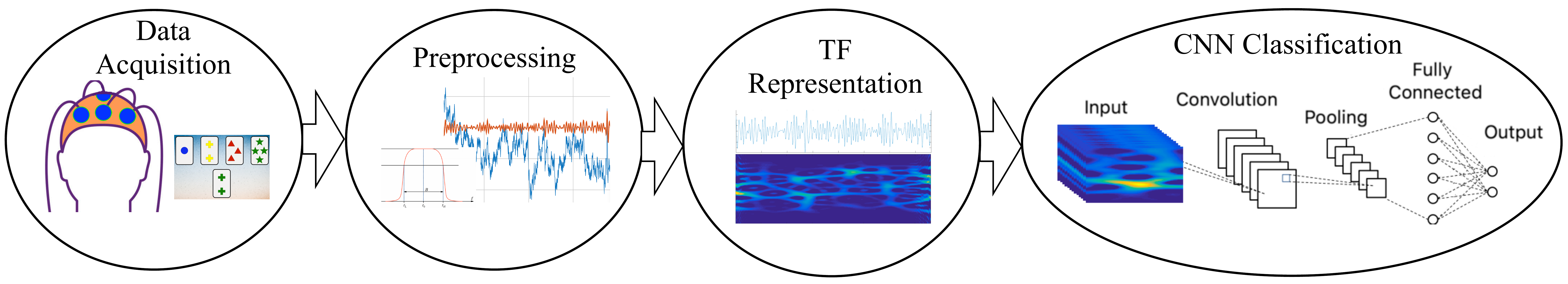}}
\caption{Data acquisition and analysis of the rule-based concept learning experiment. After acquiring the data and preprocessing, TF representation is applied to each window of the EEG signal. Then TF images with labels acquired from wrong and correct moves of the WCST experiment are fed to the CNN model for classification.}
\label{fig:flowwcsg} 
\vspace{-4mm}
\end{figure*}

\begin{itemize}[leftmargin=*]

\item Filtering \& Facial Artifacts:
EEG data is preprocessed using a high-pass filter (HPF) with a cut-off frequency of $0.5$ $Hz$ for removing trends and $40$ $Hz$ to remove the high-frequency components. To remove facial artifacts, such as jaw movement, eye blink, and eye movement, we applied independent component analysis (ICA). ICA is a computational method for separating a multivariate signal into additive subcomponents. The assumption is that the subcomponents are statistically independent of each other, which applies to EEG signals and facial artifacts~\cite{chen2021application,agarwal2021classification}.

\begin{figure*}
  \centering
  \subfigure[b][Timestamp for the actions taken  averaged across all participants during  the WCSG experiment. The linear  growth of the timing depicts the  uniformity of the time taken for subsequent actions.]{\includegraphics[scale=0.25]{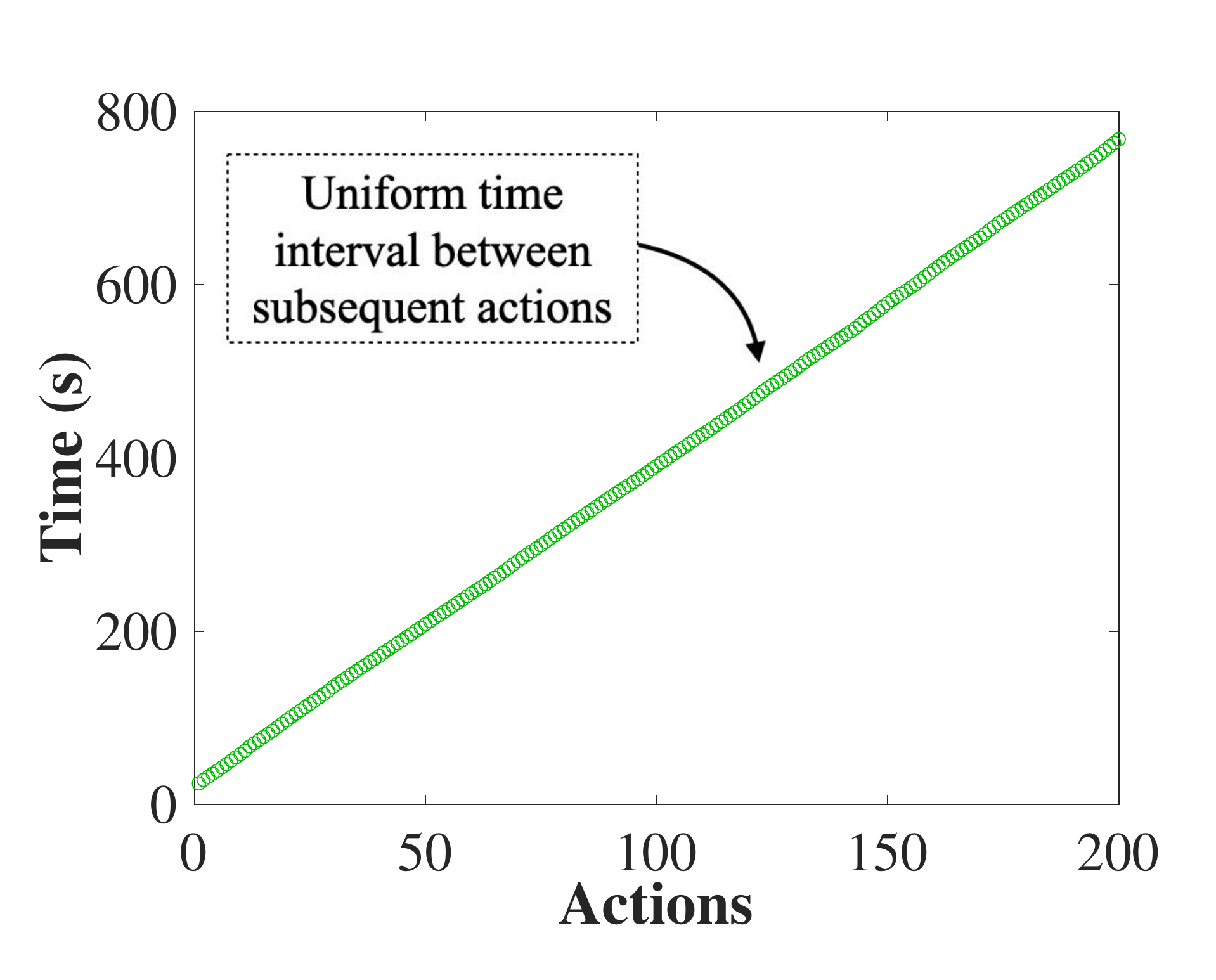}}\hspace{5mm}
  \subfigure[Histogram of the average time the participants took to take one action in WCST. The horizontal axis is the time intervals in seconds, and the vertical axis is the number of actions taken in each time interval.]{\includegraphics[scale=0.27]{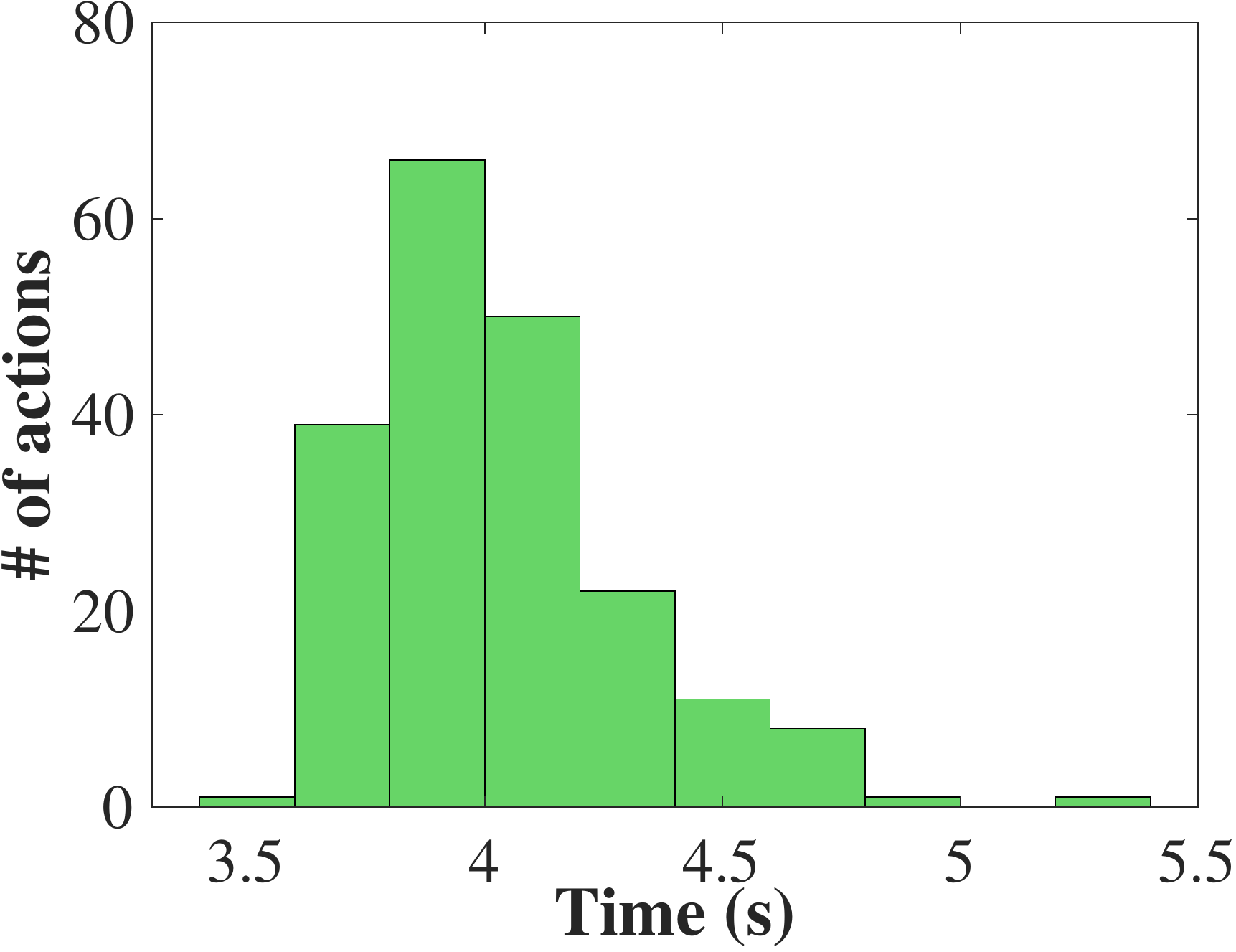}}\hspace{5mm}
    \subfigure[Time taken per action averaged across all participants in WCST.]{\includegraphics[scale=0.27]{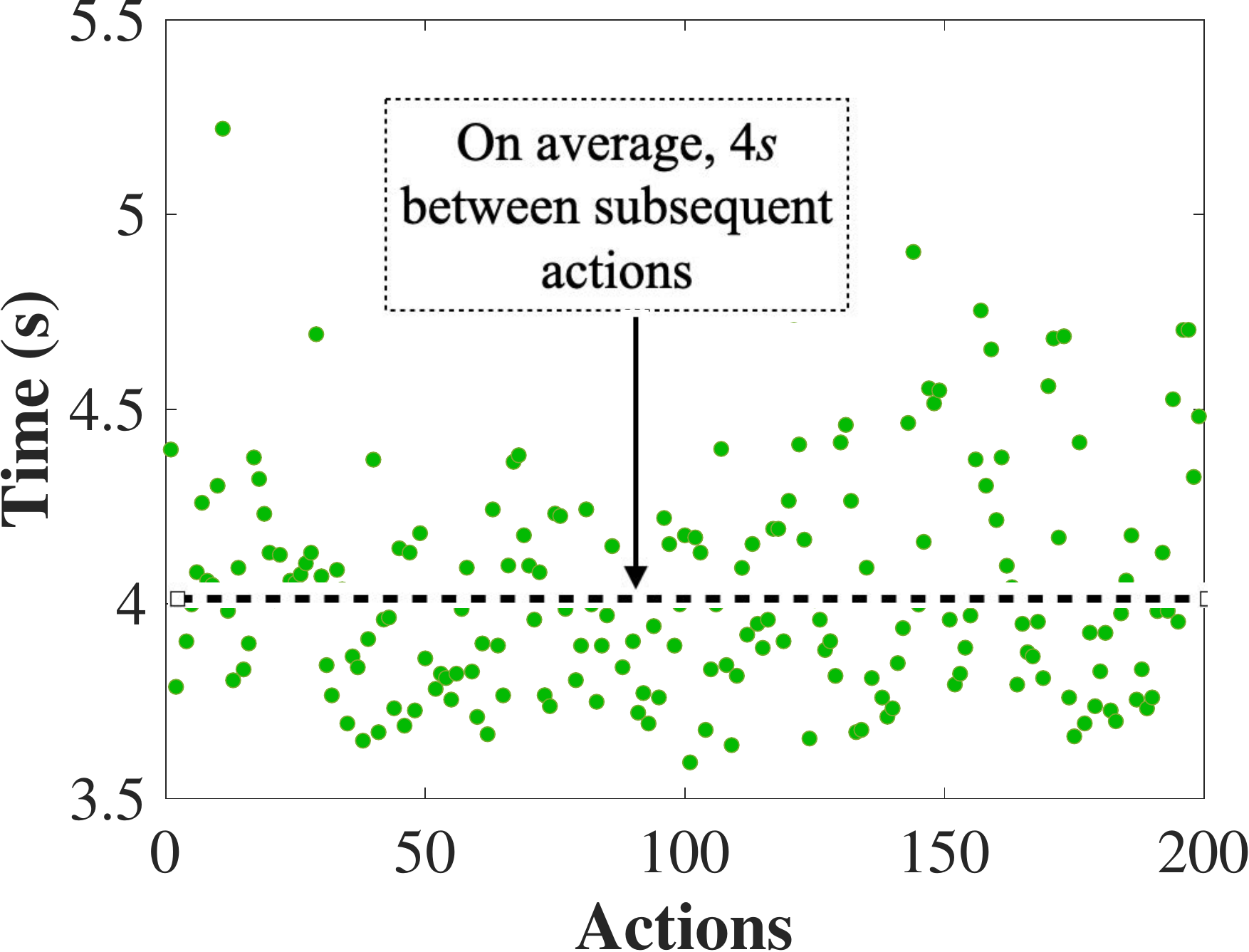}}
  \caption{Windowing size analysis averaged across all participants to segment the EEG data for WCST.}
     \label{fig:windowanalysis}
\end{figure*}

\item Windowing: Since the WCST has a very short response time, we need to capture this response event carefully. Hence, we applied a $4s$ window with $1s$ overlap. We segmented the EEG data into multiples of $4s$ data using this $4s$ window. The reason is that, on average, every participant takes one action every $4s$. Figure~\ref{fig:windowanalysis}a depicts the timestamp of the taken actions averaged across all participants. As Figure \ref{fig:windowanalysis}a shows, the timing grows linearly, which refers to the uniformity of the time a participant takes between subsequent actions. We use this outcome as the base for segmenting the EEG signal. With this window size, aligning or warping the EEG data to fit the segments is not required. Due to the uniformity of the time taken between subsequent actions, we chose a fixed window size. In particular, Figure~\ref{fig:windowanalysis}b shows the distribution of the time intervals between subsequent actions averaged across all participants where we observe more than $80\%$ of the actions are taken between $375$ and $450$ milliseconds, which, in turn, illustrates that window size $4s$ is the dominant window size for the participants. Figure~\ref{fig:windowanalysis}c shows the scatter-plot of the time all participants take between subsequent actions.

\item Spectro-Temporal representation of the EEG signal: \label{Spectro-Temporal} After preprocessing, we utilize time-frequency (TF) representation to analyze the collected EEG signals in the temporal and spectral domains. TF could be considered a non-stationary signal analysis with frequency content varying with time. TF is a suitable representation for non-stationary and multi-component signals such as EEG, which can simultaneously describe the given signal's energy distribution over time and frequency space. TF selection for EEG signal representation is the first step in designing an appropriate representation. A proper TF highlights the non-stationarities in the input signal that enables the system to discriminate between variations of the signal in both temporal and spectral domains.

We use the TF representation to investigate the spectro-temporal correlations between the learning event and EEG signals. In particular, we used the Smoothed Wigner-Ville distribution (Smoothed-WVD) to create the TF representation since, in practice, it has been shown to improve the quality of representation due to its ability to reduce the cross-term interference for a signal with multiple frequency components~\cite{boashash2015time}\footnote{Methods for TF representation can be categorized in six groups as follows: Gaussian kernel, Wigner--Ville distribution (WVD), spectrogram, modified-B, smoothed-WVD, and separable kernel. Reduced interference approaches, such as Smoothed-WVD, can improve the representation quality by reducing the effect of the cross-terms~\cite{ava}.}.  

Figure~\ref{fig:tflearn} illustrates the TF representation of the EEG signal from the $Fz$ channel for five  participants\footnote{We did the same analysis for all participants and depicted five participants in the Figure~\ref{fig:tflearn} for visualization purposes.}. In particular, the first row in Figure~\ref{fig:tflearn} depicts $4s$ duration TF representation of the EEG signal before the participants learn the rule. The second row in Figure~\ref{fig:tflearn} shows the same duration of the EEG signal after participants find out ($learns$) the rule and can take five consecutive correct actions, which is considered the duration they have completely mastered the rule. After participants learn the rule and take consecutive correct moves, we observe that the representation contains more activities in the higher spectral component. To quantify this observation,  we measured the spectral power of the high-frequency sub-bands ($10$ - $25$) $Hz$ of the EEG signal from $Fz$ channel before and after the participant learns the rule of the WCST as shown in Table~\ref{tbl:spectral}\footnote{We use $v^2{Hz}^{-1}$ unit for spectral power, as the EEG signal is defined in terms only of a voltage and there is no unique power associated with the stated amplitude.}. Results in Table~\ref{tbl:spectral} show spectral power increases on average by $76.52\%$  after participants learn the rule of the WCST.

\begin{figure*}
\centering
\begin{tabular}{l | c c c c c}
     & $P_1$ & $P_2$ & $P_3$ & $P_4$ & $P_5$ \\\hline
     \rotatebox{90}{\small{Before learning}} 
     & {\includegraphics[trim={6.2cm 6.5cm 4cm 11.5cm},clip, scale=0.25]{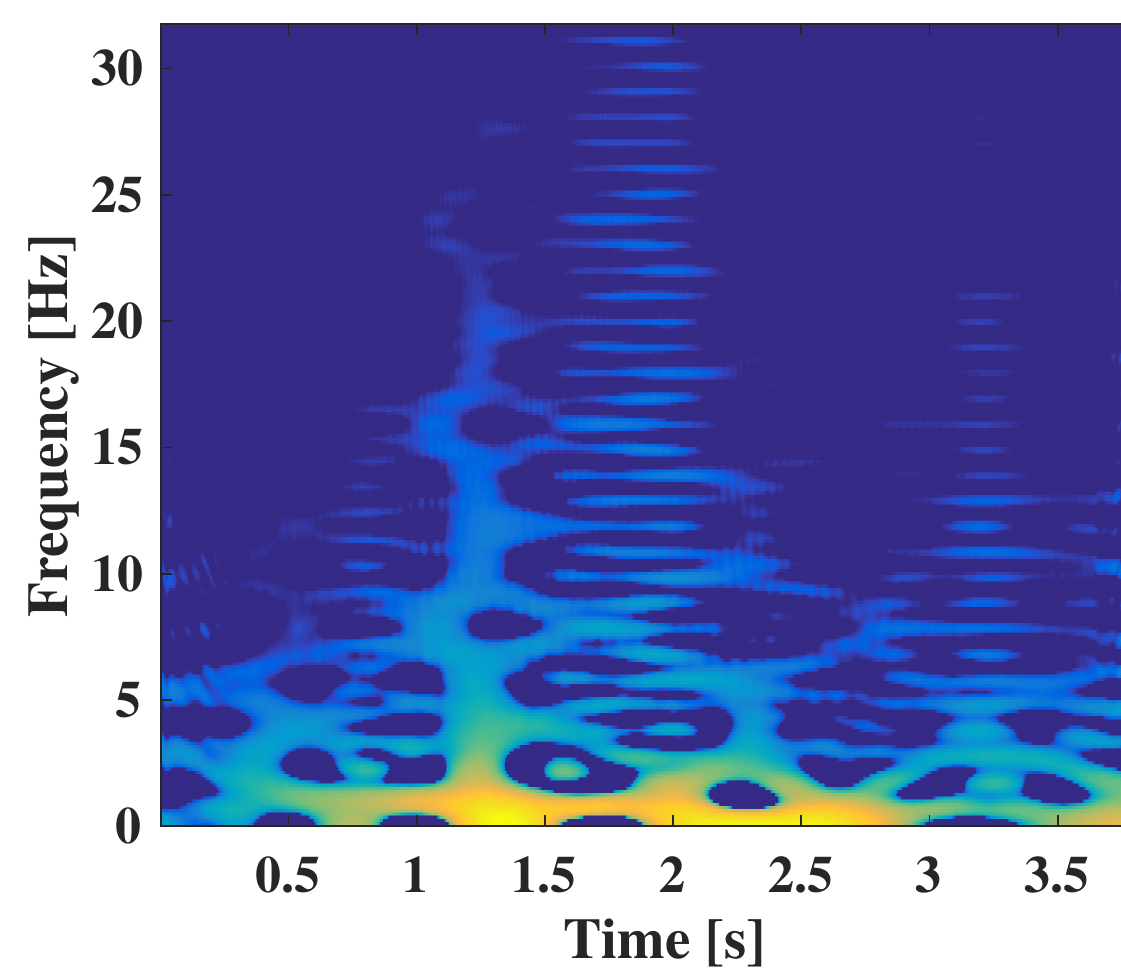}} 
     & {\includegraphics[trim={6.2cm 6.5cm 4cm 11.5cm},clip, scale=0.25]{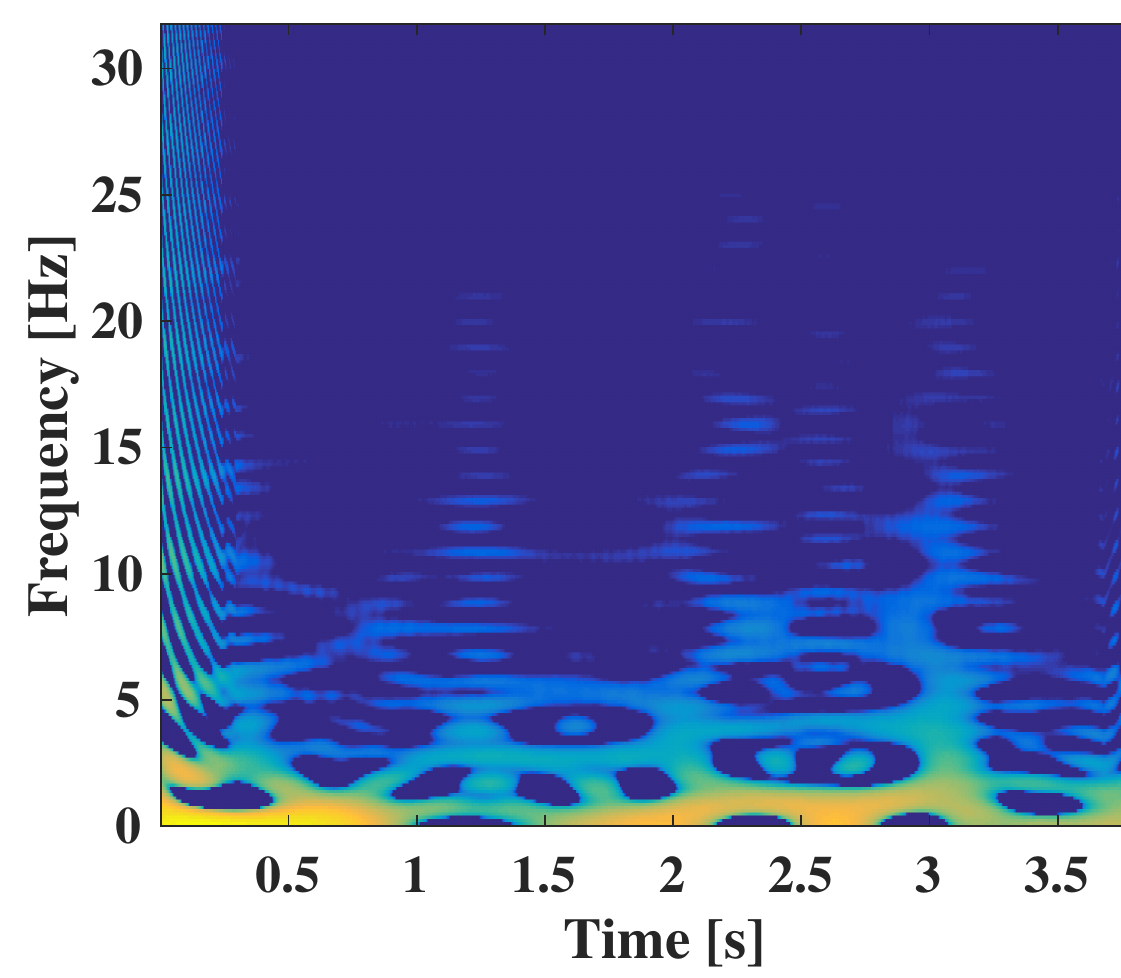}} 
     & {\includegraphics[trim={6.2cm 6.5cm 4cm 11.5cm},clip, scale=0.25]{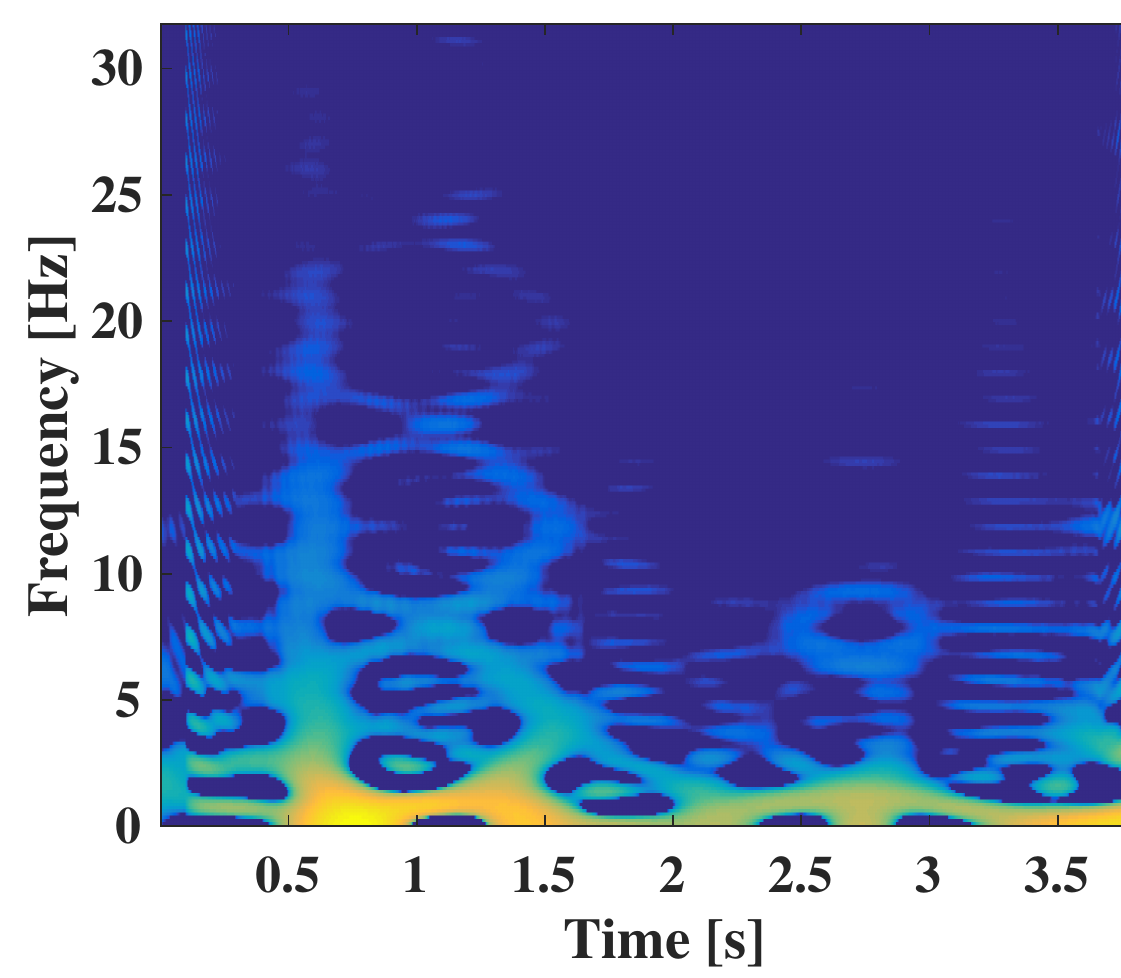} }
     & {\includegraphics[trim={6.2cm 6.5cm 4cm 11.5cm},clip, scale=0.25]{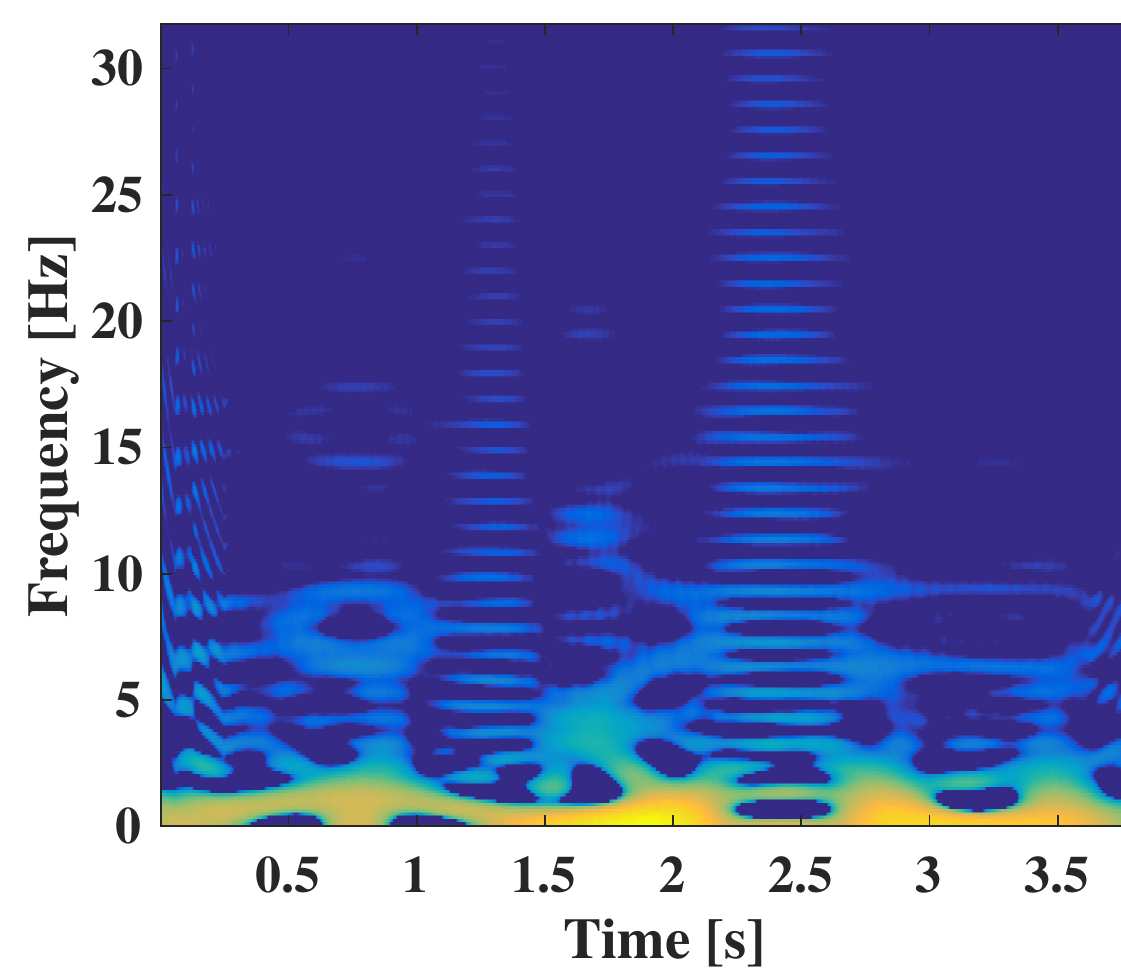} }
     & {\includegraphics[trim={6.2cm 6.5cm 3cm 11.5cm},clip, scale=0.25]{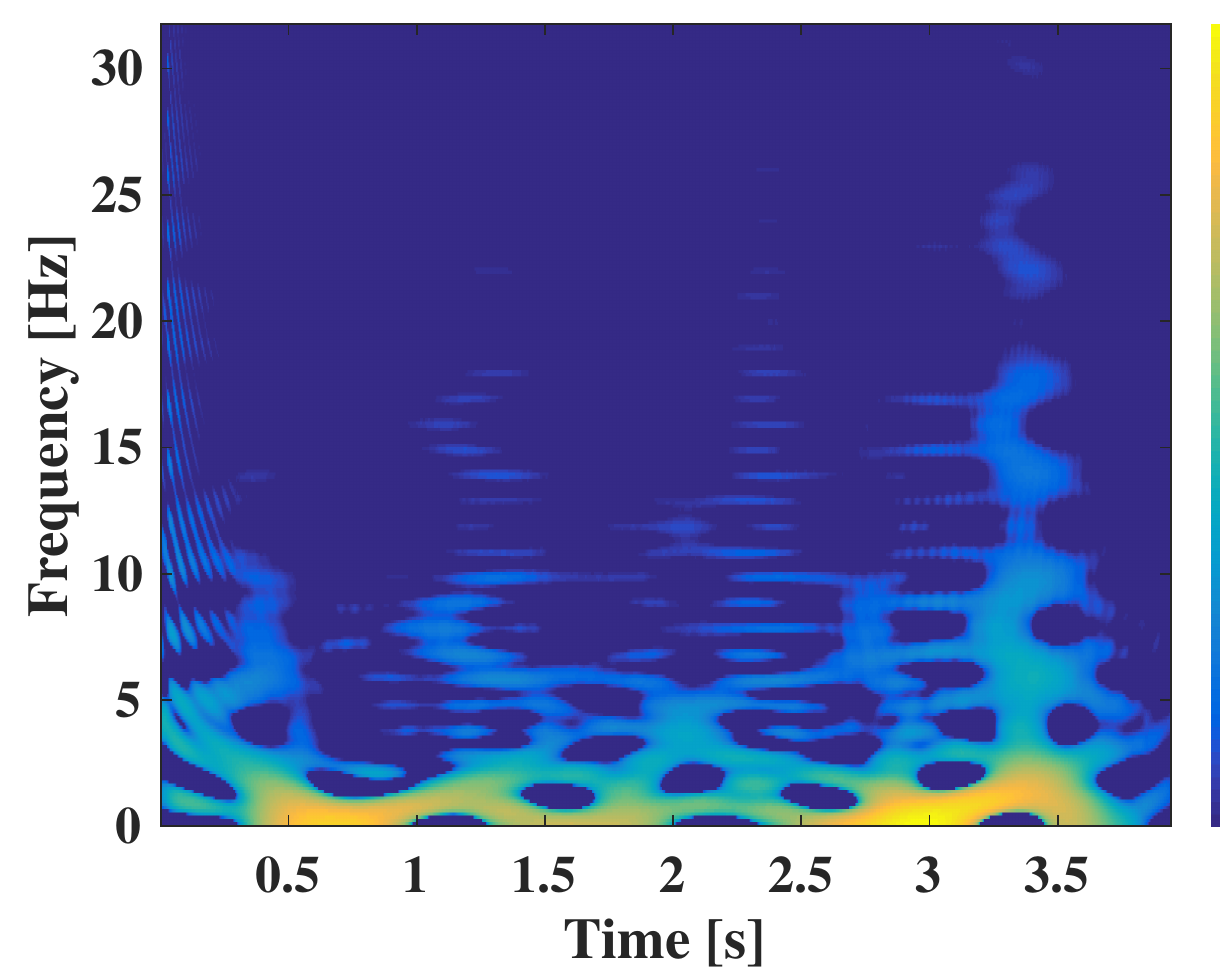}} \\\hline
     \rotatebox{90}{\small{After learning}} 
     & {\includegraphics[trim={6.2cm 6.5cm 4cm 11.5cm},clip, scale=0.25]{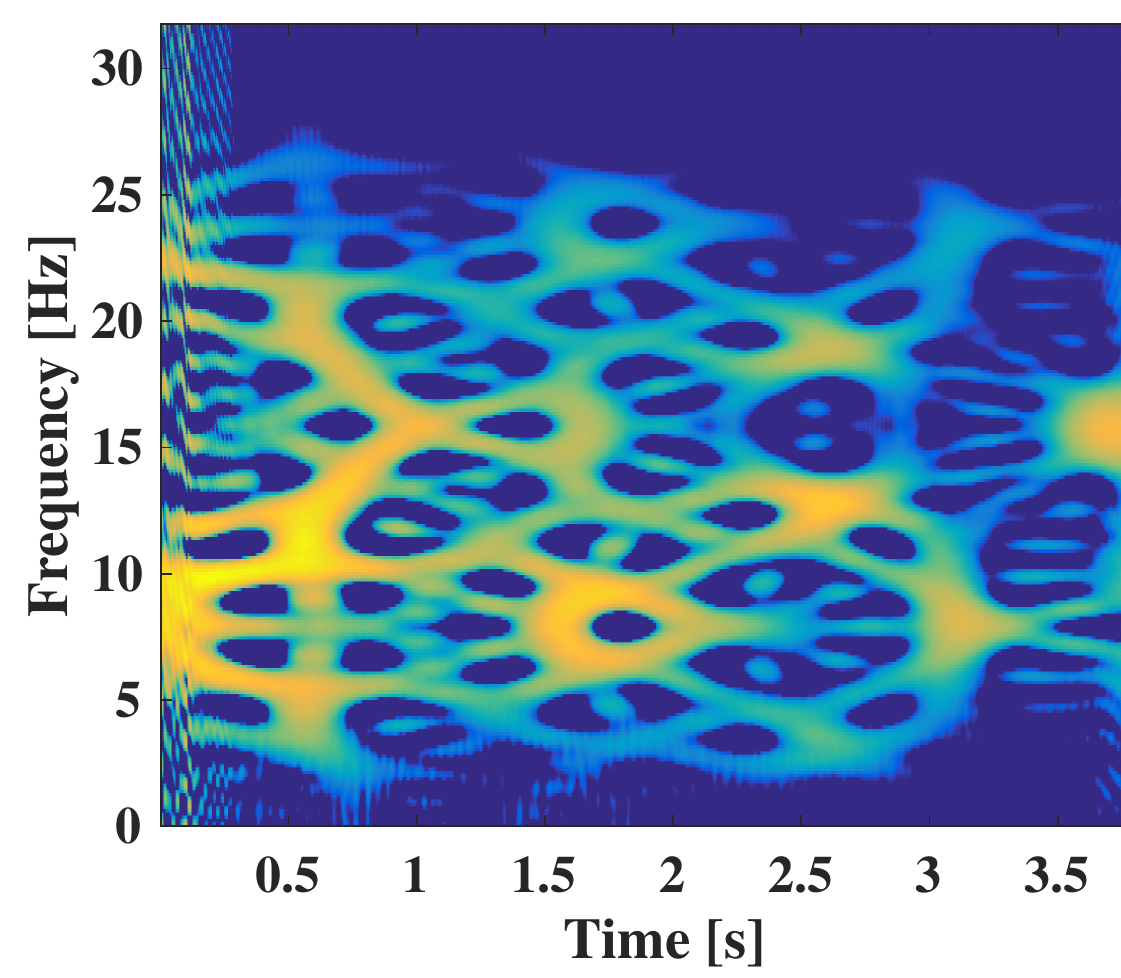} }
     & {\includegraphics[trim={6.2cm 6.5cm 4cm 11.5cm},clip, scale=0.25]{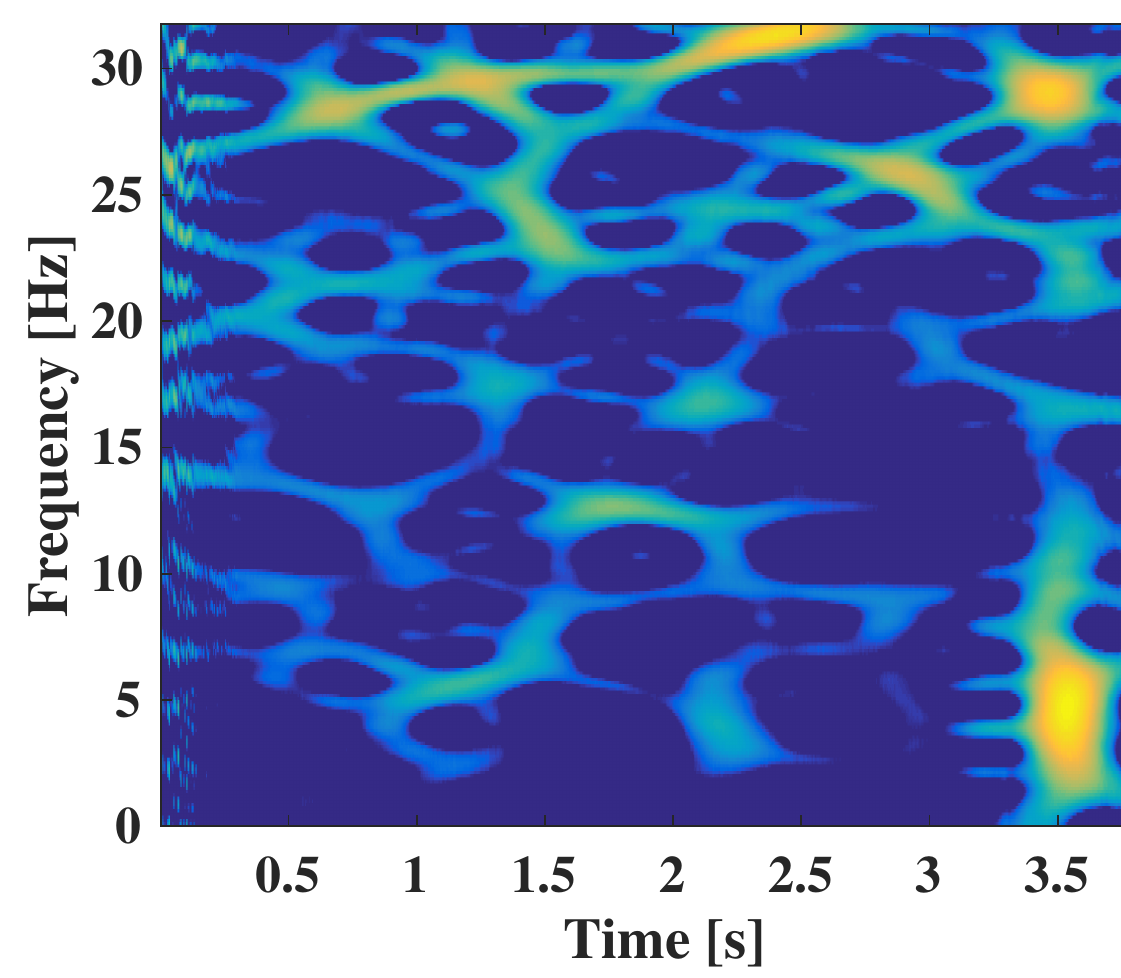}} 
     & {\includegraphics[trim={6.2cm 6.5cm 4cm 11.5cm},clip, scale=0.25]{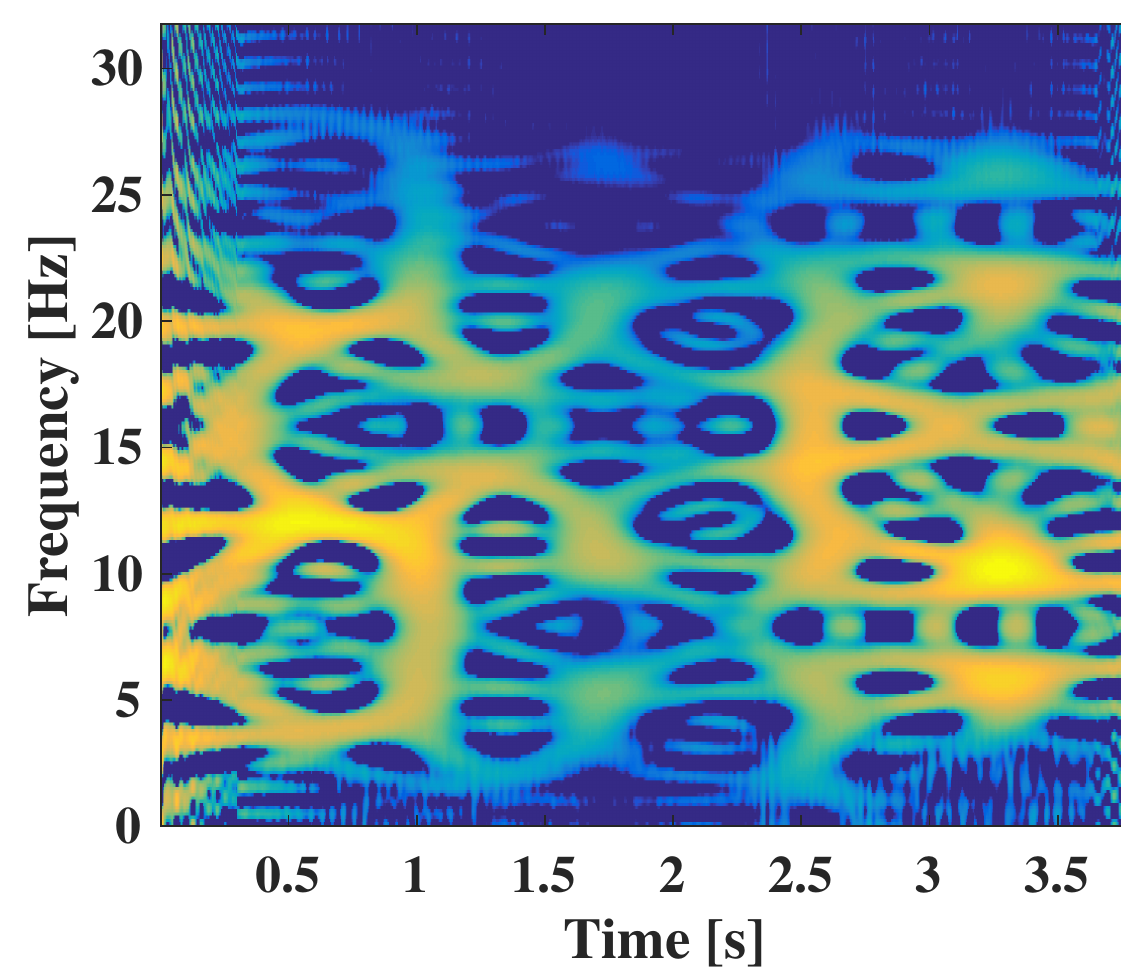} }
     & {\includegraphics[trim={6.2cm 6.5cm 4cm 11.5cm},clip, scale=0.25]{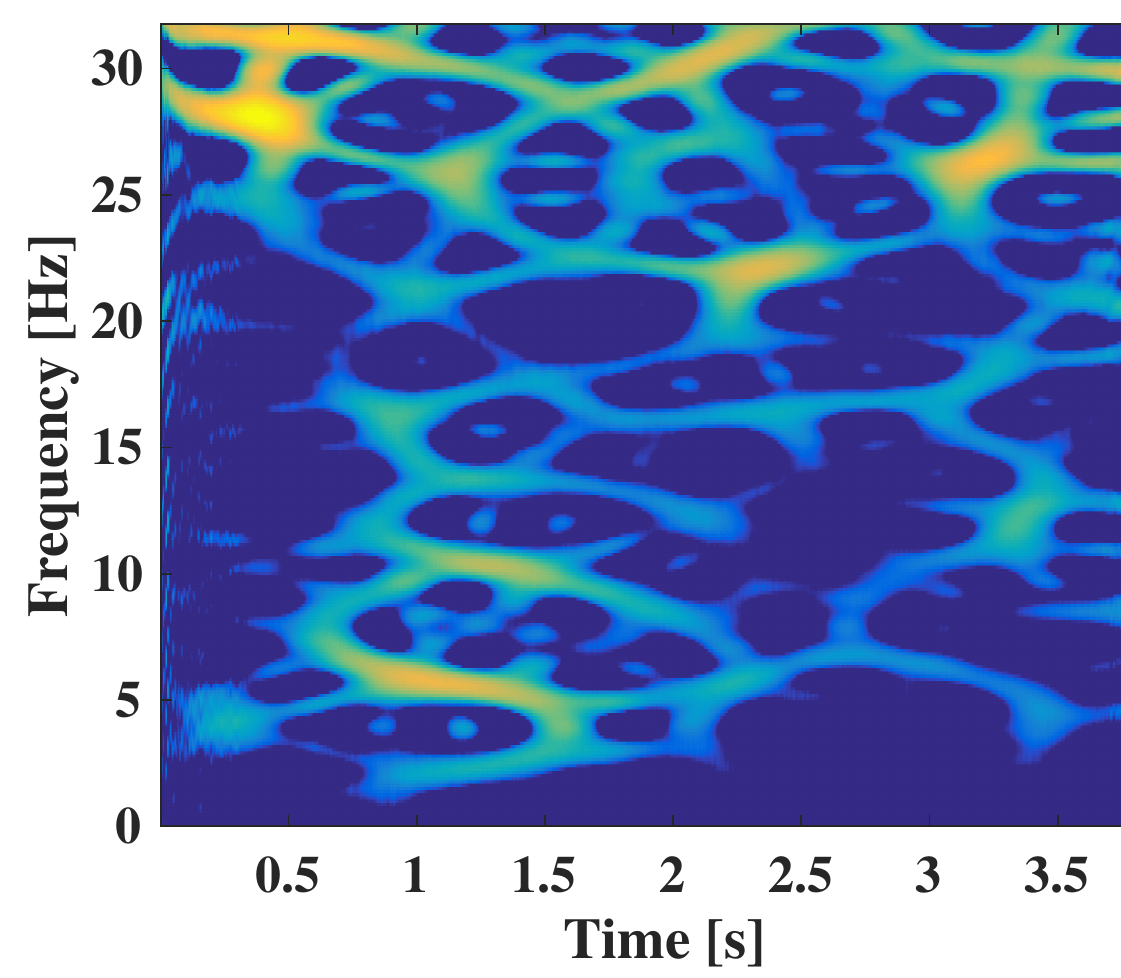} }
     & {\includegraphics[trim={6.2cm 6.5cm 3cm 11.5cm},clip, scale=0.25]{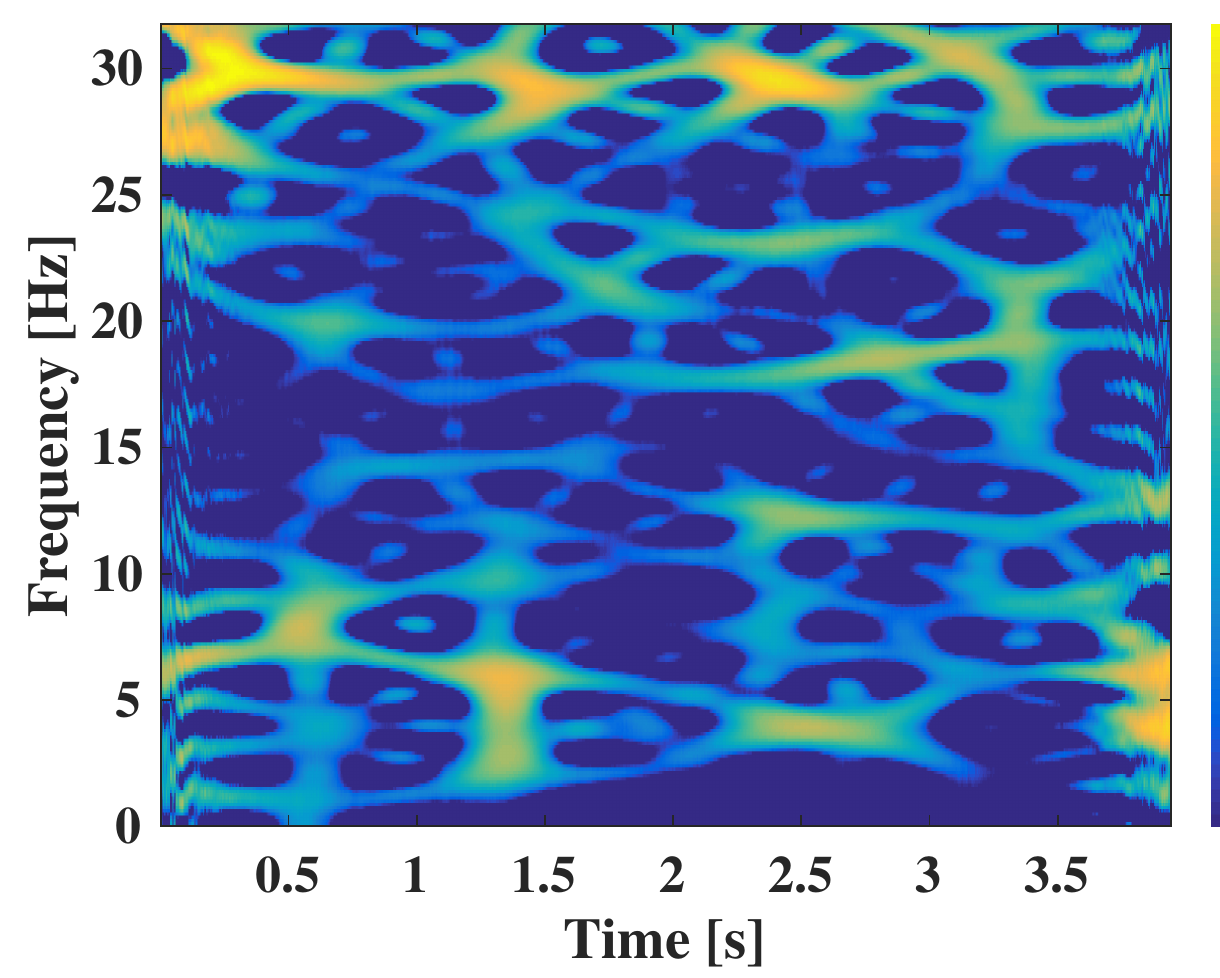}} \\\hline
\end{tabular}

\caption{Time-frequency (TF) representation using Smoothed-WVD of EEG signal from $Fz$ channel. Brighter color refers to higher energy in the associated frequency in $Hz$ and time in seconds, and darker colors mean lower energy for the associated frequency in $Hz$ in the given time in seconds. 
(a) The first row represents the TF representation of the EEG signal for five participants before they learn the rule in WCST. (b)The second row represents the TF representation of the EEG signal for the same five participants after they learn the rule in WCST by taking five correct actions.
}\label{fig:tflearn}
\end{figure*}

\begin{table}[!h]
  \small
  \centering
  \caption{Spectral power of the high-frequency sub-bands ($10$ - $25$) $Hz$ of the EEG signal from the $Fz$ channel before and after the participant learns the rule of the WCST.} \label{tbl:spectral}
  \begin{tabular}{|l||cc|cc|}
    \hline
       & \multicolumn{2}{c|}{\textbf{Before learning}}  & \multicolumn{2}{c|}{\textbf{After learning}}   \\
  \textbf{Participant} & Session 1 & Session 2 & Session 1 & Session 2\\
  &  ($v^2{Hz}^{-1}$) & ($v^2{Hz}^{-1}$) & ($v^2{Hz}^{-1}$) & ($v^2{Hz}^{-1}$) \\\midrule

 Participant 1 &  $0.094$  & $0.087$  &    $0.223$ & $0.196$  \\
 Participant 2 &   $0.079$ & $0.083$  & 	$0.193$ & $0.213$  \\
 Participant 3 &   $0.138$ & $0.112$  & 	$0.251$ & $0.217$  \\
Participant 4  &   $0.201$ & $0.189$  & 	$0.297$ & $0.257$  \\
Participant 5  &   $0.147$ & $0.183$  & 	$0.261$ & $0.239$  \\ 
Participant 6  &   $0.129$ & $0.198$  & 	$0.272$ & $0.298 $ \\
Participant 7  &   $0.085$ & $0.097$  & 	$0.179$ & $0.194$  \\ 
Participant 8  &   $0.139$ & $0.172$  & 	$0.199$ & $0.216$  \\ 
Participant 9  &   $0.091$ & $0.118$  & 	$0.183$ & $0.213$  \\
Participant 10 &   $0.169$ & $0.196$  & 	$0.214$ & $0.268$  \\\hline
 \cellgrey \textbf{Average}   &   $0.127$ & $0.144$  &    $0.223$ & $0.231$  \\\hline
  \end{tabular}
\end{table}

\end{itemize}

\subsection{Classification of Learning Events in WCST}\label{sec:wcstcalss}
As we mentioned earlier, we aim to use a representation of the learning state as a sensor modality in a learning IoT system. One approach to infer this learning state is to use a supervised learning model. In particular, we want to infer two classes; learning and not-learning events; where the ``learning'' label is assigned if there are five consecutive correct moves in the WCST, and the not-learning label is assigned if there is one wrong move. We applied three classifiers with different computation complexity to do this classification:
\begin{itemize}[itemsep=1ex,topsep=0pt]
    \item Support Vector Machine (SVM) Model: We designed an SVM discriminative linear classifier with a Gaussian kernel using the spectral and temporal features from the preprocessed EEG data. Features used to train the SVM model are (1) autoregressive (5 features), which represents an estimation of temporal characteristics of the EEG signal, (2) four features extracted from each sub-bands of the Wavelet decomposition of the EEG signal, 
    and (3) short-time Fourier transform (STFT) decomposition, including four features. We applied five levels of ``Haar''  Wavelet decomposition to each EEG window (which provides five sub-bands) and extracted four features from each sub-band. In total $29$ features were extracted. 
    \item Convolutional Neural Network (CNN): We designed a supervised CNN model to classify the $4s$ windows of the TF representation as an input image (as shown in Figure~\ref{fig:tflearn}) into two classes; learning and not-learning events.  
    We used four main building blocks in the CNN model: convolution, pooling, rectified linear unit (ReLU), and a fully connected layer. The designed model has $4$ convolutions, $3$ max-pooling, $1$ flatten, and $1$ fully connected layer as described in~\cite{dehzangi2017imu}. The inputs fed to the CNN model are the images of the TF representation of the $4s$ window with the label learning or not-learning as explained above. 
    \item Long Short-term Memory (LSTM)-based Model: We exploited a recent LSTM-based model in the literature, which is a Deep Spatio-Temporal Convolutional Bidirectional LSTM Network (DSTCLN)~\cite{jeong2019classification} to classify the EEG signal into two classes; learning and not-learning. Details of the DSTCLN model are provided in Appendix~\ref{sec:DSTCLNlearning}.    
\end{itemize}

\begin{table*}
\centering
\caption{Accuracy ($\%$) of the SVM, CNN, and DSTCLN classifiers for two separate sessions of WCST averaged across ten participants. The sensitivity of the classifiers and variance of the $10$-CV results are averaged across sessions 1 and 2. 
} 
\label{tbl:cnnres}
\small
\begin{tabular}{|c|cccc|cccc|cccc|}
\hline
  & \multicolumn{4}{c|}{\textbf{SVM}}  & \multicolumn{4}{c|}{\textbf{CNN}}   & \multicolumn{4}{c|}{\textbf{ (DSTCLN)}}   \\\hline
 \textbf{\rotatebox{90}{Channel}}  & \rotatebox{90}{Session1} & \rotatebox{90}{Session2} & \rotatebox{90}{Sensitivity} & \rotatebox{90}{Variance}  & \rotatebox{90}{Session1} & \rotatebox{90}{Session2} & \rotatebox{90}{Sensitivity} & \rotatebox{90}{Variance}  & \rotatebox{90}{Session1} & \rotatebox{90}{Session2} & \rotatebox{90}{Sensitivity} & \rotatebox{90}{Variance} \\\hline
\midrule
$Fz$   &   $58.36\%$  &   $51.53\% $   & $59.35\%$ &  $8.17\%$  &  $81.03\%$    &    $77.16\%$   &$84.27\%$  & $6.26\%$  &  $84.81\%$    &    $83.35\%$   & $86.18\%$  & $5.82\%$  \\
$F3$   &   $67.42\% $ &   $70.12\%$    & $73.64\%$ &  $11.26\%$  &  $71.92\%$    &    $72.73\% $  & $75.23\% $ & $ 7.56\%$   &  $73.23\%$    &    $75.32\%$   &$78.19\%$  & $8.12\%$ \\
$F4$   &   $59.13\%$  &   $65.89\%$    & $64.83\%$ &  $9.27\%$   &  $78.73\%$    &    $73.45\%$   & $80.34\%$  & $8.81\%$   &  $79.87\%$    &    $77.16\%$   & $80.42\%$  & $9.14\%$ \\
$C3$   &   $49.83\%$  &   $57.61\% $   & $45.72\%$ &  $13.39\%$  &  $69.81\% $   &   $ 65.32\% $  & $66.38\%$  & $11.72\%$   &  $72.63\%$    &    $70.31\% $  & $74.13\% $ & $10.21\%$ \\
$C4$   &   $56.27\%$  &   $62.18\%$    & $51.67\%$ &  $16.31\%$  &  $72.31\%$    &    $74.28\%$   & $75.13\%$ & $8.21\%$   & $ 75.18\%$    &    $77.07\% $  & $78.68\%$  & $7.62\%$ \\\hline

\end{tabular}
\end{table*}

In all these three classifiers, 10-fold cross-validation (10-CV) was applied\footnote{10-CV divides the total input data of $n$ samples into ten equal parts. In every iteration, 1 part is considered a test sample set, and the remaining $9$ parts are considered for validation and training sample sets. There is no overlap between the test sample set ($10\%$ of data) with the validation and training sample set ($90\%$ of data).}. 
We report the results of the classification accuracy of the three classifiers in Table~\ref{tbl:cnnres}. In particular, the three classifiers were used on the EEG data collected from the five channels that we considered from the frontal lobe. In Table~\ref{tbl:cnnres}, we report the average across all the participant populations in the two different sessions, as we explained before.  
As Table~\ref{tbl:cnnres} illustrates, the classification accuracy of DSTCLN outperforms the classification accuracy of SVM per EEG channel. Moreover, in  DSTCLN, the accuracy of classification using the $Fz$ channel is higher than the other channels. Table~\ref{tbl:cnnres} also illustrates the sensitivity and the variance of the $10$-CV results of the classifiers. As the classification results of Table~\ref{tbl:cnnres} show, channel $Fz$ outperforms other channels and depicts the correlation between the inferred learning class (learning or not-learning) and that region of the brain.

In this analysis, we studied the EEG channels separately and not a combination for two reasons. First, we need to localize the part of the brain related to learning new concepts. For this purpose, we considered each channel's performance independently to determine the regional correlation between EEG channels' location and firing neurons activated during the learning event. Second, we need to reduce the computation cost of processing the EEG signal as we aim to use it in IoT systems. Hence, we find the best-performing channel instead of considering all channels or a different combination.
Moreover, although the DSTCLN model performs better than the CNN model, it comes with a higher computation cost. In particular, DSTCLN has $\approx 30$ million parameters compared to $\approx 28$ million parameters in the CNN model.  
More details on the timing analysis and the processing overhead of the EEG data will be discussed in Section~\ref{sec:evaluation}.

\subsection{Insights from Rule-Based Concept Learning}\label{insightRule}
From the rule-based concept learning experiment, we gained three main insights:
\begin{itemize}[leftmargin=*, noitemsep]
    \item Using TF representation, we observed that consecutive correct moves correlate with higher spectral activities in the higher frequency ranges, as shown in Figure~\ref{fig:tflearn}. 
    Moreover, the spectral power of the EEG signal from the $Fz$ channel shows an increase of $76.52\%$  when the learning event occurs (five consecutive moves) as reported in Table~\ref{tbl:spectral} (addressing \textbf{Q1} in the introduction).
    \item Results from three different classifiers (DSTCLN, CNN, and SVM) across two different sessions of WCST averaged on ten participants showed that the $Fz$ channel can be used to classify the learning versus not learning events (addressing \textbf{Q1} in the introduction).  
    \item The total correct moves by each participant in two sessions shows no experience or memory involved in the results. As illustrated in Figure~\ref{fig:demog}, there is on average $\pm 4 \%$  change between sessions in the total number of the correct moves for all participants. 
\end{itemize}

We will use the insight we gained from the rule-based learning experiment that learning event correlates positively with the high-frequency sub-bands of the frontal lobe channels of the EEG signal to study the explanation-based concept learning as explained in Section~\ref{sec:exp-based}.

\section{Explanation-based Concept Learning}\label{sec:exp-based}

While the rule-based concept learning experiment gave us the necessary insights to classify the learning event, there are two pivotal points that we need to consider. First, humans can learn through techniques other than rule-based concept learning. Second, although we needed to use a high accuracy EEG cap (Figure~\ref{fig:EEGdevice}b) to exploit the feasibility of classifying the learning events, we need to consider more socially acceptable wearable devices as we envision human-in-the-loop IoT learning system~\cite{elmalaki2021towards}. 

As highlighted in the background section (Section~\ref{sec:related}), we exploit another concept learning approach called ``explanation-based concept learning''. Explanation-based learning suggests that humans learn new concepts by experiencing examples and forming a basic outline. 
Accordingly, explanation-based learning can be viewed as presenting concepts in some form or ``lecturing'' new materials to humans. Hence, in this section, we pick teaching materials explained through a video and present them to different participants to understand the effect of explanation-based concept learning on brain activity in different presentation modalities.

\subsection{Presentation Modalities for Learning}
By the definition of this learning methodology, no specific event indicates the learning event akin to choosing the correct rule in rule-based concept learning. Hence, we need to assess whether the human retained and learned the content of the videos through some questionnaires while monitoring the EEG signals under different presentation modalities. In particular, we used two modalities of the same teaching material: traditional modality and immersive modality. The traditional modality version of videos is recorded under the traditional two-dimensional ($2$D) videos. The immersive modality version of the videos is recorded as three-dimensional ($3$D) videos. The $3$D video content is created by converting video from $2$D to $3$D using tools such as the Ani3D~\cite{Ani3D}, which creates imagery for each eye from one $2$D image. This $2$D-to-$3$D conversion adds the binocular disparity depth cue to digital images perceived by the brain, thus, significantly improving the immersive effect when broadcasted in Virtual Reality (VR).  

While previous research has explored the effect of $3$D and, more specifically, the impact of VR on brain engagement~\cite{babini2020physiological}, what we aim to do through this experiment is to be able to quantify the learning event in the explanation-based concept learning using a wearable device in the two different presentation modalities (traditional and immersive).

\subsection{Data Collection Using Wearable Device}
We recruited $15$ healthy participants for this experiment.
Since caffeine affects brain activity and causes inconsistent results, participants were asked not to consume beverages containing caffeine within $24$ hours before the experiments~\cite{siepmann2002effects}.

We chose different topics depending on the participant's backgrounds to ensure they did not have prior knowledge of the presented material in the video. In particular, we picked videos covering topics on biology~\cite{videobio}, modern architecture~\cite{videomodernarch}, and space~\cite{videospace}. The duration of these videos is $\approx 7.5$~minutes, $\approx 7$~minutes, and $\approx 12$~minutes, respectively. We divided each video into two parts and converted them into $3$D, meaning that we have $2$ versions of the same video content in $2$D and $3$D. Hence, we have $6$ videos ($2$D and $3$D) and $6$ sets of a questionnaire for each video. 

\begin{figure}
  \centering
  \subfigure[EMOTIV Epoc\_x and Oculus devices are worn by a participant. The left screen in the figure casts the view of the participant on the Oculus device, and the right screen presents the EEG signal collected from the EMOTIV device.]{\includegraphics[scale=0.036]{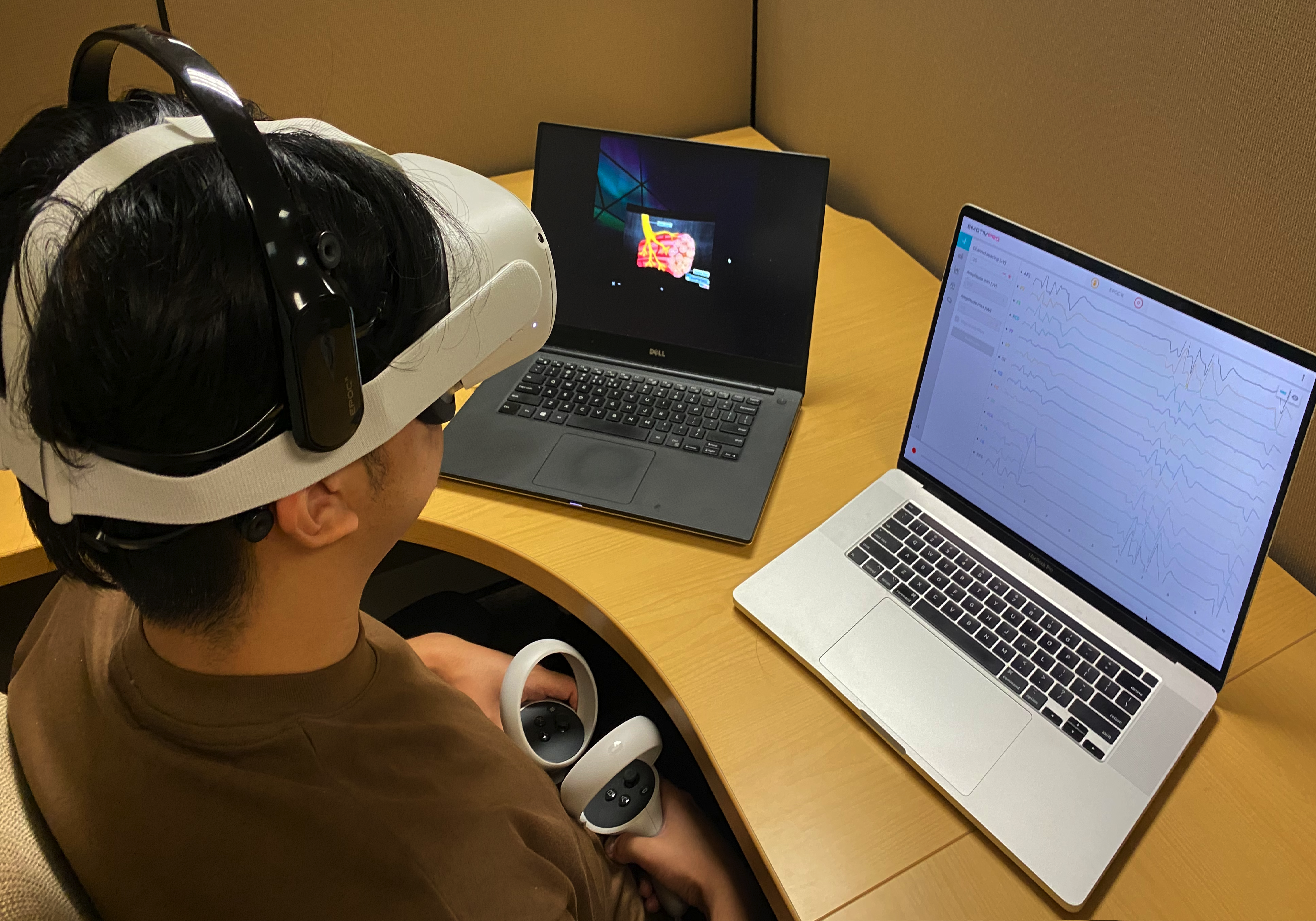}}\quad
  \subfigure[Screenshot of the view of the participant while watching biology content on an Oculus device with an office background.]{\includegraphics[scale=0.134]{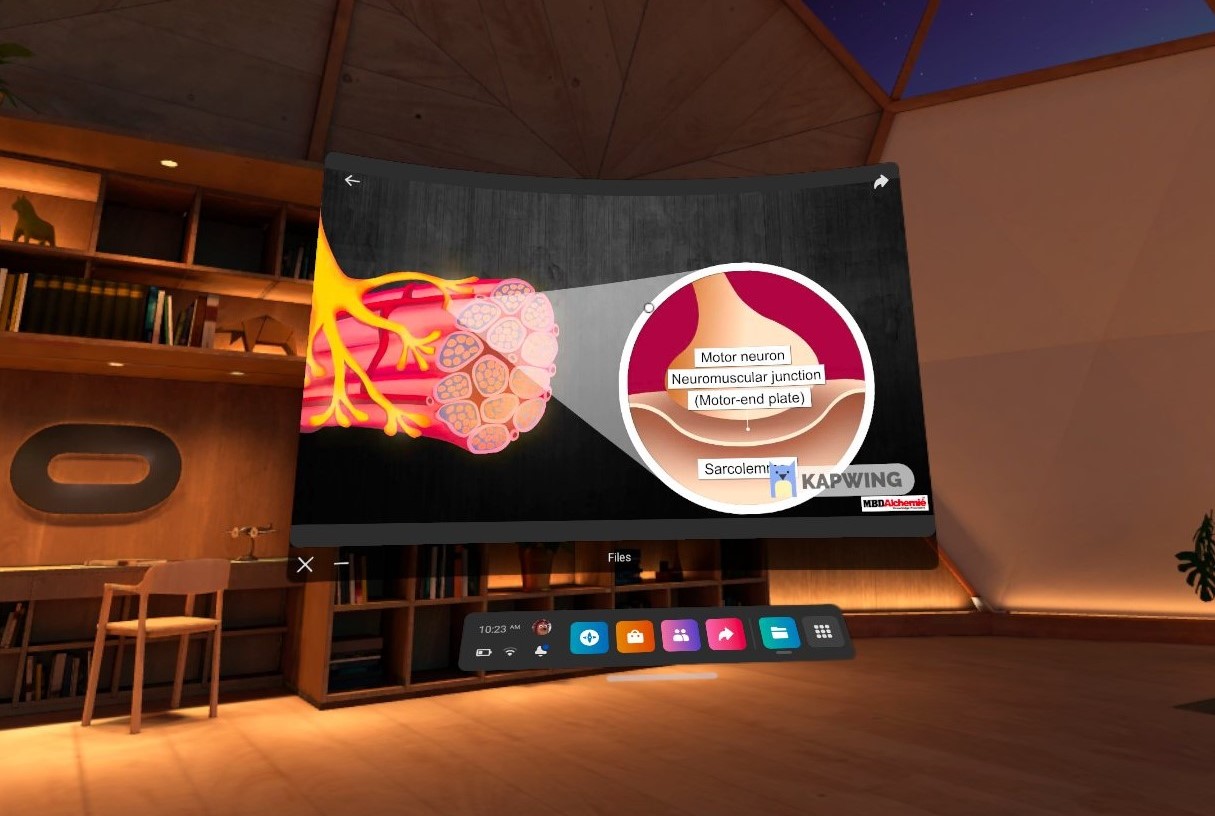}}
  \caption{EMOTIV Epoc\_x and Oculus devices worn by a participant for Virtual Reality (VR) learning environment.}
     \label{fig:VR}
\end{figure}

We recorded the EEG signal using EMOTIV EPOC\_X~\cite{emotiv}, a commodity EEG wearable device. We used the $14$ Channel EEG headset with a sampling frequency of $256$ $Hz$. Figure~\ref{fig:VR}a shows the wearable EMOTIV EPOC\_X device placed on the participant's head with the Oculus device. Electrode arrangement was a $14$-channel subset of $24$-channel $10-20$ systems where the P3 and P4 channels are the ground references. Table \ref{tbl:paramEMOTIV} provides the details and settings of the EMOTIV EPOC\_X EEG device. For the $3$D videos, we used the Oculus VR setup and streamed the $3$D video on it. Figure~\ref{fig:VR}b shows an example of what a participant visualizes in the VR environment while watching one of the $3$D videos.

Before the experiment, we introduced the $3$D video presentation broadcasted on the VR device (Oculus) to participants to familiarize them with the technology to reduce any possible effects of facing new technology in the EEG signal.  
The experiment protocol started by asking the participants to watch the traditional $2$D video or the immersive VR $3$D video. Afterward, the participants were asked to answer three questions related to the content. After a $1$-minute rest period, the participants switched to the other modality ($2$D to VR or VR to $2$D) of the same content. Three questions were then asked again, followed by another rest period for $1$~minute. The order of presenting the $2$D and VR videos is random. There is no emphasis on delivering one modality first, meaning that we show the $2$D video first and then the VR video; in other cases, it was vice versa. We continued this protocol to collect EEG signals from the participants (along with the responses to content-related questions) with the rest of the $6$ sets of videos, providing the participants with $1$~minute of rest between watching the videos. To ensure the repeatability of the result, we made two sessions separated by one month for the $15$ participants and changed the questionnaire in the second session.

\begin{table}[!t]
 \centering
 \caption{Parameters of EMOTIV EPOC\_X EEG setup for explanation-based concept learning.}
\label{tbl:paramEMOTIV}
\small
  \begin{tabular}{|p{0.4\columnwidth} | p{0.5\columnwidth}|}
    \hline
\textbf{ Parameter}     & \textbf{Description}  \\\hline\hline
   Amplifier     & $14$-channel device (EMOTIV~\cite{emotiv})\\
   Sampling frequency  & $256$ $Hz$      \\
   Bandwidth    & $0.16$ – $43$ $Hz$  \\
   Electrode arrangement    & $14$-channel subset of $24$-channel $10-20$ systems (Figure~\ref{fig:VR}a)      \\
  Ground reference  & $P3$, $P4$\\
 Electrode type	     & Ag/AgCl       \\
    \hline
  \end{tabular}

\end{table}

\subsection{Analysis}
Unlike rule-based concept learning, no correct/wrong moves during the experiment can be used as classification labels in explanation-based concept learning. Hence, we need another approach to quantify the learning event in the EEG data. One approach is to compare the complexity of brain activity during the different types of videos and their correlation with the performance in the questionnaire.    

\subsubsection{Brain Engagement Using Fractal Dimension} One method that has been implemented to explore and compare the complexity of the EEG signal during the learning experiments is fractal dimension~\cite{foroutan1999advances}\footnote{Fractal dimension, as the primary quantitative measure of fractal theory, indicates the complexity of the process in which greater values of a fractal dimension reflect greater complexity of the object.}. Various methods have been developed to calculate the fractal dimension, mainly based on the entropy concept. In this experiment, we used the box-counting method to calculate the fractal dimension~\cite{foroutan1999advances} on the recorded time-series EEG data. In particular, we used the $F3$ and $F4$ channels and averaged their values for our analysis\footnote{EMOTIV device does not have an $Fz$ channel, so we relied on the neighbor channels $F3$ and $F4$ and averaged their output signal to have an approximation of the region.}.

\begin{table}
\small
\centering
\caption{Fractal dimension of EEG signals with two presentation modalities (traditional ($2$D) and immersive (VR)) averaged on all $15$ participants. The lecture contents are biology, architecture, and space.}\label{tbl:fract}
\begin{tabular}{ *{3}{|c||cc}}
\hline
 \textbf{Content and } &  \multicolumn{2}{c|}{\textbf{Fractal dimension  }}  \\ 
\textbf{Presentation Modality}   & \textbf{Session 1} & \textbf{Session 2}  \\\midrule
Biology ($2$D)  &  $ 1.7319 $&   $1.7311$   \\
 Biology (VR)  & $1.7393$  &    $1.7439$  \\
 Architecture ($2$D)    & $1.7302$    &     $1.7312$     \\
  Architecture (VR)   & $1.7405$    &     $1.7398$     \\
  Space ($2$D)  & $1.7339$    &       $1.7311$   \\
    Space (VR)     & $1.7421$    &     $1.7409$    \\\hline
\end{tabular}
\end{table}

\begin{figure*}
\centering
\begin{tabular}{l l l l l}
    {\includegraphics[trim={2cm 2.9cm 12.6cm 2cm},clip, scale=0.23]{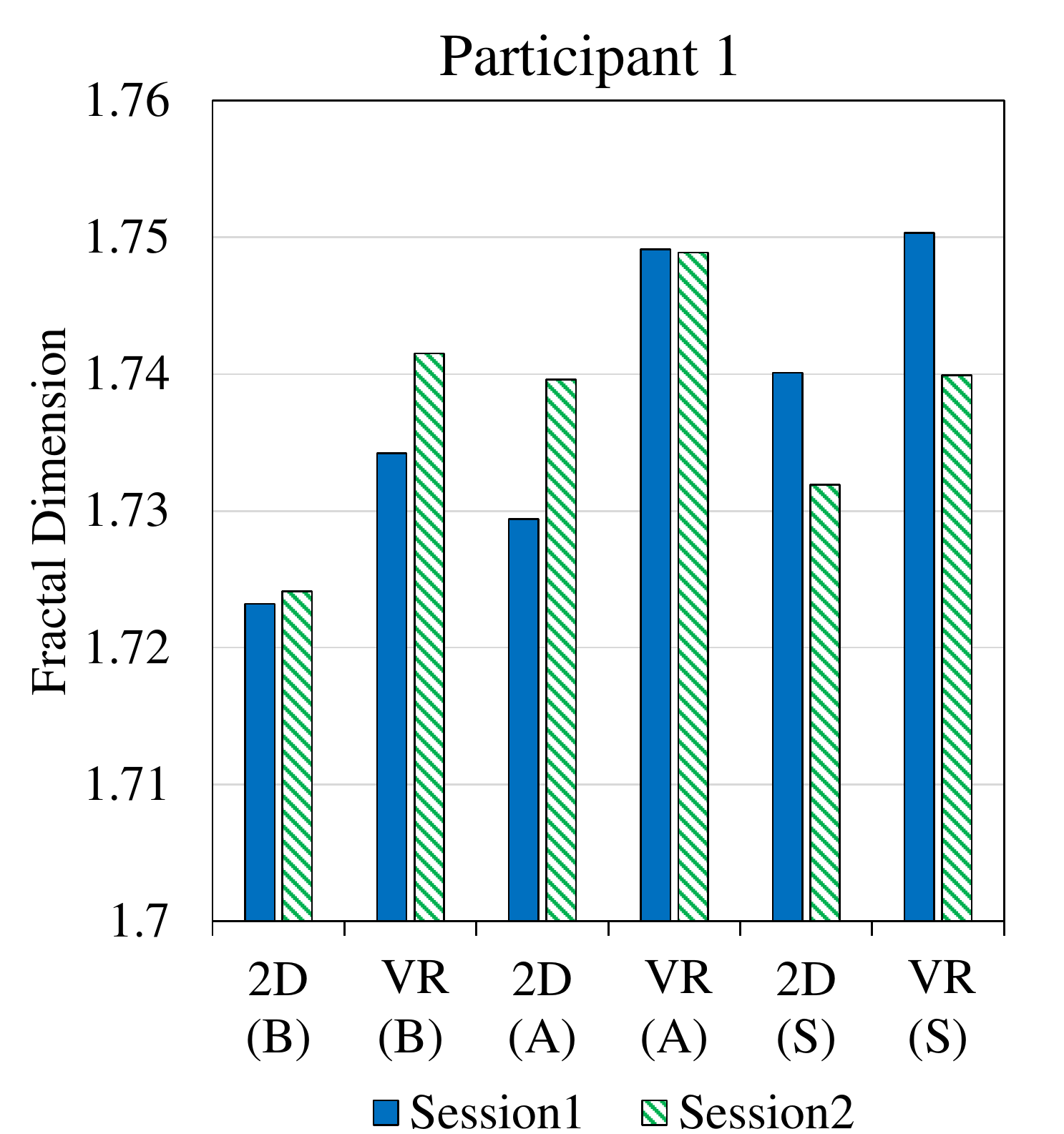} }
     {\includegraphics[trim={2cm 2.9cm 12.6cm 2cm},clip, scale=0.23]{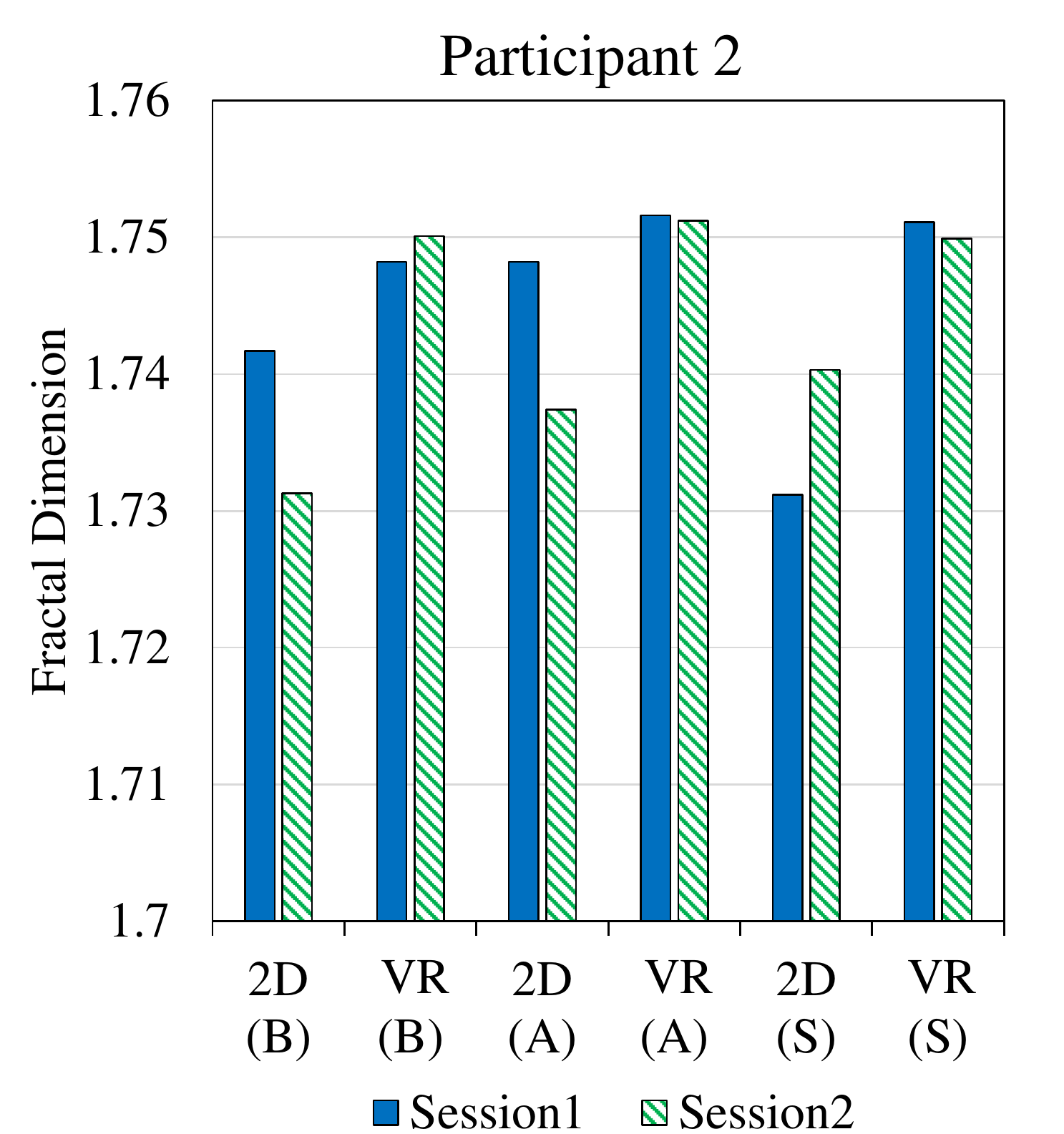} }
     {\includegraphics[trim={2cm 2.9cm 12.6cm 2cm},clip, scale=0.23]{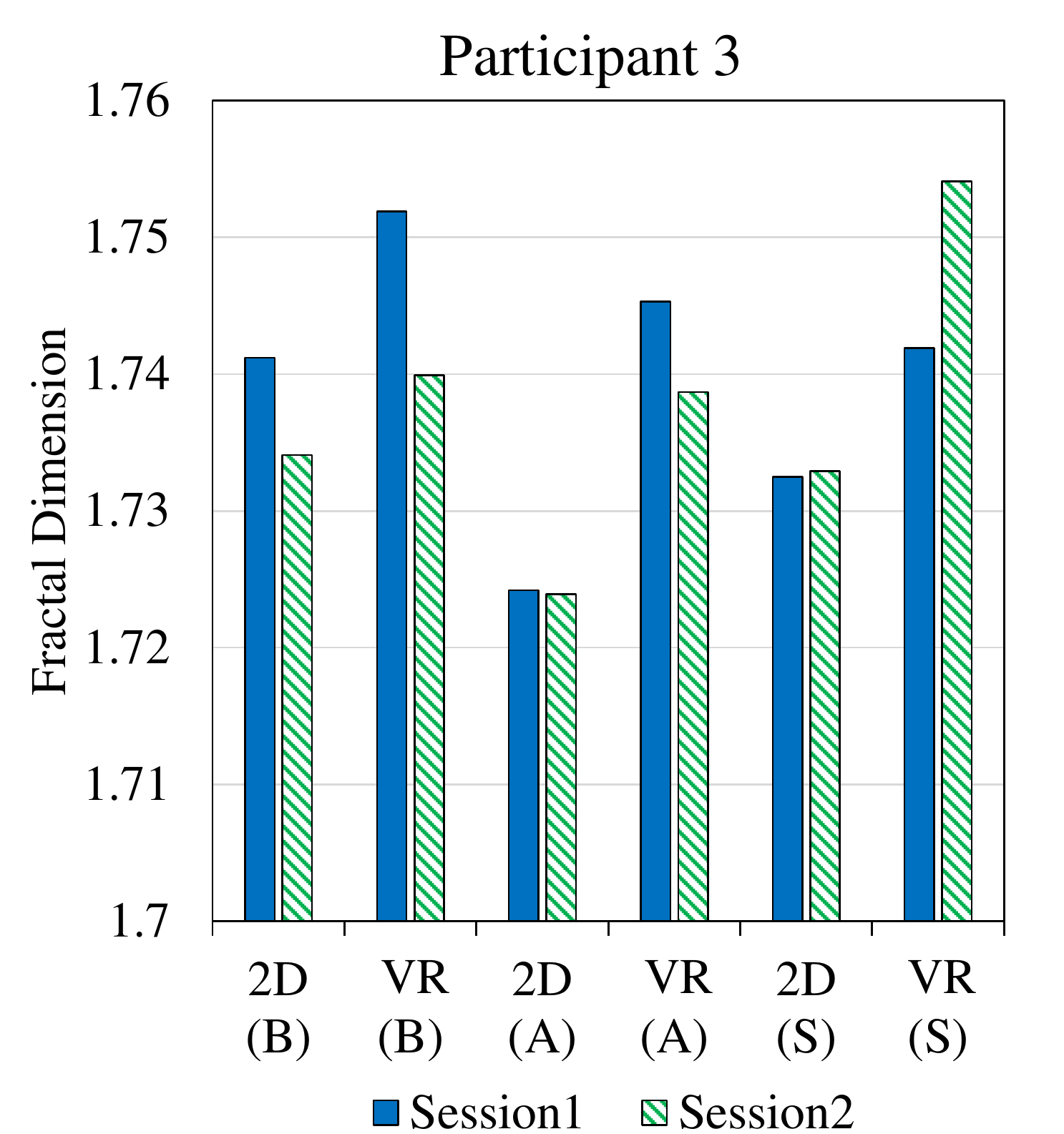} }
     {\includegraphics[trim={2cm 2.9cm 12.6cm 2cm},clip, scale=0.23]{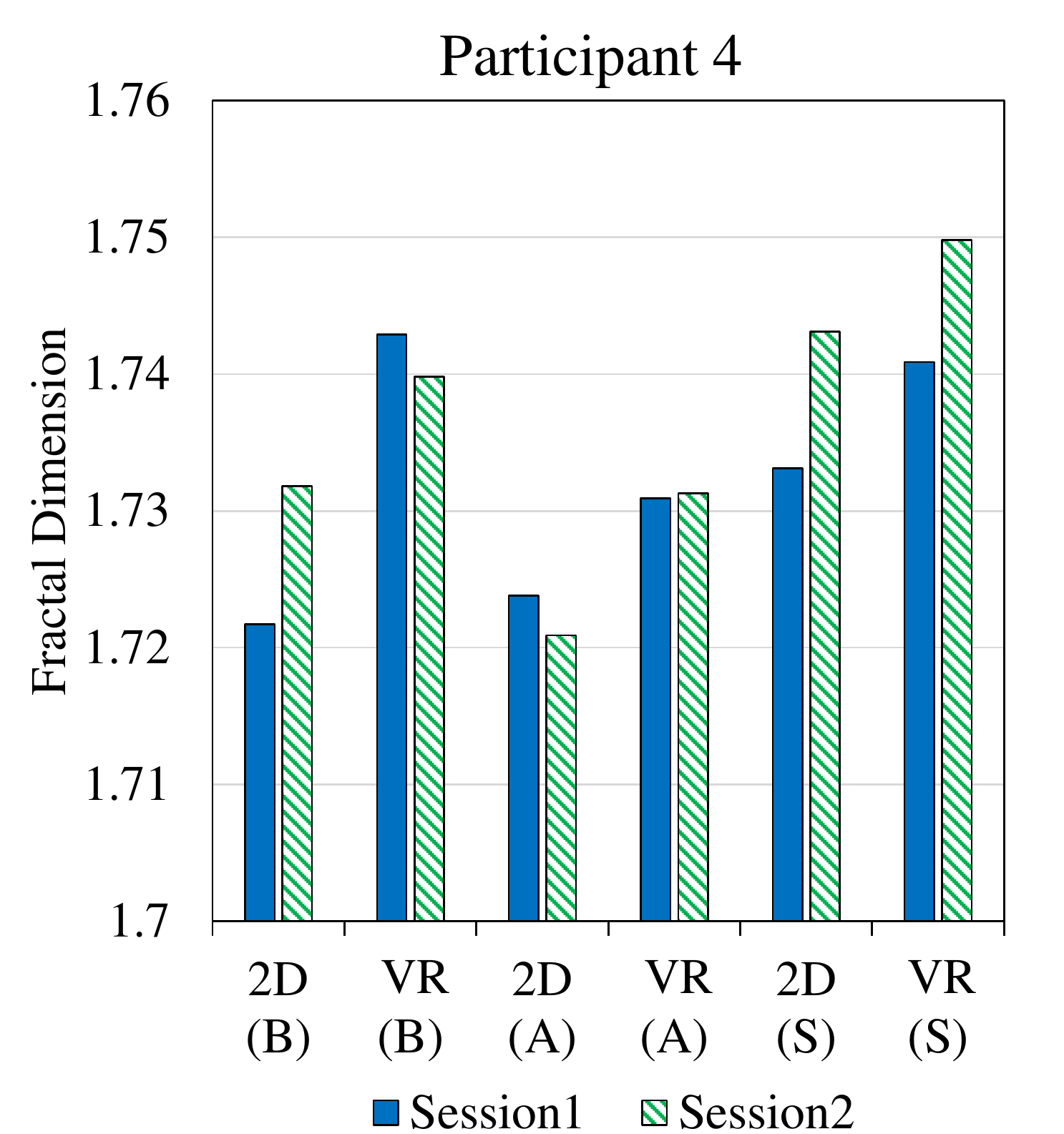} }
     {\includegraphics[trim={2cm 2.9cm 12.6cm 2cm},clip, scale=0.23]{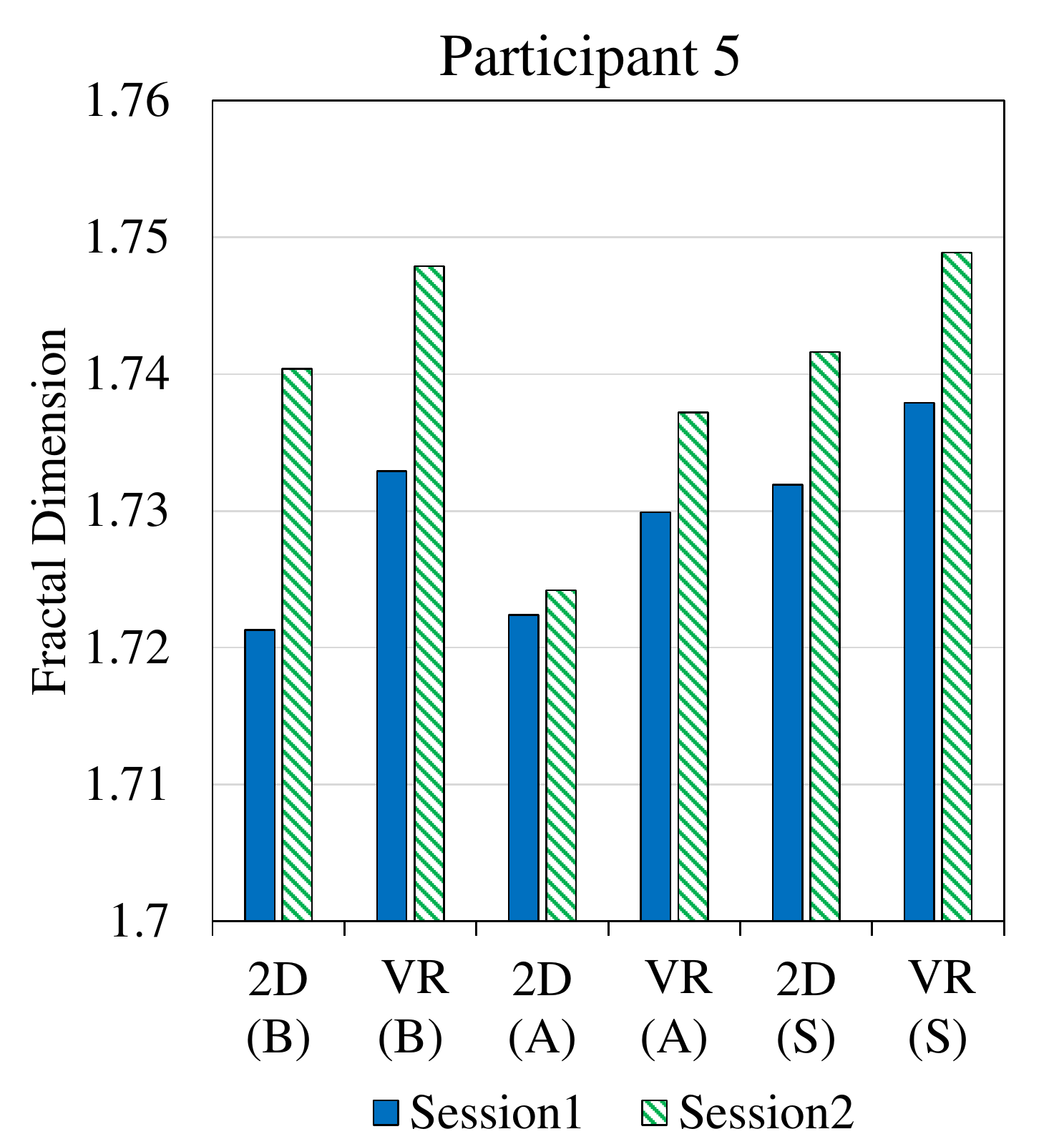}}
     \\
     {\includegraphics[trim={2cm 2.9cm 12.6cm 2cm},clip, scale=0.23]{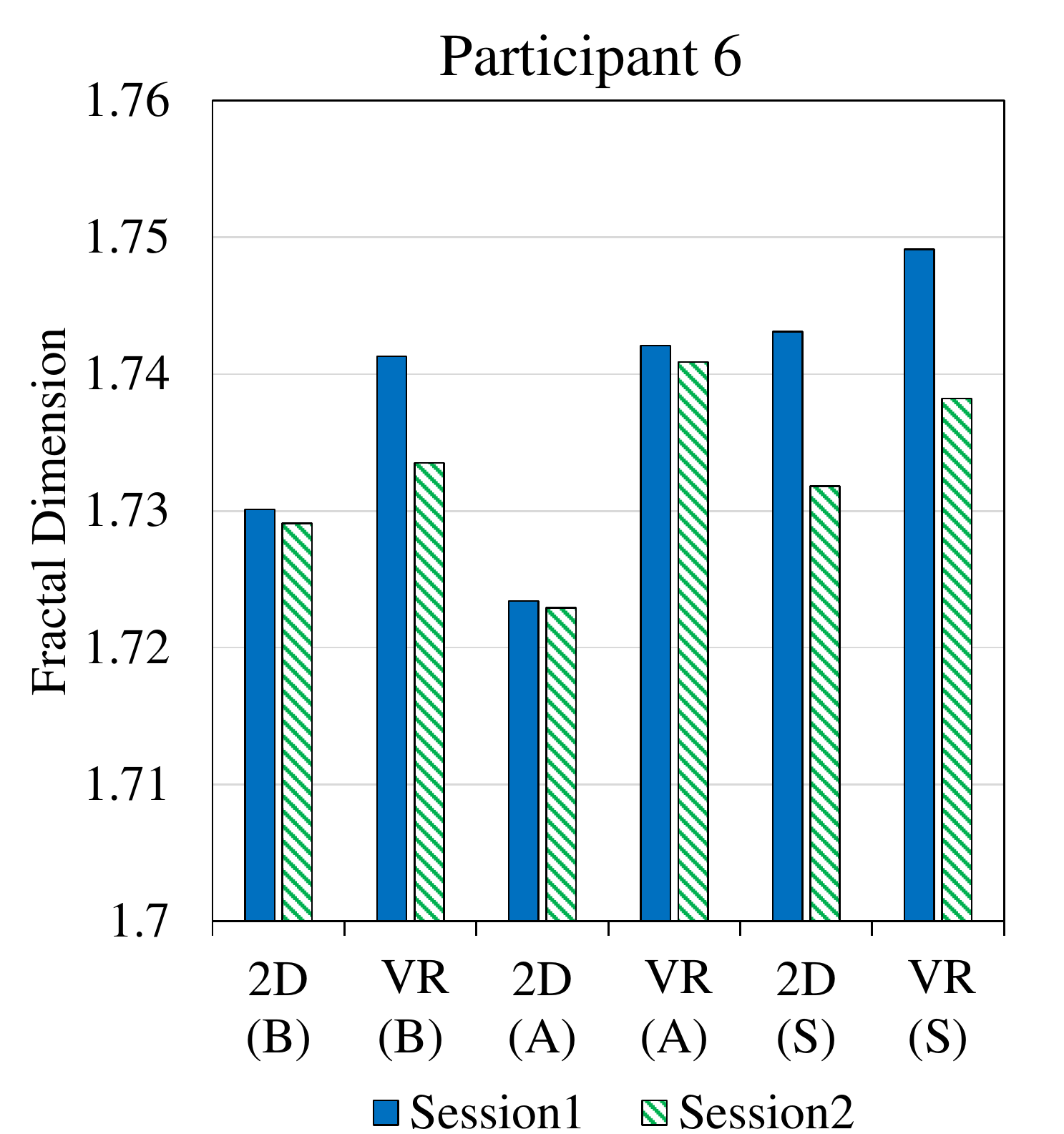} }
     {\includegraphics[trim={2cm 2.9cm 12.6cm 2cm},clip, scale=0.23]{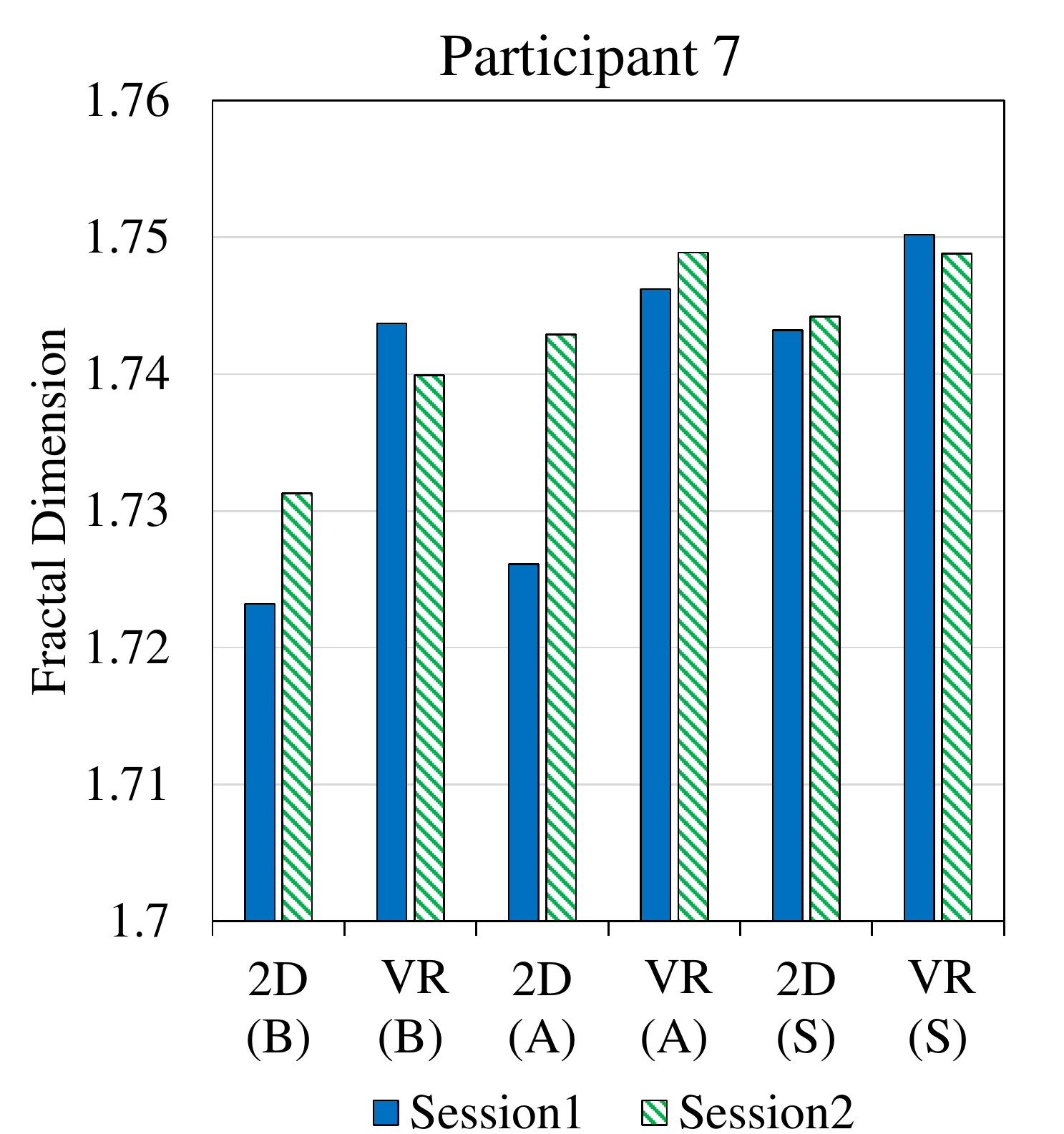}}
     {\includegraphics[trim={2cm 2.9cm 12.6cm 2cm},clip, scale=0.23]{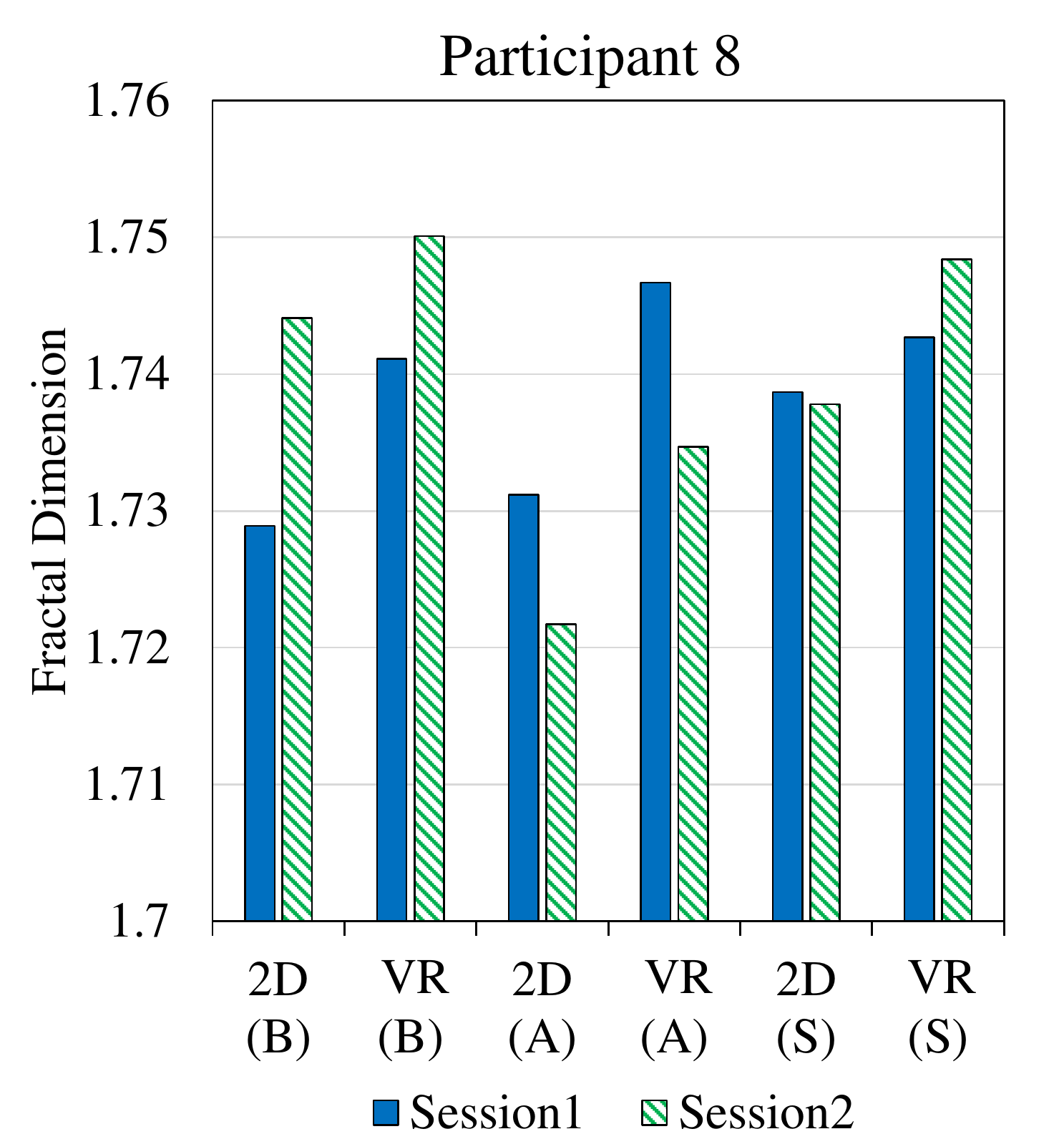} }
     {\includegraphics[trim={2cm 2.9cm 12.6cm 2cm},clip, scale=0.23]{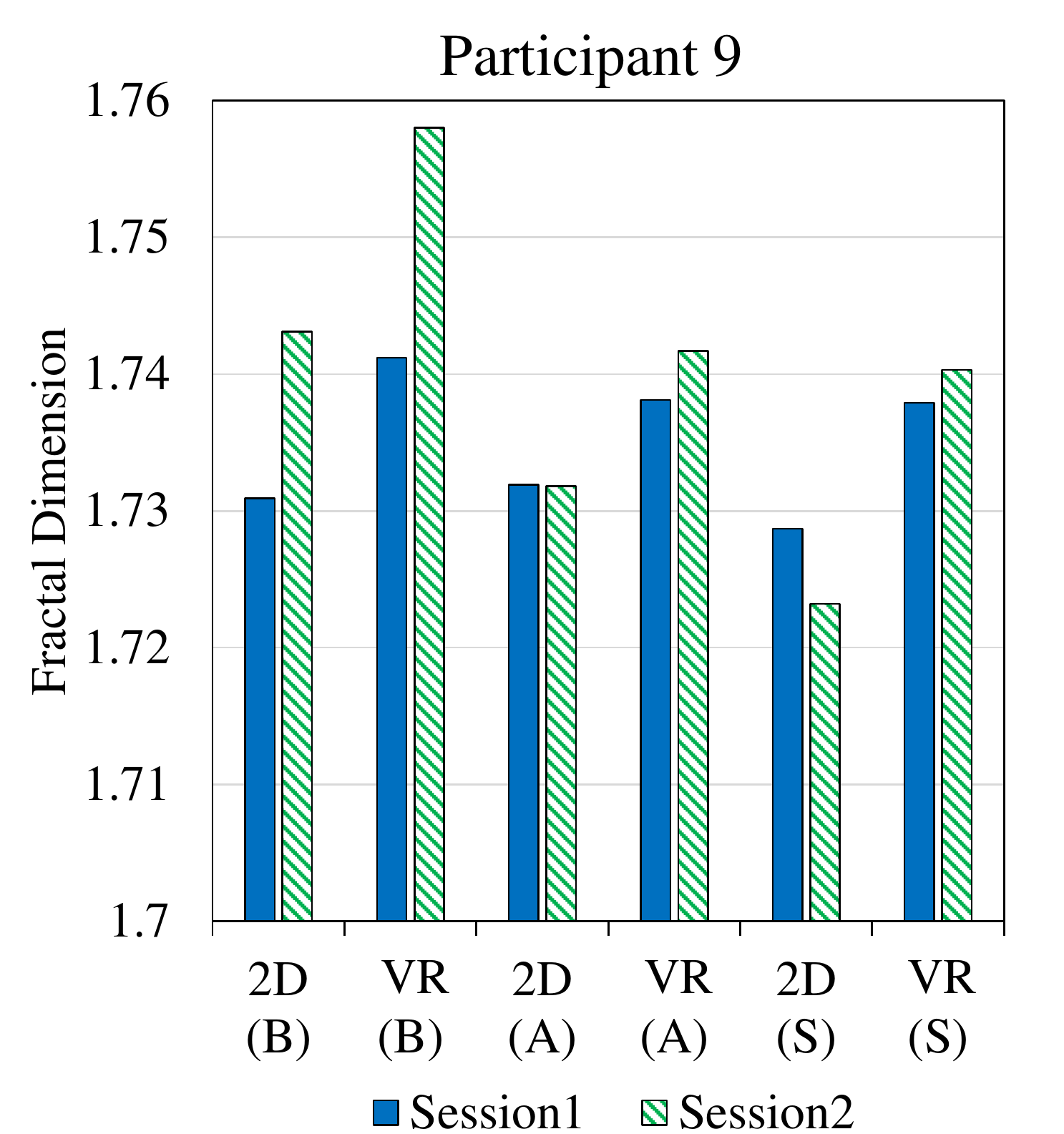} }
     {\includegraphics[trim={2cm 2.9cm 12.6cm 2cm},clip, scale=0.23]{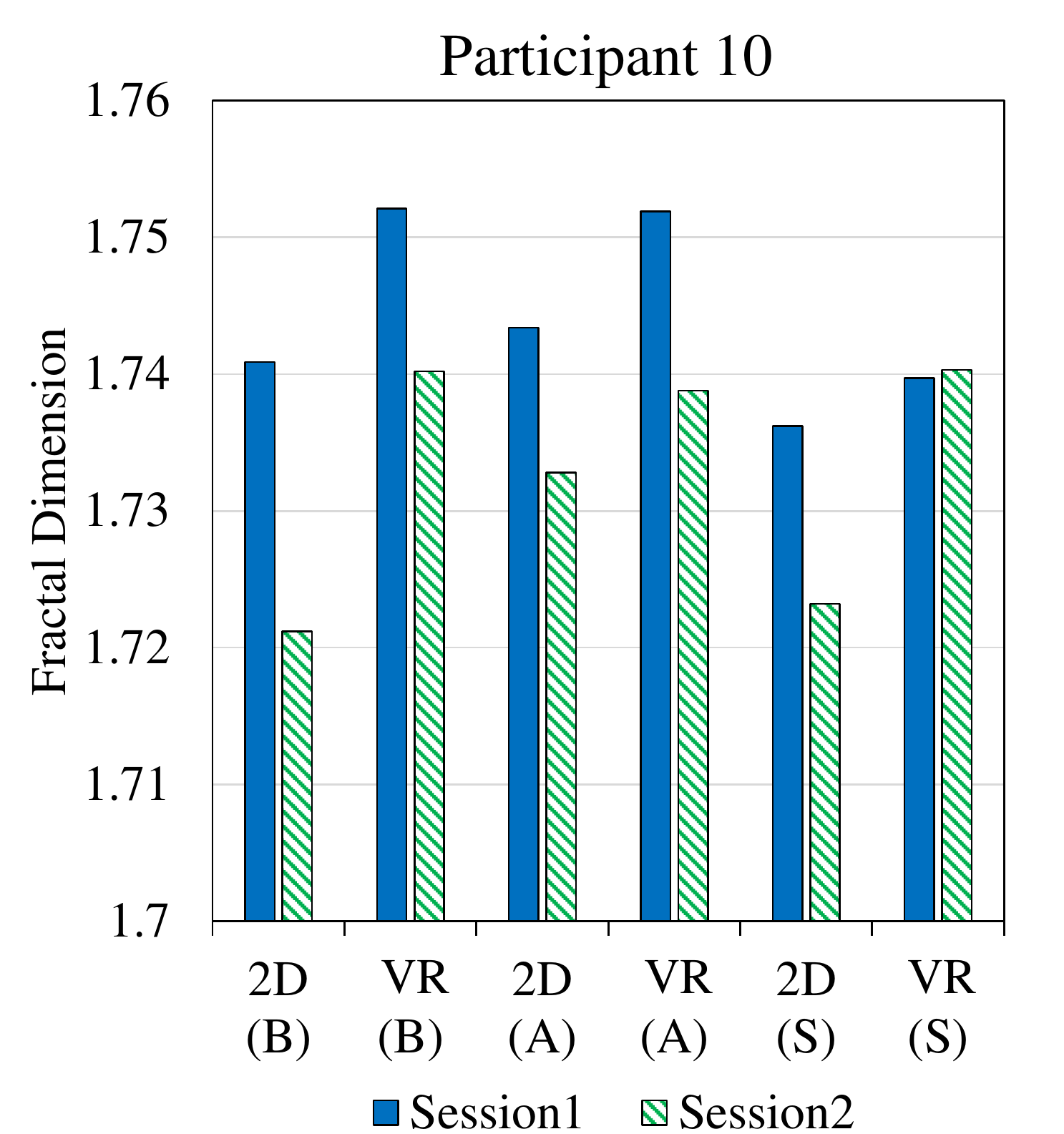}}
     \\
     {\includegraphics[trim={2cm 2.9cm 12.6cm 2cm},clip, scale=0.23]{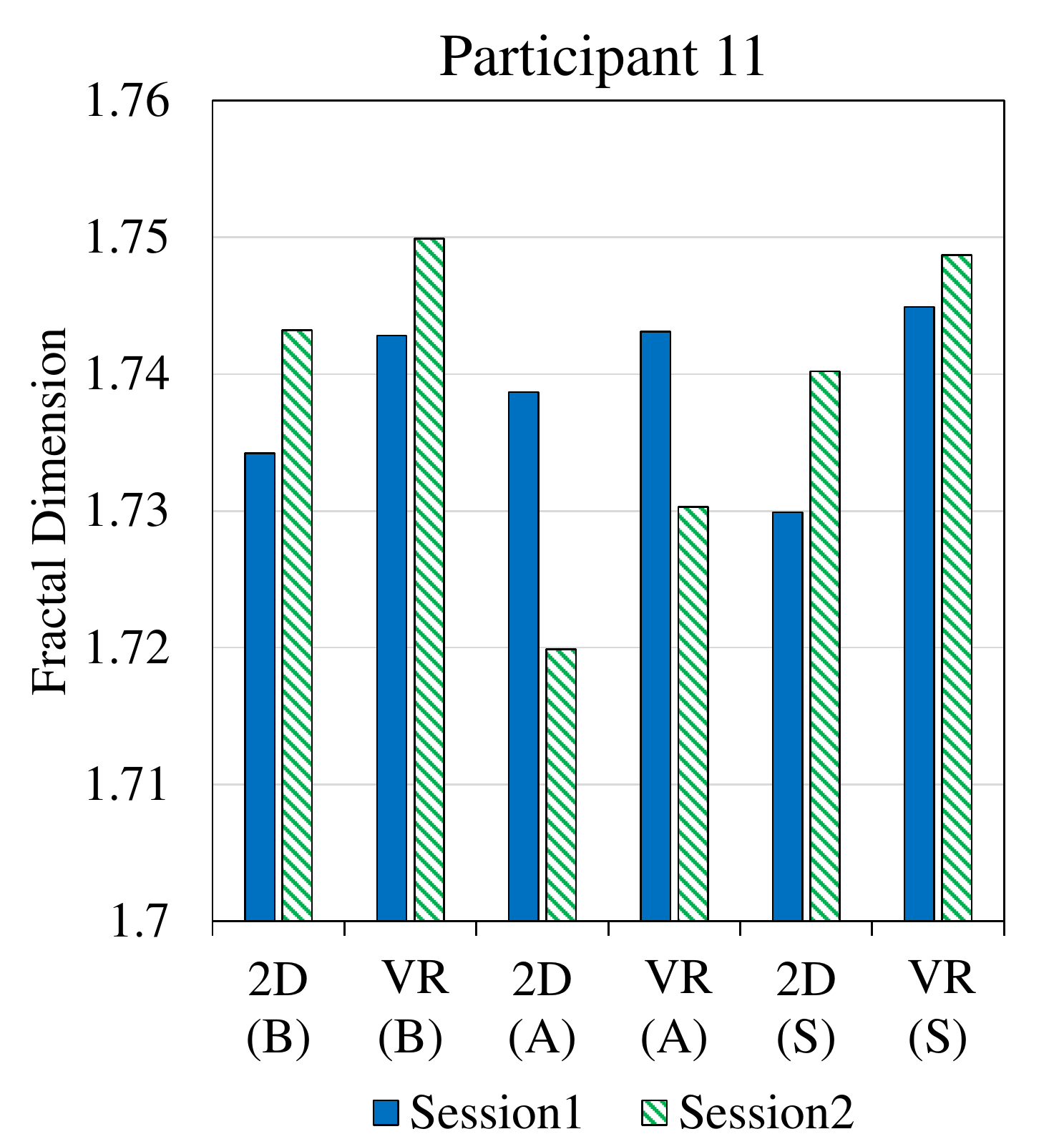} }
     {\includegraphics[trim={2cm 2.9cm 12.6cm 2cm},clip, scale=0.23]{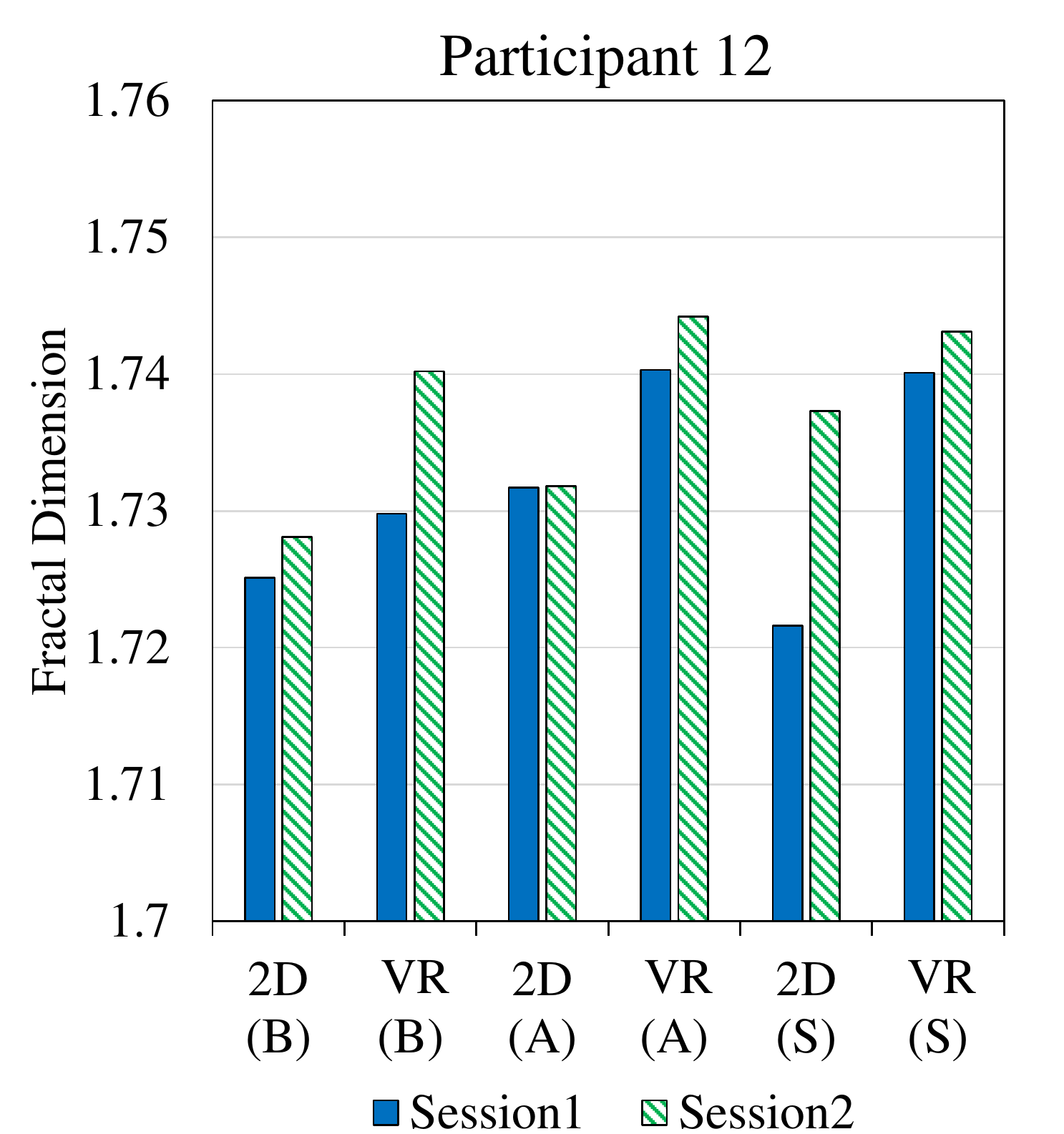} }
     {\includegraphics[trim={2cm 2.9cm 12.6cm 2cm},clip, scale=0.23]{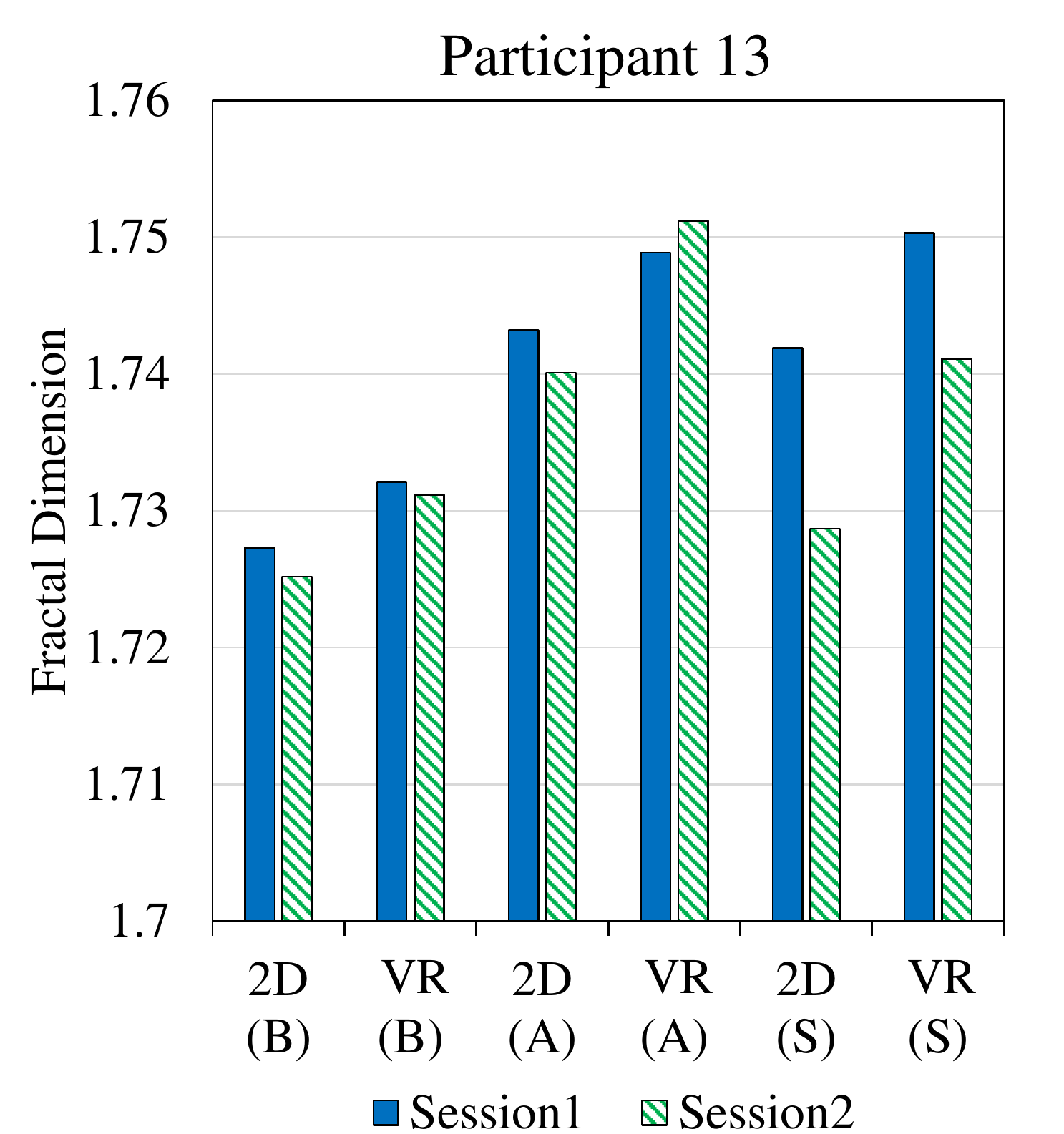} }
     {\includegraphics[trim={2cm 2.9cm 12.6cm 2cm},clip, scale=0.23]{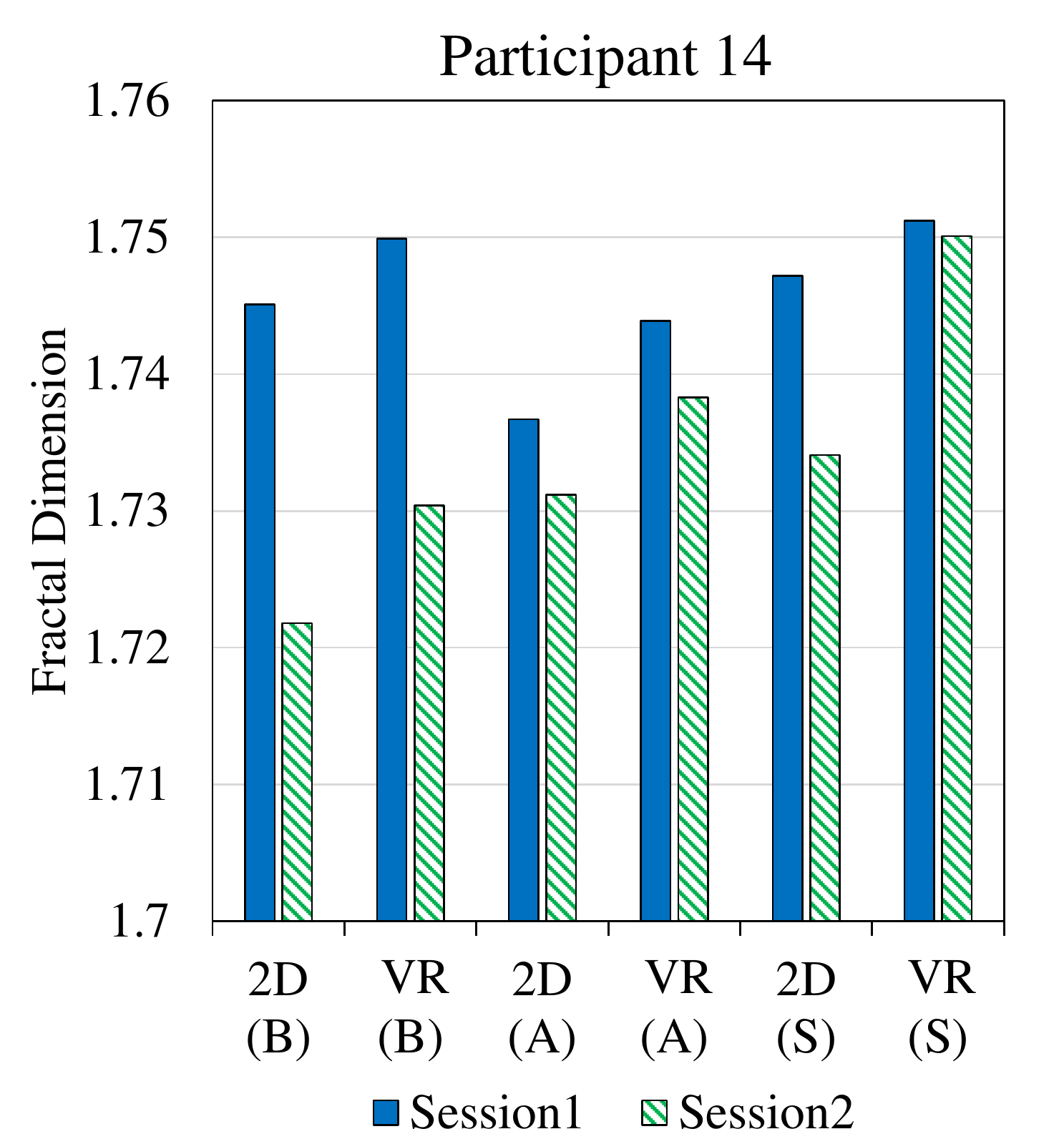} }
     {\includegraphics[trim={2cm 2.9cm 12.6cm 2cm},clip, scale=0.23]{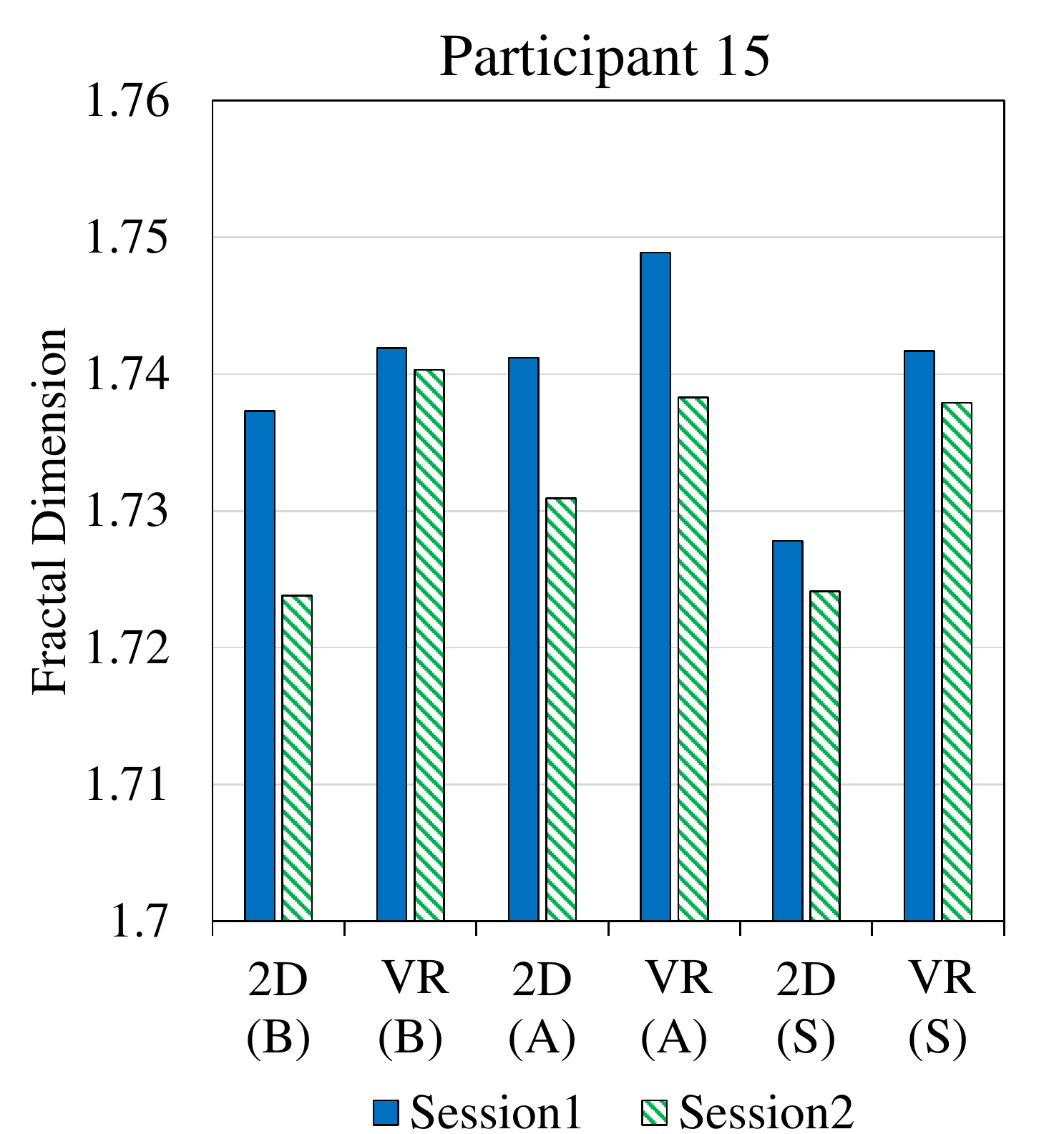}}
\end{tabular}
\caption{Fractal dimension of EEG signal of $15$ participants exposed to two presentation modalities (traditional $2$D and immersive VR). Each chart represents a participant, and two sessions are depicted for each participant. B: Biology, A: Modern Architecture, and S: Space. }
\label{fig:allsubFD}
\end{figure*}

After preprocessing (filtering and noise removal including high-pass filter and ICA as explained in Section~\ref{sec:analysis_rule},  
we applied the box-counting algorithm and examined the fractal dimension of all the $15$ participants across both sessions using the two presentation modalities. In this step, we do not window the EEG signal because there is no exact label for learning events. Thus, we process the EEG signal as a whole. In Table~\ref{tbl:fract}, we report the average fractal dimension across all $15$ participants, and Figure~\ref{fig:allsubFD} shows the fractal dimension for the $15$ participants across the two sessions. In general, we observe that the fractal dimension of the EEG signal is higher in the immersive VR modality compared to the traditional $2$D videos. Since the fractal dimension reflects the complexity of the signal, this result indicates that the EEG signal is more complex in response to VR visual stimuli than traditional $2$D visual stimuli. In other words, the human brain becomes more engaged with a stimulus when presented in the immersive VR compared to the traditional $2$D.

While the fractal dimension analysis gives us insights about brain engagement as a \emph{comparative} approach between the traditional $2$D and the immersive VR during the learning, we continue the analysis by investigating whether the same insights from the rule-based concept learning can be applied here.

\subsubsection{Correlation of Spectral Power of the EEG Signal and the Learning Performance}\label{sec:explanationSeptral-temporal}
We use the insight from the rule-based learning experiment that the learning event correlates positively with the high-frequency sub-bands of the frontal lobe channels of the EEG signal. In particular, we want to investigate whether the same insight can be applied to explanation-based concept learning. To this end, we measure the power of the spectral component of the EEG signal ($10-25$ $Hz$ band of the channels $F3$ \& $F4$) in the two presentation modalities and its correlation with the participants' answers in the questionnaire as our ground truth. Table~\ref{tbl:exp-power} provides the spectral power of the high-frequency sub-bands of the EEG signal in the two sessions of the experiment. As Table~\ref{tbl:exp-power} shows, there is an increase in the spectral power from the traditional $2$D presentation mode to the immersive VR mode in different video contents and sessions. The increase in the spectral power of the high-frequency sub-band is aligned with the increase of the performance ($\approx20\%$) provided by the questionnaire at the end of each content. 

\begin{table}
  \small
  \centering
    \caption{Spectral power of the high-frequency sub-bands ($10$ - $25$ $Hz$) of the EEG signal for two presentation modalities in the explanation-based concept learning experiment. Each column is averaged on $3$ video content, including biology, modern architecture, and space. 
  }\label{tbl:exp-power}
  \begin{tabular}{|l||cc|cc|}
    \hline
       & \multicolumn{2}{c|}{\textbf{ Traditional $2$D}}  & \multicolumn{2}{c|}{\textbf{Immersive VR}}   \\
  & Session 1& Session 2  & Session 1 & Session 2\\
  \textbf{Participant}  & ($v^2 Hz^{-1}$)&  ($v^2 Hz^{-1}$)&  ($v^2 Hz^{-1}$) &  ($v^2 Hz^{-1}$)\\
\midrule
Participant 1   &   $0.075$ & $0.059$  &   $0.102$ & $0.102$ \\
Participant 2   &   $0.078$ & $0.082$  &	$0.111$ & $0.129$ \\
Participant 3   &   $0.069$ & $0.071$  &	$0.130$ & $0.119$ \\
Participant 4   &   $0.087$ & $0.073$  &	$0.121$ & $0.131$ \\
Participant 5   &  $0.081$ & $0.097$  &   $0.118$ & $0.129$ \\
Participant 6   &   $0.083$ &$ 0.065$  &   $0.133$ & $0.102$ \\
Participant 7   &   $0.065$ & $0.062$  &	$0.102$ & $0.095$ \\
Participant 8   &   $0.089$ & $0.075$  &	$0.121$ & $0.114$ \\
Participant 9   &   $0.084$ & $0.091$  &	$0.121$ & $0.126$ \\
Participant 10   &   $0.087$ & $0.083$  &  $ 0.119$ & $0.131$ \\
Participant 11   &  $ 0.091$ & $0.088$  &   $0.112$ & $0.136$ \\
Participant 12   &   $0.061$ & $0.077$  &	$0.113$ & $0.125$ \\
Participant 13   &   $0.083$ & $0.089$  &	$0.141$ & $0.127$ \\
Participant 14   &   $0.069$ & $0.095$  &	$0.110$ & $0.130$ \\
Participant 15   &   $0.091$ & $0.097$  &   $0.118$ & $0.129$ \\\hline\hline
\textbf{\cellgrey Average}     &   $0.079$ & $0.081$  &	$0.118$ & $0.122$ \\\hline
  \end{tabular}

\end{table}

\subsection{Insights from Explanation-based Concept Learning}\label{insightExplain}
By conducting this experiment, we aimed to investigate the correlations between the EEG signal and explanation-based concept learning.
The questionnaire served as the ground truth and showed that participants performed better during the immersive VR presentation (by~$20\%$), which correlates with an increase in fractal dimensions and spectral power of the higher spectral components of the EEG signals from the frontal lobe. This insight aligns with the insights from rule-based learning.

The rest of the paper will discuss how we utilize the insights gained from rule-based and explanation-based experiments about learning to design \sysname, a human-in-the-loop IoT learning system.

\section{\sysname: Human-in-the-loop IoT learning system}\label{sec:erudite}

In the previous sections, we gained some insights into how the human brain can react to learning events in two different learning approaches, rule-based concept learning, and explanation-based concept learning. We will utilize these collected insights on cues in learning, where the correlation occurs between EEG dynamics and learning to design \sysname, a human-in-the-loop IoT learning system that can adapt the learning environment based on the current human learning state. To design \sysname, we need to consider the following aspects:

\begin{figure*}[!t]
\centering
{\includegraphics[scale=0.27]{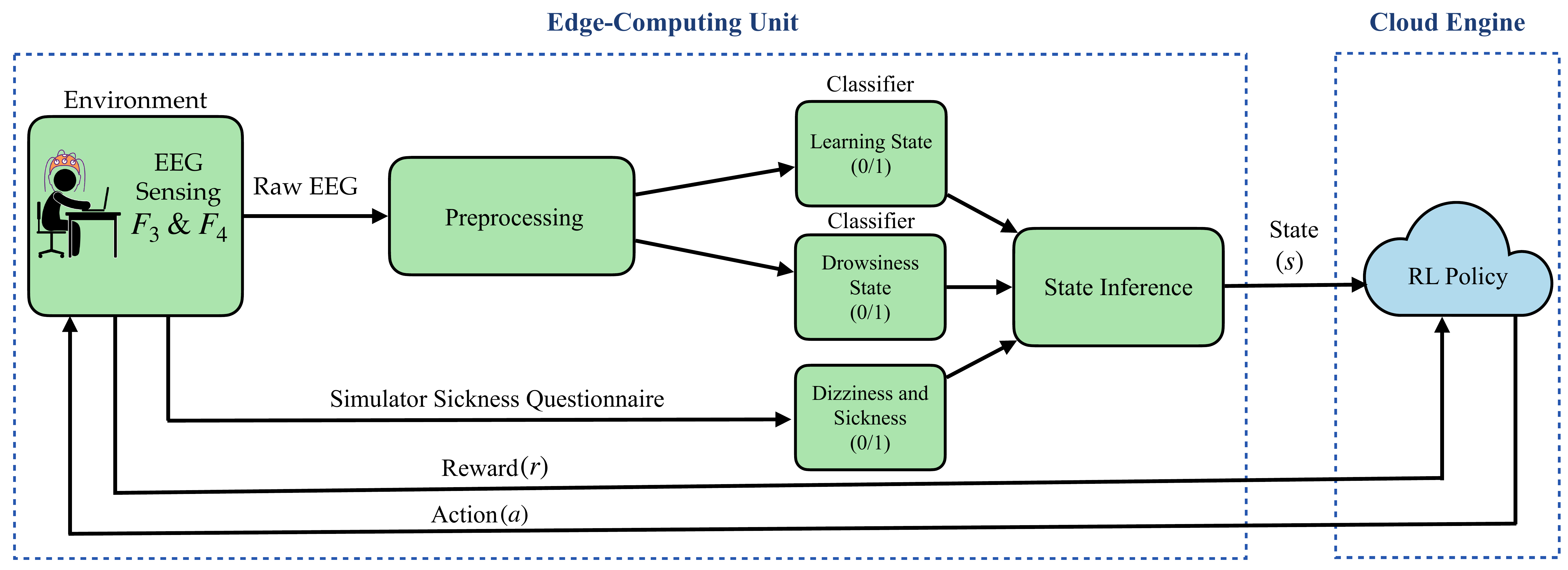}}
\caption{\sysname framework for learning IoT system. EEG data is collected using a wearable device to infer the human state ($s$). The human state is determined through three features; the learning state ($LS$), the drowsiness state ($DS$), and the simulator sickness scale ($SSQ$). The current human state $s$ is used as an input to the \sysname RL engine, which selects an adaptation action $a$ to enhance the human learning experience. This adaptation action is tuned based on a feedback reward ($r$) received from the human-in-the-loop environment.} 
\label{fig:policy} 
\end{figure*}

\begin{itemize}[noitemsep, leftmargin=*,  topsep=0pt]
\item Different mental states can affect human learning ability, such as drowsiness and stress~\cite{owens2001sleep, hwang2009automated, ha2021wistress}. 
\item Adapting the learning environment can have different effects on different humans. For example, some humans can experience dizziness and simulator sickness if exposed to an elongated VR environment~\cite{kolasinski1995simulator}. 
\item To achieve the goal of a human-in-the-loop IoT learning system, the design of \sysname needs to be scalable and easy to deploy on edge devices. 
\end{itemize}

\subsection{IoT System Design}

The goal of \sysname is to provide personalized adaptation actions in the learning environment that best fit the human to enhance the learning experience. In particular, the current human state will be used as an input to an adaptation engine to determine the correct adaptation actions. These personalized adaptations will be tuned based on the human state while interacting in the learning environment. Accordingly, there are $4$ essential components for \sysname. First, we need to infer the human state in the learning environment. Afterward, we design an adaptation engine to provide the correct adaptation actions. Then, we need to be able to assess these adaptation actions if they enhance the learning experience. Ultimately, we need to ensure that that \sysname computation model can be scalable and easily deployed on edge devices. 

Figure~\ref{fig:policy} illustrates \sysname framework. Below we will explain the different components of this framework in detail.

\subsection{Human State in a Learning Environment}
Humans can lose focus or get drowsy during elongated teaching or training periods, especially in online or remote learning environments~\cite {terai2020detecting}, which can decline human learning performance. 
Moreover, as we incorporate this new reality in the post-COVID-19 era, new technologies are being adopted in learning environments. In particular, while Virtual Reality/Mixed Reality (VR/MR) technologies have been used heavily in the gaming and entertainment industry~\cite{baldauf2013augmented},
recent years showed that these new technologies would have a significant impact on the learning~\cite{ibanez2014experimenting}, working and training~\cite{irizarry2013infospot} sectors. Accordingly, we exploit these technologies during learning (as we showed in Section~\ref{sec:exp-based}). However, humans react differently to the VR environment. Some humans report dizziness and cybersickness symptoms during exposure to VR~\cite{brunnstrom2020latency}, which can affect the learning experience.

In \sysname, we focus on the current human learning state, the drowsiness state, and the dizziness state to infer the current human state. 

\subsubsection{\textbf{Learning State (LS)}}\label{sec:lindex}
Based on the insights from rule-based and explanation-based concept learning, we concluded that the event of human learning correlates positively with the high-frequency sub-bands ($10$ - $25$ $Hz$) as discussed in Section~\ref{Spectro-Temporal} and~\ref{sec:explanationSeptral-temporal}. To infer the human learning state at run-time ($LS$), we use the DSTCLN classification model described in Section~\ref{sec:wcstcalss} to classify EEG signal into two classes of learning $(1)$ and not-learning $(0)$. In  \sysname, we designed the experiment stage to be $10$ minutes. To measure the $LS$ for each stage, we segment the $10$ minutes of the EEG signal into $4$ second windows. Then, we estimate the $LS$ for each of $4s$ windows and use the majority voting among the $4s$ windows to determine the $LS$ for the duration of $10$ minutes.

We collected the EEG data from $F3$ \& $F4$ channels using the same EMOTIV\_X device setup (as described in Table~\ref{tbl:paramEMOTIV}) and averaged them for $LS$ inference. 
Accordingly, we can then measure EEG data at run-time and use it as a sensor modality to infer the human learning state.

\subsubsection{\textbf{Drowsiness State (DS)}}\label{sec:drowsy}
While different approaches in the literature used several sensor modalities to detect drowsiness levels, such as heart rate, respiration rate, and eyelid movement~\cite{solaz2016drowsiness,rostaminia2017ilid}, we can exploit the same EEG signal ($F3$ \& $F4$ channels) that we record to infer the drowsiness state. To quantify the drowsiness state using the EEG signal, we designed another DSTCLN model~\cite{jeong2019classification} to classify the EEG signal into two classes; drowsy $(0)$ and alert $(1)$. Details of the DSTCLN model are provided in Appendix~\ref{sec:DSTCLNdrowsy}.

\subsubsection{\textbf{Cybersickness and Dizziness}}\label{sec:dizzy}
The most commonly reported measure of cybersickness symptoms is the Simulator Sickness Questionnaire ($SSQ$). 
The $SSQ$ was derived directly from the Pensacola Motion Sickness Questionnaire (MSQ)~\cite{golding1998motion}. The MSQ consists of a list of $25$ to $30$ symptoms, such as spinning, tired/fatigued, and may vomit. Symptoms severity are rated on four levels, ``none'' (0), ``slight'' (1), ``moderate'' (2), and ``severe'' (3). A total score was computed by summing item scores. The highest score was determined to specify emesis as the worst case of sickness. In particular, the $SSQ$ is a selection of $16$-items as in Table~\ref{tbl:ssq} from the MSQ with a different scoring scheme.  
Based on four main subfactors of cybersickness, including Nausea ($N$), Oculomotor ($O$), and Disorientation ($D$), related scores for the symptoms for the specific subfactor are calculated as shown in Equation~\ref{totfact}. Afterward, a Total Score ($TS$) is computed, representing the overall severity of cybersickness experienced by the users of VR systems. 

Each subfactor is scaled to have a standard deviation of $15$ for all observations. Subfactor scores can range from $0$ to $30.54$ ($N$), $28.58$ ($O$), and $34.92$ ($D$). A score of total severity ($TS$) is derived by summing the raw (unscaled) subfactor scores, then multiplying that sum by $3.74$ as in Equation~\ref{totfact}. In particular, $TS$ can range from $0$ to $235.62$~\cite{stone2017psychometric}.

\begin{table}[!t]
\begin{center}
\caption{Calculations in the Simulator Sickness Questionnaire. Total is the sum obtained by adding the symptoms scores. Omitted scores are zero.
}\label{tbl:ssq}
\scalebox{0.78}{
\begin{tabular}{|l ||c|c|c|}
\hline    
     \multirow{2}{*}{\textbf{Symptoms}}   & \multicolumn{3}{c|}{\textbf{Subfactor}} \\\cline{2-4}
    & \textbf{Nausea (N)}     & \textbf{Oculomotor (O)}     & \textbf{Disorientation (D)} \\\hline
General discomfort  &  $1$   & $1$    &            \\
Fatigue  &     & $1$    &             \\
Headache  &    &  $1$   &          \\
Eye strain   &     &  $1$   &            \\
Difficulty focusing  &     & $1$    &     $1$  \\
Increased salivation   &   $1$  &     &       \\
Sweating  &   $1$  &     &       \\
Nausea  &  $1$   &     &   $1$    \\
Concentrating  &  $1$   &  $1$   &       \\
Fullness of head  &     &     &      $1$ \\
Blurred vision  &     & $1$    &     $1$  \\
Dizzy (eyes open)   &     &     &    $1$   \\
Dizzy (eyes closed)   &     &     &      $1$ \\
Vertigo   &     &     & $1$      \\
Stomach awareness   &  $1$   &     &       \\
Burping   &  $1$   &     &       \\\hline\hline

\textbf{\cellgrey Subfactor Total}  & \cellgrey N\_T   & \cellgrey O\_T     & \cellgrey D\_T       \\\hline
\end{tabular}
}
\end{center}

\end{table}

\begin{equation}
\label{totfact}
\begin{aligned}
\text{N} &= N\_T \times 9.54\\
O &= O\_T \times 7.58\\
D &= D\_T \times 13.92\\
TS   &= (N\_T \ + O\_T + D\_T) \times 3.74
\end{aligned}
\end{equation}

We use the reported value of $TS$ to monitor the humans' simulator sickness level. In particular, participants are asked to fill out the $SSQ$ questionnaire based on Table \ref{tbl:ssq}. Using this questionnaire, we calculate the $TS$ based on Equation~\ref{totfact}. The threshold for $TS$ to consider it a simulator sickness depends on the application~\cite{stone2017psychometric}.

It is recommened to choose $\delta_{SSQ}$ to be $\frac{TS_{max}}{4}$ where $TS_{max}=235.62$ as mentioned above~\cite{stone2017psychometric}. If the measured $TS$ is bigger than the $\delta_{SSQ}$ the $SSQ$ state is classified as dizzy $(0)$, and if it is less than $\delta_{SSQ}$ the $SSQ$ state is classified as is not-dizzy $(1)$. At the end of each experiment stage, participants are asked to fill the $SSQ$ based on Table \ref{tbl:ssq}. The result of the questionnaire is used to calculate the $TS$ based on Equation~\ref{totfact}.

\subsection{\sysname Adaptation Engine}

The human state ($s$), a tuple of ($LS$, $DS$, $SSQ$), is used to choose the best adaptation action to enhance the learning experience for the human. For example, if the human is drowsy and the learning state is decreasing, then one possible action could be to change the presentation modality to VR. However, this can cause dizziness/simulation sickness, as we explained before. Hence, another possible action could be to switch back to the traditional $2$D presentation or give a break. However, every human may react differently, and there are intrinsic inter- and intra- human variability. In particular, the same human within the same state can prefer an adaptation action at one time and prefer another adaptation action at another time (intra-human variability).
Moreover, different humans can have other preferences even if they are within the same state (inter-human variability)~\cite{elmalaki2018sentio, elmalaki2021fair, elmalaki2022maconauto}. Accordingly, the adaptation engine has to monitor the human's preference and the effect/response of the adaptation action on the learning experience to provide personalized adaptation actions. This state-action-response interaction to tune the adaptation engine fits perfectly within the Reinforcement Learning (RL) paradigm. In particular, in the RL paradigm, an agent interacts with an environment by observing the state of the environment and applying an adaptation action. The RL agent then learns if this action is a good or bad action through a notion of a reward. If the agent chooses a wrong action at a particular state, the agent receives a negative reward from the environment. In contrast, if the RL agent chooses a good action, the agent receives a positive reward. Through this interaction, the RL agent converges into a policy that determines the best action per state.

By designing this RL agent, we will answer the questions (\textbf{Q2} and \textbf{Q3}) we posed in Section~\ref{sec:intro}.

\subsubsection{Reinforcement Learning (RL)}
Due to the intrinsic variability in modeling the human preferences to particular adaptation actions, the Q-learning algorithm can be used to model this uncertainty. Based on the current state $s$, Q-learning chooses an action $a$. Learning the optimal policy $\pi(s, a)$ ---action per state that maximizes the total reward--, by applying an action ($a$) in a particular state ($s)$ and observing the next state ($s'$), the RL converges to the optimal policy that maximizes the total reward. With every interaction, the RL agent updates a value for every state-action pair ($Q(s, a)$) and receives a reward ($r(s, a)$) as follows:

\begin{equation}\label{eq:qupdate}
\setlength\abovedisplayskip{0pt}
\belowdisplayshortskip=-1pt
Q(s, a) \leftarrow Q(s, a) +  \alpha[r(s,a) + \gamma \max_a Q(s', a) - Q(s, a)] \vspace{-2mm}
\end{equation}

The hyperparameters $\gamma$ and $\alpha$ are known as the discount factor and the learning step size, respectively. 
To choose an action, $a$ at each state $s$ from the possible set of actions, an $\epsilon$-greedy policy can be adopted. We discuss the details of hyperparameter selection in Appendix~\ref{sec:hyper}.

\subsubsection{State Space}\label{sec:stateinferrence}

\begin{table}[!t]
    \caption{Human mental state in \sysname is one of 8 states depending on the binary classification of the learning state, drowsiness state, and simulator sickness (SSQ). } 
    \label{tab:state}
    \centering
    \begin{tabular}{|c||c|c|c|c|c|c|c|c|}
    \hline
         \cellgrey \textbf{$LS$} & \cellone & \cellone & \cellone & \cellone & \cellzero & \cellzero & \cellzero & \cellzero \\ \hline
         \cellgrey \textbf{$DS$} & \cellone & \cellone & \cellzero & \cellzero & \cellone & \cellone & \cellzero & \cellzero \\\hline
         \cellgrey \textbf{$SSQ$} & \cellone & \cellzero & \cellone & \cellzero & \cellone & \cellzero & \cellone & \cellzero \\\hline \hline
         \cellgrey \textbf{State} & $s_8$ & $s_7$ & $s_6$ & $s_5$ & $s_4$ & $s_3$ & $s_2$ & $s_1$ \\\hline
    \end{tabular}

\end{table}

In \sysname, human state refers to a combination of $3$ binary features, the learning state ($LS$), the drowsiness state  ($DS$), and the simulator sickness score ($SSQ$). We chose binary features to reduce our state space. In particular, the learning state is classified as learning $(1)$ versus not-learning $(0)$, the drowsiness state is classified as alert $(1)$  versus drowsy $(0)$, and $SSQ$ is classified as not-dizzy $(1)$ versus dizzy $(0)$. Table~\ref{tab:state} shows the state space that we consider in \sysname. Accordingly, the best human state is $s_8$, where the learning state of the human is high, the human is alert, and is not experiencing cybersickness. In contrast, the worst human state is $s_1$, where the human learning state is low and is experiencing drowsiness and cybersickness. Indeed, humans can transition between any of these states.

\subsubsection{Action Space}\label{sec:policy}
Based on the current human state, \sysname takes appropriate action to enhance the human learning experience. The action space in \sysname includes the following five actions. 
\begin{itemize}[leftmargin=*,  noitemsep, topsep=0pt]   
    \item \textbf{$a_1$}: Give a break to the human.
    \item \textbf{$a_2$}: Enable VR mode by switching from $2$D to $3$D\footnote{Switching from $2$D to $3$D means changing the presentation mode from $2$D video on laptop screen to a $3$D video presentation using VR device.}.
    \item \textbf{$a_3$}: Disable VR mode by switching from $3$D to $2$D.
    \item \textbf{$a_4$}: Changing the content of the presentation.    
    \item \textbf{$a_5$}: No change to the learning environment.
\end{itemize}

In particular, enabling the VR mode increases brain engagement and enhances learning performance, as shown in Section~\ref{sec:exp-based}. However, some humans may experience cybersickness with exposure to VR; hence \sysname may need to switch back to the traditional $2$D to reduce cybersickness symptoms. Moreover, a break during a learning session may also be needed to minimize drowsiness or cybersickness symptoms.

\subsubsection{Reward}\label{sec:reward} 
At the end of each experiment stage, the participant is asked to take a quiz about the presented lecture. The reward value is based on the participant's performance on the quiz and the transition state. The human performance in a quiz dictates the reward value after every learning module, where the score in this quiz is measured as a percentage. Furthermore, the next state, determined at the end of each stage of the experiment ($10$ minutes), is included in the reward as follows in Eq.~\ref{eq:reward}. States with higher learning performance, lower drowsiness, and lower $SSQ$ measures are considered better states with higher rewards. It means that state $s_8$ ($LS$=$1$, $DS$=$1$, and $SSQ$=$1$) and state $s_1$ ($LS$=$0$, $DS$=$0$, and $SSQ$=$0$) are the best and worst states in terms of the reward value, respectively. 

Quantification of the state transition for reward is linear in the $[0,100]$ range with the best state allocated value $100$. For instance, state $s_8$ and $s_1$ receive reward $r=100$ and $r=0$, respectively. For other states, reward values are distributed evenly (reward unit $=100/8$). Transitions to higher states (i.e., $s_6$ to $s_7$) receive one positive unit of reward, and transitions to lower states (i.e., $s_6$ to $s_5$) receive one negative unit of reward. 

The quiz result is in the range of $[0,100]$. Quiz quantification applies to $10$ multiple-choice questions uniformly as the $10/10$ and $0/10$ scores receive rewards $100$ and $0$, respectively. Similarly to the state transitions, quiz performance improvement (i.e., score $6/10$ to score $7/10$) receives one unit of positive reward ($+10$), and a drop in quiz performance (i.e., score $8/10$ to score $7/10$) receives one unit of negative reward ($-10$).

\begin{equation}\label{eq:reward}
r(s,a) = Quiz_{(grade)} +  State \  Improvement_{(s \rightarrow s')}  
\end{equation}

\subsection{\sysname IoT Design Choice}
As we envision \sysname to be used in the future smart classroom, we designed \sysname as an edge-cloud system. In particular, for a smart classroom, \sysname needs to consider all students' states to ensure fairness in the classroom and to enhance system performance through scalability. 
Indeed, sharing the human high-level state ($LS$, $DS$, and $SSQ$) can be a privacy concern. However, recent work in the literature showed that even if adaptation policies run on the edge and only the adaptation action is shared with the cloud, there can still be privacy leaks through monitoring the time-series of adaptation actions~\cite{taheri2023, elmalaki2019spycon, elmalaki2022vindico}. Hence, the decision to run the RL agent in the cloud is driven by the goal of enabling the tradeoff between privacy protection, fairness of adaptation, and efficacy within \sysname. 
We discuss some of these privacy and fairness concerns in the discussion in Section~\ref{sec:disc}.

\section{Evaluation}\label{sec:evaluation}

We designed a learning environment to evaluate the proposed \sysname framework. We used lecture contents from Khan Academy  
with quizzes that cover topics on biology~\cite{KhanBiology}, chemistry~\cite{KhanChemistry}, and physics~\cite{KhanPhysics}. We asked $15$ participants (eight male and seven female) to watch these lectures, all within the age range of $20-30$. The age average and standard deviation of the participants were $26$ and $2.16$, respectively. Each lecture is stand-alone and does not require any prior knowledge from the participants to be understood. The participants' main task was to watch the lecture and pay attention to answer the questions regarding the content at the end of the lecture.  
Originally, each lecture is $\approx 55$ minutes, in narrative style, and does not include any quizzes in the middle of them.

\subsection{System Implementation} \label{sec:sysimpl}

We used an EMOTIV EPOC\_X portable EEG device (as described in Table~\ref{tbl:paramEMOTIV}) to collect the EEG data from participants. Before the experiment, we present a $10$ minutes $2$D video presentation to measure the baseline for each participant. Every $10$ minute, the raw EEG  data are processed to measure the $LS$, and $DS$. $SSQ$ is calculated using an online questionnaire based on Table~\ref{tbl:ssq}. 
After measuring the baseline, we presented the lectures (selected from Khan Academy lectures) which were divided into $10$ minutes videos. Each $10$-minute video is called one stage. Overall,  $5$ stages (or chunks of $10$ minute durations) for each lecture\footnote{The average attention span of the human is $10$ to $15$ minutes~\cite{mckeachie2006teaching}.}. Between every stage, the human state is inferred as explained in~Section~\ref{sec:lindex}. For each stage of the experiment ($10$ minutes), we process every $4$ seconds and use the majority voting to infer the human state at the end of the stage.
Then, the \sysname policy runs to determine the adequate adaptation action as explained in Section~\ref{sec:policy}. In particular, the participant starts the video in the traditional $2$D presentation, and depending on \sysname policy, the learning environment can switch to VR, or the participant can take a break, or change the content, or the learning environment remains in the traditional $2$D presentation. Between every stage, \sysname evaluates the participant's state and applies an adaptation action.

\subsection {Results}
Table~\ref{tbl:state.action} shows the optimal action for each state for each participant for $5$ stages of the experiment (each stage duration is $10$ minutes). Each column represents the current state and each row represents the optimal action for the current state. State/action ($a/s$) pairs in each cell represent the action and future state (transitioned).
For example, when participant $3$ is in state $2$ (it is highlighted in green in Table~\ref{tbl:state.action}), the optimal action is $a_2$. Taking action  $a_2$ results in a transition from state $s_2$ to state $s_7$. In state $s_2$, the participant experiences low learning ($LS=0$), drowsy ($DS=0$), and no cybersickness ($SSQ=1$) (according to the Table~\ref{tab:state}). Based on the RL model, the optimal action for state $s_2$ is action $a_2$. Action $a_2$ is enabling the VR mode by switching from $2$D to $3$D presentation. As the participant$\ 3$ (in the state $s_2$) is not experiencing cybersickness but suffering from drowsiness and a bad learning state, switching to VR presentation mode is a reasonable action.

In another instance, participant $2$ is in state $s_6$ (it is highlighted in blue in Table~\ref{tbl:state.action}). Optimal action in state $s_6$ is action $a_4$. Taking action $a_4$ results in a transition to either state $s_8$ or state $s_7$ with an equal transition probability. In state $s_6$ participant $2$ suffers from drowsiness. VR mode is already enabled and participant $2$ is comfortable with it (in the state $s_6$, $SSQ=1$).  
In this situation, the RL model takes action $a_4$ as an optimal action which is changing the content of the lecture. By taking action $a_4$, in the next transitioned state, either the participant $2$ is yet comfortable with the VR mode (which transitions from $s_6$ to $s_8$) or it shows signs of cybersickness (which transitions from $s_6$ to $s_7$).
In both transitioned states ($s_7$ and $s_8$) participant $2$ is not experiencing drowsiness which was the problem in state $s_6$ and action $a_4$ aimed to address it. In the next part of the results, we show the adaptability of the \sysname to differences among the participants. 

\begin{figure*}[!t]
  \centering
  {\includegraphics[scale=0.4]{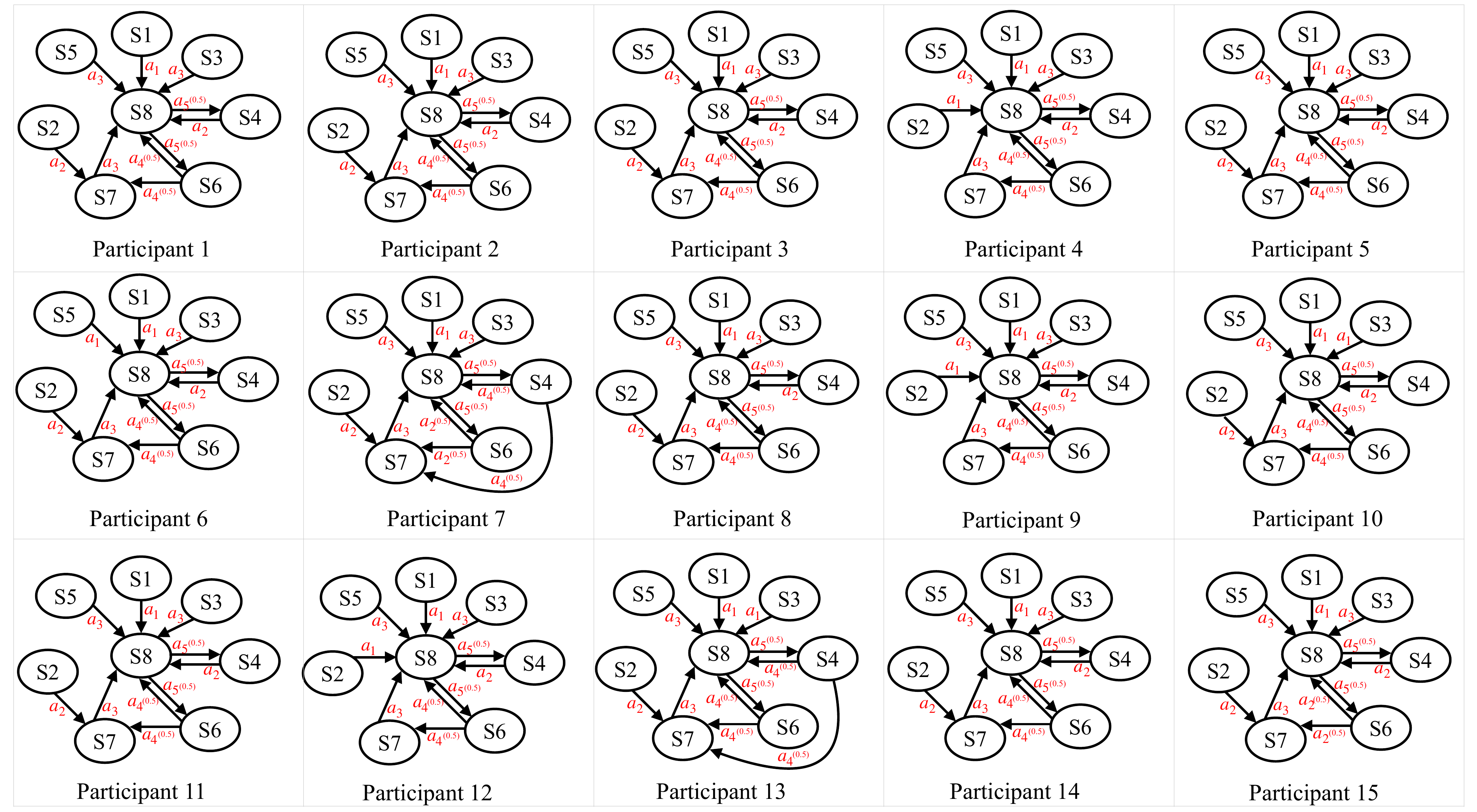}}
  \caption{Representatives of the state spaces of the participants for visualization. Each sub-figure depicts the states and converged actions and their associated transition probability. Probability $1$ is not displayed. Green highlight depicts the variations among the participants and shows how \sysname adapts itself to these variations. Each column includes some representatives for that state space.}
  \vspace{-2mm}
  \label{fig:statespace}
\end{figure*}

Figure \ref{fig:statespace} presents state space models for representative participants in the \sysname evaluation for visualization. Each circle represents a state with an arrow presenting the optimal action for that state. For some actions, the transitioned state might be more than one state. The transition probability for each action is provided in each action arrow, otherwise, the transition probability is $1$. For instance, for participant $6$, action $a_5$ transitions state $s_8$ to either state $s_6$ or state $s_4$ with the equal probability of $0.5$.

Different humans may react differently to the same state. Figure \ref{fig:statespace} highlights the adaption capability of the \sysname by presenting the differences among the state spaces of the participants in green highlights. For instance, state space for participant $6$ shows that in the state $s_5$ the optimal action is $a_1$ (highlighted in green). On the other hand, the optimal action in the state $s_5$ for participants $4$ and $9$ is action $a_3$. This difference depicts the adaptability of the \sysname  to different humans. 

We also quantified the improvement in the learning experience of the participants. To quantify the learning improvement, we measured the overall improvement in the quiz results and state transitions. 
At each stage, taken action transitions \sysname to the next state. The next state can be either a better state (with a positive reward) or a worse state (with a negative reward), depending on the current state. Quantification of the state transition and quiz performance is as explained in Section~\ref{sec:reward}. For $5$ stages of the experiment, $5$ state transitions receive a transition reward each. Also, each stage receives a quiz performance reward. The baseline (quiz score, $LS$, $DS$, and $SSQ$) is compared with the new states to evaluate the  \sysname performance. For instance, participant $1$ started the experiment with the baseline state $s_8$ and after one stage of the experiment, it transitioned to state $s_4$. Transitioning from state $s_8$ to state $s_4$ receives a negative reward of $-40$ as explained in Section~\ref{sec:reward}. Also, participant one's performance on the quiz decreased by $20\%$ which receives the quiz performance reward of $-20$.

The last column of Table~\ref{tbl:state.action} illustrates the overall performance of the \sysname for each participant. For instance, for participant $1$, the ``Improvement'' indicates that at the end of the experiment, the overall learning performance of this particular participant increased by $21\%$. This value is compared with the baseline state (state and quiz score) to calculate the improvement. 
As Table~\ref{tbl:state.action} illustrates, on average \sysname improves the overall performance by $26\%$ over the participants' population.

\begin{table*}[!t]
 \small
  \centering
  \caption{Optimal action for different states for $15$ participants. For each state, each row presents the optimal action. The experiment includes one session for baseline (to determine the initial state) and then five stages where at each stage, $LS$, $DS$, and $SSQ$ are calculated to determine the current state. }
  \label{tbl:state.action}
  \begin{tabular}{|c | c  c c c c c c c c|}
\toprule
 \textbf{Participant} & \textbf{State1} & \textbf{State2} & \textbf{State3} & \textbf{State4} & \textbf{State5} & \textbf{State6} & \textbf{State7} & \textbf{State8} & \textbf{Improvement} \\\cline{1-10}
{\textbf{P1}}   
	 &  $a_1/s_8$   & $a_2/s_7$  & $a_3/s_8$ & $a_2/s_8$  & $a_3/s_8$ & $a_4/(s_8|s_7)$ &  $a_3/s_8$ &  $a_5/(s_6|s_4)$ & $21\%$ \\ \hline
{\textbf{P2}} 
	 &  $a_1/s_8$   & $a_2/s_7$  & $a_3/s_8$ & $a_2/s_8$  & $a_3/s_8$ & \cellcolor{blue!20}$a_4/(s_8|s_7)$ &  $a_3/s_8$ &  $a_5/(s_6|s_4)$ & $27\%$ \\ \hline
{\textbf{P3}} 
	 &  $a_1/s_8$   & \cellcolor{green!25}$a_2/s_7$  & $a_3/s_8$ & $a_2/s_8$  & $a_3/s_8$ & $a_4/(s_8|s_7)$ &  $a_3/s_8$ &  $a_5/(s_6|s_4)$  & $24\%$ \\ \hline
{\textbf{P4}}  
	 &  $a_1/s_8$   & $a_1/s_8$  & $a_3/s_8$ & $a_2/s_8$  & $a_3/s_8$ & $a_4/(s_8|s_7)$ &  $a_3/s_8$ &  $a_5/(s_6|s_4)$ & $24\%$ \\ \hline
{\textbf{P5}}  
	 &  $a_1/s_8$   & $a_2/s_7$  & $a_3/s_8$ & $a_2/s_8$  & $a_3/s_8$ & $a_4/(s_8|s_7)$ &  $a_3/s_8$ &  $a_5/(s_6|s_4)$ & $26\%$ \\ \hline
{\textbf{P6}}  
	 &  $a_1/s_8$   & $a_2/s_7$  & $a_3/s_8$ & $a_2/s_8$  & $a_1/s_8$ & $a_4/(s_8|s_7)$ &  $a_3/s_8$ &  $a_5/(s_6|s_4)$  & $19\%$ \\ \hline
{\textbf{P7}}  
	 &  $a_1/s_8$   & $a_2/s_7$  & $a_3/s_8$ & $a_4/(s_7|s_8)$  & $a_3/s_8$ & $a_2/(s_8|s_7)$ &  $a_3/s_8$ &  $a_5/(s_6|s_4)$  & $26\%$ \\ \hline
{\textbf{P8}} 
	 &  $a_1/s_8$   & $a_2/s_7$  & $a_3/s_8$ & $a_2/s_8$  & $a_3/s_8$ & $a_4/(s_8|s_7)$ &  $a_3/s_8$ &  $a_5/(s_6|s_4)$  & $31\%$ \\ \hline
{\textbf{P9}}  
	 &  $a_1/s_8$   & $a_1/s_8$  & $a_3/s_8$ & $a_2/s_8$  & $a_3/s_8$ & $a_4/(s_8|s_7)$ &  $a_3/s_8$ &  $a_5/(s_6|s_4)$ & $24\%$ \\ \hline
{\textbf{P10}}  
	 &  $a_1/s_8$   & $a_2/s_7$  & $a_1/s_8$ & $a_2/s_8$  & $a_3/s_8$ & $a_4/(s_8|s_7)$ &  $a_3/s_8$ &  $a_5/(s_6|s_4)$ & $35\%$ \\ \hline
{\textbf{P11}}  
	 &  $a_1/s_8$   & $a_2/s_7$  & $a_3/s_8$ & $a_2/s_8$  & $a_3/s_8$ & $a_4/(s_8|s_7)$ &  $a_3/s_8$ &  $a_5/(s_6|s_4)$ & $26\%$ \\ \hline
{\textbf{P12}} 
	 &  $a_1/s_8$   & $a_1/s_8$  & $a_3/s_8$ & $a_2/s_8$  & $a_1/s_8$ & $a_4/(s_8|s_7)$ &  $a_3/s_8$ &  $a_5/(s_6|s_4)$ & $22\%$ \\ \hline
{\textbf{P13}}   
	 &  $a_1/s_8$   & $a_2/s_7$  & $a_1/s_8$ & $a_4/(s_7|s_8)$  & $a_3/s_8$ & $a_4/(s_8|s_7)$ &  $a_3/s_8$ &  $a_5/(s_6|s_4)$ & $29\%$ \\ \hline
{\textbf{P14}}  
	 &  $a_1/s_8$   & $a_2/s_7$  & $a_3/s_8$ & $a_2/s_8$  & $a_3/s_8$ & $a_4/(s_8|s_7)$ &  $a_3/s_8$ &  $a_5/(s_6|s_4)$ & $25\%$ \\ \hline
{\textbf{P15}} 
	 &  $a_1/s_8$   & $a_2/s_7$  & $a_3/s_8$ & $a_2/s_8$  & $a_3/s_8$ & $a_2/(s_8|s_7)$ &  $a_3/s_8$ &  $a_5/(s_6|s_4)$ & $29\%$ \\\hline
{\textbf{Average}} 	 & - &-&-&-&-&-&-&-& $26\%$\\ \hline
 \end{tabular}

\end{table*}

\subsection{Deployment and Execution Timing Analysis}
The DSTCLN models were trained on a MacBook with a $2.3$ GHz $8$-Core Intel Core i$9$ M$1$ processor. The trained models are then deployed and executed on an edge prototyping platform. In our evaluation, we used an ARMv8-based prototyping platform to ensure the scalability of the \sysname in terms of time, memory, and power consumption.

To collect the EEG signal and process it at run-time on the edge efficiently, we use Cortex API~\cite{cortex} which is provided by EMOTIV for brain-computer interactions and is available for macOS, Android, iOS, Windows, and various microcontrollers.

To accommodate the edge device constraints, our implementation is designed to process a 4-second window of the EEG data to infer the drowsiness and the learning state at run-time from the trained DSTCLN models as explained in Sections~\ref{sec:drowsy} and~\ref{sec:wcstcalss}, respectively. The classification results of multiple 4-second windows are then collected for $10$ minutes, then we take the majority vote as the final state inference. After we determine the learning state, drowsiness, and SSQ states on the edge device (ARMv8-based prototyping platform), the tuple ($LS$, $DS$, $SSQ$) state is shared with the cloud that runs the RL agent to select the proper adaptation action as depicted in Figure~\ref{fig:policy}.

\paragraph{\textbf{Memory footprint}}
In terms of memory footprint, the trained DSTCLN models for drowsiness and learning states occupy $\approx$ $100$ MB of memory of the ARMv8-based prototyping platform. We utilized SLOCCount~\cite{SLOCCount} for analyzing significant lines of code (SLOC) for the deployed DSTCLN models (written in $C++$) on the microcontroller. It generated an output of $118$ SLOC in total for both the deployed DSTCLN models for drowsiness and learning states. Moreover, as explained in Section~\ref{sec:sysimpl}, we only keep a 4-second window of EEG data to use for the DSTCLN models. The size of a 4-second window is approximately $2$kB (256 sampling rate and 16 bits).

\paragraph{\textbf{Execution time}}
The execution path starts by sending the EEG data from EMOTIV to the ARMv8-based prototyping platform to infer the human state. The human state is then shared with the cloud server to take the proper action using the RL policy. As explained in Section~\ref{sec:sysimpl}, the EEG data for the 4-second window are passed through the DSTCLN models to infer the drowsiness state and learning state. We collect these states for $10$ minutes and then use majority voting to infer the final human state at the end of the $10$ minutes period. This final state is the one shared with the cloud server. To ensure the responsiveness of the system, during the last 4-second window in the $10$ minutes period, we start to share the final inferred state with the cloud server, where the \sysname RL policy is executed to choose the appropriate adaptation action. Hence, the last 4-second window is not used in the majority voting.

Accordingly, the main computation task that executes on the ARMv8-based prototyping platform is executing the already trained DSTCLN models, which takes approximately $1$ and $1.15$ seconds on average to infer the drowsiness and learning states, respectively. In terms of communication with the ARMv8-based prototyping platform, EMOTIV shares the EEG data via Bluetooth version 4.2 and the size of the data is approximately $2$kB (256 sampling rate, 16 bits, and 4 seconds duration). For a Bluthooth type 4.2 it takes $16$ milliseconds to transfer $2$kB of data. 
On the cloud server which is a MacBook with a $2.3$ GHz $8$-Core Intel Core i$9$ M$1$ processor, it takes $0.12$ second to run the \sysname RL policy to select the proper action. Since each window duration is a 4-second window, this provides enough time for the communication, EEG processing on the edge device, and RL policy on the cloud server to finish their execution.

\paragraph{\textbf{Power consumption}}
ARMv8 is engineered for low power consumption, typically eliminating the need for heat sinks, which makes it a viable choice for edge implementations. Using ARMv8-based prototyping platform, \sysname consumes less than $75$~mW of power on average for processing each 4-second window of EEG signal during the execution of the DSTCLN models.

\section{Discussion and Future Work}\label{sec:disc}

In this paper, we proposed an IoT system for a human-in-the-loop learning environment. We tackled the problem of learning state monitoring by exploiting two approaches of concept learning theory (rule-based learning and explanation-based learning) and showed that it is possible to infer the learning state at run-time using wearable devices by decoding the EEG signal. Furthermore, we integrated this learning state as a sensor modality in the learning environment to provide a personalized real-time adaptation using reinforcement learning to improve the learning state of a human. Indeed, \sysname can be extended to address more challenges in designing an IoT learning system. Below we list our future work to extend \sysname.

\paragraph{\textbf{False Negatives in Detecting Learning State}}
Concept learning perception may vary between individuals. It has been known that an individual's perception of learning influences their motivation and capability to learn. For example, if during a presentation a human believes they learned the material, it will affect their motivation and capability to learn the rest of the presentation~\cite{postman1962perception}. This could signal a false negative in detecting the learning state. In particular, the $LS$ will be classified as ``not-learning'' while it should be ``learning''. In \sysname, we address this by using a quiz that can ensure that the human actually learned the material. Indeed, this can be extended by fusing more sensor modalities to hinder these false negatives.

\paragraph{\textbf{Generalization of the Learning Measurement}}
To generalize \sysname, we need to consider the factors affecting the system. Different factors can affect human concept learning, including age, and health conditions. 
For instance, on the one hand, younger people may comfortably use VR devices, and wearing the device may improve their concept learning performance. On the other hand, wearing VR devices by adults may increase their drowsiness and simulator sickness, leading to a worse learning state. The other factor that can influence learning ability is pre-health conditions. For instance, neurodivergent populations may require different learning environments and adaptations. Considering these factors requires more samples from each of these groups and studying the proper adaptation actions in the learning environment that these groups may prefer.

\paragraph{\textbf{Privacy Concerns}}
The brain's frontal lobe is responsible for several essential tasks, such as cognitive functions, voluntary movement or activity, consciousness, memory, attention, and motivation. Collecting and decoding the EEG data from the frontal lobe (channels $F3$, $Fz$, and $F4$) might reveal several other critical states of the participant, which brings up the privacy problems~\cite{debie2020privacy,xiao2019can}. While in \sysname, we shared the high-level state ($LS$, $DS$, and $SSQ$) instead of sharing the raw EEG data with the RL adaptation engine in the cloud, sharing the high-level states such as $LS$ can still be invasive. Moreover, even if the high-level states are protected, spyware monitoring the actions taken by the adaptation engine can still be used to infer the private states~\cite{elmalaki2019spycon, elmalaki2022vindico}. We can extend \sysname including the trade-off between the utility of learning adaptation and the private state of the brain as part of our future work~\cite{taheri2023}.

\paragraph{\textbf{Fairness Concerns}}
In this paper, we described a learning environment where the adaptation action is personalized. However, we may have multiple humans sharing the same learning environment such as workforce training. When multiple humans share the same learning environment where one adaptation policy is applied to all of them, a fairness concern may arise. In particular, a system that monitors the aggregate learning performance across multiple humans to provide one adaptation to the learning environment may suffer from the ``Matthew effect'', which is summarized as the rich get richer and the poor get poorer~\cite{bol2018matthew}. One approach to address this is to integrate a fairness constraint in \sysname for the multi-human learning environment through minimizing the covariance between the learning rates across all the humans instead of considering the aggregate learning performance~\cite{elmalaki2021fair, zhao2023fairo}.

\paragraph{\textbf{WCST Limitations}}
While the WCST has been a valuable tool for assessing prefrontal-lobe functioning and executive functions, it is important to recognize that it may not provide as specific insights into these cognitive processes as initially believed~\cite{poreh2012cleveland,miles2021considerations}. The WCST involves multiple neural circuits and may not fully capture the nuanced aspects of prefrontal-lobe functioning. Despite these limitations, we opted to use the WCST due to its established utility and the available data on its performance. Nevertheless, future research endeavors may benefit from considering alternative assessment tools and methodologies such as Tower of Hanoi~\cite{kim2016performance,bustini1999tower,mitani2022brain} to gain a more comprehensive understanding of prefrontal-lobe functioning and executive functions.

\paragraph{\textbf{Applications Beyond Education}} The potential of \sysname extends beyond the domain of education and personal state detection systems. Our IoT system can find applications in various systems, including neuro-recommendation systems. By adapting the concept learning theory-based approach and EEG signal decoding to different contexts, such as recommendation systems, we can enable more personalized and effective user experiences~\cite{panda2023eeg,chang2017personalized}. Additionally, exploring the integration of \sysname in other personal state detection systems opens up opportunities for enhancing well-being, productivity, and user satisfaction in a wide range of applications such as personalized medicine~\cite{arns2012eeg} and health monitoring~\cite{peng2021personalized}.

\section{Conclusion}\label{conc}
Understanding how the human brain learns new concepts by decoding the EEG signals from wearable technology can open the gate to many applications in education, workforce training, and human-machine symbiosis. These applications can incorporate human learning as another sensor modality to provide a personalized learning experience. In this paper, we proposed \sysname, a human-in-the-loop IoT learning system built upon insights from concept-learning theory and exploited the EEG signals to adapt to the learning environment. \sysname was evaluated across participants in a learning environment and showed that by using the brain signals as a sensor modality to infer the human learning state and providing personalized adaptation to the learning environment, the participants' learning performance increased by $26.39\%$ on average. We implemented \sysname on an edge-based prototype. Evaluation of hardware shows that the proposed IoT system can be implemented for devices such as ARM$v8$ with a minimum RAM of $100$ MB.

\appendices

\section{DSTCLN Model for Learning}\label{sec:DSTCLNlearning}
We tuned $2$ DSTCLN models for learning and drowsiness classification tasks. For learning classification, the input data is the TF image extracted from the $4$s window of the EEG data. In the DSTCLN, the spectro-temporal CNN architecture included a hierarchical CNN divided into five convolutional blocks to extract high-level features. We tuned this model and used a batch size of $16$, filter size $5\times 5$,  stride size of $2$, dropout ratio of $0.25$, and maximum pooling. Exponential linear units (ELUs) are applied as activation functions in the convolutional blocks.

\section{DSTCLN Model for Drowsiness}\label{sec:DSTCLNdrowsy}
The second DSTCLN model was tuned for drowsiness state classification. The input data was composed of the TF image extracted from a $4s$ window of the EEG data. 
In the DSTCLN, the spectro-temporal CNN architecture included a hierarchical CNN divided into five convolutional blocks to extract high-level features. Each convolutional block includes one batch normalization layer and two convolutional layers. The final model architecture used a batch size of $32$,  filter size $5\times 5$,  stride size of $1$, dropout ratio of $0.5$, and maximum-pooling worked best. Exponential linear units (ELUs) are applied as activation functions in convolutional blocks. For the rest of the architecture, we used the same Bi-LSTM network structure with $4$ layers with $256$ hidden units and one dropout layer. Other details of the model are explained in~\cite{jeong2019classification}.   

To train this model, we conducted an experiment and collected the EEG data of $5$ participants watching a $2D$ presentation which lasted for $50$ minutes~\cite{50min}. We used the Karolinska Sleepiness Scale ($KSS$) as ground truth to label the data for the level of drowsiness~\cite{shahid2011karolinska}. $KSS$ includes a nine-point scale ranging from 1 (Extremely alert) to 9 (Extremely sleepy). Before beginning the presentation, participants are asked to rate their drowsiness level based on the $KSS$. Every $10$ minute of the presentation, the participants are asked to choose from the $KSS$ to update the drowsiness reference. The participants' labeling serves as the ground truth to infer the drowsiness level. We divided the $KSS$ $1-6$ and $7-9$ as $Alert (1)$ and $Drowsy (0)$, respectively. Then, we used the designed DSTCLN model to classify the EEG data into these two classes.

\section{RL Hyperparameters}\label{sec:hyper}
In Eq.\ref{eq:qupdate}, hyperparameters $\gamma$ and $\alpha$ are known as the discount factor and the learning step size, respectively. 

\begin{equation}\label{eq:qupdate}
\setlength\abovedisplayskip{0pt}
\belowdisplayshortskip=-1pt
Q(s, a) \leftarrow Q(s, a) +  \alpha[r(s,a) + \gamma \max_a Q(s', a) - Q(s, a)] \vspace{-2mm}
\end{equation}

To choose an action, $a$ at each state $s$ from the possible set of actions, an $\epsilon$-greedy policy can be adopted. In the $\epsilon$-greedy policy, the RL agent chooses the action that it believes has the best long-term effect with probability $1$ - $\epsilon$, which is the maximum value of $Q(s, a)$, and with probability $\epsilon$, it picks an action uniformly at random.
This hyperparameter $\epsilon$ (the exploration vs. exploitation parameter) controls how much the RL agent is willing to explore new actions that were not taken before versus relying on the best action learned. By updating $Q(s, a)$, it is guaranteed that the optimal policy $\pi$ will converge to a deterministic action $a$ per state $s$ that provides the maximum reward $r(s, a)$ in a finite time steps $T$~\cite{sutton2018reinforcement}.   In this study, every time \texttt{q\_value} is updated, we gradually lower $\epsilon$ following an exponential decay of $0.01$.

In Equation~\ref{eq:qupdate}, $\gamma$ determines how much the RL agent cares about rewards it receives in the distant future relative to the immediate reward. In our design, a low discount factor $\gamma=0.001$ is selected. 
Also, in Equation~\ref{eq:qupdate}, $\alpha$ (learning rate) is a hyperparameter that controls how much an agent updates its estimates of the optimal policy or value function in response to new experiences or data.
A low learning rate means the agent will be slow to adapt to new information, while a high learning rate may result in unstable convergence or overfitting to noisy data. The optimal value of $\alpha$  may vary depending on the specific problem and is often determined through experimentation and tuning. In this study, we chose $\alpha = 0.05$.

\section*{Acknowledgment}
This research was partially supported by NSF awards CNS-2105084 and CCF-2140154. Any opinions, findings, conclusions, or recommendations expressed in this material are those of the authors and do not necessarily reflect the views of our funding agency.

\ifCLASSOPTIONcaptionsoff
  \newpage
\fi

\bibliographystyle{IEEEtran}
\bibliography{sample-base}

\begin{IEEEbiographynophoto}{Mojtaba Taherisadr} 
received his master’s degree in
computer engineering from the University of Central Florida, Orlando, Florida, USA. He is currently pursuing a Ph.D. degree in computer engineering
with the University of California at Irvine, CA, USA.
His research is focused on the design and analysis of personalized context-aware machine learning models for healthcare systems with a focus on performance, privacy, and adaptability.
\end{IEEEbiographynophoto}

\begin{IEEEbiographynophoto}{Mohammad Abdullah Al Faruque} (Senior Member, IEEE) received a Ph.D. degree in computer science from Karlsruhe Institute of Technology, Karlsruhe, Germany, in 2009. He is currently with the University of California at Irvine, Irvine, CA, USA, as a Full Professor and Directing the Embedded and Cyber–Physical Systems Lab. His current research is focused on the system-level design of embedded and cyber–physical systems (CPS) with a special interest in low-power design, CPS security, and data-driven CPS design. Dr. Al Faruque has received four Best Paper awards (ACSAC 2022, DATE 2016, DAC 2015, and ICCAD 2009) and many Best Paper award nominations. Dr. Al Faruque has been an IEEE CEDA Distinguished Lecturer since 2022 and an ACM Senior Member.
\end{IEEEbiographynophoto}

\begin{IEEEbiographynophoto}{Salma Elmalaki} received a Ph.D. degree in electrical and computer engineering from the University of California, Los Angeles, in 2018. She is currently with the University of California at Irvine, CA, USA, as an Assistant Professor and directing the Pervasive Autonomy Lab. Her current research is focused on cyber-physical systems (CPS) with a special focus on human adaptability, fairness, and privacy guarantees. Dr. Elmalaki is the recipient of the Best Paper award and Best Community Paper award at MobiCom'15. 
\end{IEEEbiographynophoto}

\vfill

\end{document}